\documentclass[twocolumn,tighten]{aastex63}

\usepackage{color, hyperref}
\usepackage[normalem]{ulem}
\usepackage{amsmath}
\usepackage{relsize}
\usepackage{CJKutf8}

\DeclareMathAlphabet\mathbfcal{OMS}{cmsy}{b}{n} 

\newcommand{\sn}[2]{\ensuremath{#1 \times 10^{#2}}} 

\newcommand{\Normal}[2]{\ensuremath{\mathcal{N}\left[{#1}, {#2} \right]}} 

\newcommand{\meanwrt}[2]{\ensuremath{\mathbb{E}_{{#2}}\left[{#1}\right]}}

\newcommand{\params}{\ensuremath{\boldsymbol\theta}}
\newcommand{\eparams}{\ensuremath{\boldsymbol\phi}}
\newcommand{\flux}{\ensuremath{\mathbf{F}}}
\newcommand{\mags}{\ensuremath{\mathbf{m}}}
\newcommand{\absmag}{\ensuremath{\mathbf{M}}}
\newcommand{\errors}{\ensuremath{\boldsymbol\sigma}}
\newcommand{\feh}{\ensuremath{{\rm [Fe/H]}}}
\newcommand{\afe}{\ensuremath{{\rm [\alpha/Fe]}}}
\newcommand{\rvec}{\ensuremath{\mathbf{R}}}
\newcommand{\dvec}{\ensuremath{\mathbf{D}}}

\newcommand{\data}{\ensuremath{\mathbf{D}}}

\newcommand{\likelihood}{\ensuremath{\mathcal{L}}}
\newcommand{\prior}{\ensuremath{\pi}}

\newcommand{\cov}{\ensuremath{\mathbf{C}}}
\newcommand{\meanvec}{\ensuremath{\boldsymbol{\mu}}}
\newcommand{\T}{\ensuremath{\mathrm{T}}}

\usepackage{graphicx}
\DeclareGraphicsExtensions{.png,.pdf}
\graphicspath{{./}{figures/}}

\newcommand{\ms}{\textsc{Minesweeper}}
\newcommand{\brutus}{\textsc{brutus}}
\newcommand{\starhorse}{\textsc{StarHorse}}

\newcommand{\python}{\textsc{Python}}
\newcommand{\astropy}{\textsc{astropy}}
\newcommand{\healpy}{\textsc{healpy}}

\newcommand{\dustmaps}{\textsc{dustmaps}}
\newcommand{\mesa}{\textsc{mesa}}
\newcommand{\atlas}{\textsc{ATLAS12}}
\newcommand{\synthe}{\textsc{SYNTHE}}
\newcommand{\pytorch}{\textsc{PyTorch}}
\newcommand{\scipy}{\textsc{SciPy}}
\newcommand{\numpy}{\textsc{NumPy}}
\newcommand{\matplotlib}{\textsc{matplotlib}}
\newcommand{\hdbscan}{\textsc{hdbscan}}
\newcommand{\corner}{\textsc{corner}}

\newcommand{\bayestarmap}{\texttt{Bayestar19}}
\newcommand{\mist}{\texttt{MIST}}
\newcommand{\bayestar}{\texttt{Bayestar}}
\newcommand{\ctk}{\texttt{C3K}}

\shorttitle{Deriving Stellar Properties with {\brutus}}
\shortauthors{Speagle et al.}
\begin{document}

\title{Deriving Stellar Properties, Distances, and Reddenings
using Photometry and Astrometry with BRUTUS}

\author[0000-0003-2573-9832]{Joshua S. Speagle}
\affiliation{Department of Statistical Sciences, University of Toronto, Toronto, ON M5S 3G3, Canada}
\affiliation{David A. Dunlap Department of Astronomy \& Astrophysics, University of Toronto, Toronto, ON M5S 3H4, Canada}
\affiliation{Dunlap Institute for Astronomy \& Astrophysics, University of Toronto, Toronto, ON M5S 3H4, Canada}
\affiliation{Data Sciences Institute, University of Toronto, 17th Floor, Ontario Power Building, 700 University Ave, Toronto, ON M5G 1Z5, Canada} 
\affil{Center for Astrophysics\:\textbar\:Harvard \& Smithsonian, 
60 Garden St., Cambridge, MA, USA 02138}
\email{j.speagle@utoronto.ca}

\author[0000-0002-2250-730X]{Catherine Zucker}
\affil{Center for Astrophysics\:\textbar\:Harvard \& Smithsonian, 
60 Garden St., Cambridge, MA, USA 02138}
\affil{Space Telescope Science Institute, 3700 San Martin Drive, Baltimore, MD 21218, USA}

\author[0000-0002-8658-1453]{Angus Beane}
\affil{Center for Astrophysics\:\textbar\:Harvard \& Smithsonian, 
60 Garden St., Cambridge, MA, USA 02138}

\author[0000-0002-1617-8917]{Phillip A. Cargile}
\affil{Center for Astrophysics\:\textbar\:Harvard \& Smithsonian, 
60 Garden St., Cambridge, MA, USA 02138}

\author[0000-0002-4442-5700]{Aaron Dotter}
\affil{Center for Astrophysics\:\textbar\:Harvard \& Smithsonian, 
60 Garden St., Cambridge, MA, USA 02138}

\author[0000-0003-2808-275X]{Douglas P. Finkbeiner}
\affil{Center for Astrophysics\:\textbar\:Harvard \& Smithsonian, 
60 Garden St., Cambridge, MA, USA 02138}
\affil{Department of Physics, 
Harvard University, 17 Oxford St, Cambridge, MA 02138}

\author[0000-0001-5417-2260]{Gregory M. Green}
\affil{Max-Planck-Institut f{\"u}r Astronomie, K{\"o}nigstuhl 17, 
D-69117 Heidelberg, Germany}

\author[0000-0002-9280-7594]{Benjamin D. Johnson}
\affil{Center for Astrophysics\:\textbar\:Harvard \& Smithsonian, 
60 Garden St., Cambridge, MA, USA 02138}

\author[0000-0002-3569-7421]{Edward F. Schlafly}
\affil{Lawrence Livermore National Laboratory, 
7000 East Avenue, Livermore, CA 94550, USA}
\affil{Space Telescope Science Institute, 3700 San Martin Drive, Baltimore, MD 21218, USA}

\author[0000-0002-7846-9787]{Ana Bonaca}
\affil{Center for Astrophysics\:\textbar\:Harvard \& Smithsonian, 
60 Garden St., Cambridge, MA, USA 02138}
\affil{The Observatories of the Carnegie Institution for Science, 813 Santa Barbara Street, Pasadena, 91101, CA, USA}

\author[0000-0002-1590-8551]{Charlie Conroy}
\affil{Center for Astrophysics\:\textbar\:Harvard \& Smithsonian, 
60 Garden St., Cambridge, MA, USA 02138}

\author[0000-0003-3734-8177]{Gwendolyn Eadie}
\affiliation{Department of Statistical Sciences, University of Toronto, Toronto, ON M5S 3G3, Canada}
\affiliation{David A. Dunlap Department of Astronomy \& Astrophysics, University of Toronto, Toronto, ON M5S 3H4, Canada}

\author[0000-0002-2929-3121]{Daniel J. Eisenstein}
\affil{Center for Astrophysics\:\textbar\:Harvard \& Smithsonian, 
60 Garden St., Cambridge, MA, USA 02138}

\author[0000-0003-1312-0477]{Alyssa A. Goodman}
\affil{Center for Astrophysics\:\textbar\:Harvard \& Smithsonian, 
60 Garden St., Cambridge, MA, USA 02138}
\affil{Radcliffe Institute for Advanced Study, 
Harvard University, 10 Garden St, Cambridge, MA 02138}

\author[0000-0002-6800-5778]{Jiwon Jesse Han}
\affil{Center for Astrophysics\:\textbar\:Harvard \& Smithsonian, 
60 Garden St., Cambridge, MA, USA 02138}

\author[0000-0001-5625-5342]{Harshil M. Kamdar}
\affil{Center for Astrophysics\:\textbar\:Harvard \& Smithsonian, 
60 Garden St., Cambridge, MA, USA 02138}


\author[0000-0003-3997-5705]{Rohan Naidu}
\affil{Center for Astrophysics\:\textbar\:Harvard \& Smithsonian, 
60 Garden St., Cambridge, MA, USA 02138}
\affil{Kavli Institute for Astrophysics and Space Research, Massachusetts Institute of Technology, 70 Vassar Street, Cambridge, MA 02139, USA}

\author[0000-0003-4996-9069]{Hans-Walter Rix}
\affil{Max-Planck-Institut f{\"u}r Astronomie, K{\"o}nigstuhl 17, 
D-69117 Heidelberg, Germany}

\author[0000-0002-6561-9002]{Andrew K. Saydjari}
\affil{Department of Physics, 
Harvard University, 17 Oxford St, Cambridge, MA 02138}
\affil{Center for Astrophysics\:\textbar\:Harvard \& Smithsonian, 
60 Garden St., Cambridge, MA, USA 02138}
\affil{Department of Astrophysical Sciences, Princeton University, Princeton, NJ 08544, USA}

\author[0000-0001-5082-9536]{Yuan-Sen Ting 
(\begin{CJK*}{UTF8}{bsmi}丁源森\ignorespacesafterend\end{CJK*})}
\affil{Institute for Advanced Study, Princeton, NJ 08540, USA}
\affil{Department of Astrophysical Sciences, Princeton University, 
Princeton, NJ 08544, USA}
\affil{Observatories of the Carnegie Institution of Washington, 
813 Santa Barbara Street, Pasadena, CA 91101, USA}
\affil{Research School of Astronomy and Astrophysics, 
Australian National University, Cotter Road, ACT 2611, Canberra, Australia}
\affil{Research School of Computer Science, 
Australian National University, Acton ACT 2601, Australia}
\affil{Department of Astronomy, The Ohio State University, Columbus, OH 43210, USA}

\author[0000-0002-7588-976X]{Ioana A. Zelko}
\affil{Center for Astrophysics\:\textbar\:Harvard \& Smithsonian, 
60 Garden St., Cambridge, MA, USA 02138}
\affil{Canadian Institute for Theoretical Astrophysics, University of Toronto, 60 St George Street, Toronto, M5S 3H8, Ontario, Canada}
\affil{Department of Physics and Astronomy, University of California-Los Angeles, 475 Portola Plaza, Los Angeles, CA 90095, USA}

\begin{abstract}
We present {\brutus}, an open source {\python} 
package for quickly deriving stellar properties,
distances, and reddenings to stars based on grids of stellar models
constrained by photometric and astrometric data.
We outline the statistical framework for deriving these quantities, 
its implementation, and various Galactic priors over the 3-D distribution of
stars, stellar properties, and dust extinction (including $R_V$ variation). 
We establish a procedure to empirically calibrate {\mist} v1.2 isochrones by using
open clusters to derive corrections to the effective temperatures
and radii of the isochrones, which reduces systematic errors on the lower main sequence.
We also describe and apply a method to estimate photometric offsets between
stellar models and observed data using nearby, low-reddening field stars.
We perform a series of tests on mock and real data to examine
parameter recovery with {\mist} under different modeling assumptions,
illustrating that {\brutus} is able to recover distances and
other stellar properties using optical to near-infrared photometry and astrometry.
The code is publicly available at \url{https://github.com/joshspeagle/brutus}.
\end{abstract}


\section{Introduction} \label{sec:intro}

One of the central challenges in Galactic astronomy is to convert
the projected 2-D positions of sources on the sky 
into 3-D maps that we can use to infer properties about the
Milky Way. This challenge has only accelerated in recent
years as large datasets have become publicly available
from large projects such as the ground-based
Sloan Digital Sky Survey \citep[SDSS;][]{york+00} and
the space-based \textit{Gaia} mission \citep{gaia+16}.
Together, these observational efforts promise to provide new, much sharper
maps of the stellar components of the Galaxy using billions of individual sources. 

Many recent and potential discoveries concerning the 
structure and evolution of the Milky Way depend upon reliable 3-D maps.
Past work with large photometric datasets have discovered large collections
of streams \citep[e.g.,][]{belokurov+06} and mapped out broad components of
Milky Way structure \citep[e.g.,][]{juric+08}.
More recent work has uncovered the remnants of a major 
merger $\sim 10\,{\rm Gyr}$ ago, referred to as 
``Gaia-Enceladus'' or the ``Sausage'' 
\citep[e.g.,][]{koppelman+18,belokurov+18,helmi+18,naidu+21}
and a phase-space ``spiral'' \citep[e.g.,][]{antoja+18}.

Large spectroscopic surveys of stellar chemistry will shed 
light on the role of hierarchical assembly and radial migration
in the present-day distribution of stellar populations 
\citep[e.g.,][]{roskar+08}.
In the halo, accurate phase-space maps of stellar streams 
will constrain the potential of the Galaxy 
\citep[e.g.,][]{johnston+99,lawmajewski10,bonacahogg18,greenting20}
and probe the existence of a thick dark disk expected 
in $\Lambda$CDM \citep{read+08}, 
with the latter having strong implications for the interpretation of direct 
detection experiments of dark matter \citep{read14}. 
Measuring the radial profile of the inner dark matter halo through 
maps of dynamical tracers can constrain its 
accretion history \citep{wechsler+02}.
The key to all these discoveries is the need for a robust statistical
framework to infer 3-D properties of a large number of stars.

In order to build these maps, the raw observations from 
large scale surveys need to be converted into physical quantities
such as 3-D positions and velocities, effective temperatures, surface gravities,
metallicities, $\alpha$-enhancements, masses, and ages.
Many of these quantities are reliably
estimated from spectroscopy using a combination of
empirical relations, 
theoretical stellar atmosphere models \citep[e.g., {\atlas} and
{\synthe};][]{kurucz70,kuruczavrett81,kurucz93},
or some combination of the two \citep[e.g.,][]{ness+15}.
Comparisons to the observed flux densities, combined with estimates of foreground
extinction and the properties of Galactic dust (i.e. the ``reddening''), 
then enable a measurement of distance \citep[see, e.g.,][]{green+14}.

Most sources ($\sim$ 99\%) seen in large photometric surveys, however, do not have measured
spectra. Instead, they only have spectral energy distributions (SEDs) that are
comprised of flux densities estimated across a range
of broad-band and narrow-band photometric filters. 
More recently, \textit{Gaia} DR2 \citep{gaia+18} and EDR3 \citep{gaia+21}
has also provided astrometric parallax measurements for
many of these sources, giving independent constraints on the distance.
Mapping out the Milky Way in detail and at scale thus requires effective utilization
and joint analysis of all of these datasets. Most importantly, it will require
robust modeling of stellar SEDs, which will continue to outpace the supply of
high-quality parallax measurements and far outpace the supply of even moderate-quality
spectra for the foreseeable future.

In recent years there has been extensive work towards this goal
from a wide variety of researchers in areas from 3-D dust mapping
\citep[e.g.,][]{rezaeikh+18,leikeenslin19,lallement+19,green+19,leike+20}
to stellar parameter estimation
\citep[e.g.,][]{ness+15,garciaperez+16,cargile+20,anders+19,xiang+19}.
We add to these efforts through
{\brutus}\footnote{Available online at: \url{https://github.com/joshspeagle/brutus}.}, 
a public, open source {\python} package for quickly 
and robustly deriving stellar properties, distances, and
reddenings to stars with astrometric and/or photometric data.
{\brutus} is designed to be well-documented, user-friendly, and highly
modular, with various components that can be used for 
individual stellar parameter estimation,
analysis of co-eval stellar populations, and 3-D dust mapping within
an internally-consistent statistical framework. The code has also already
been used in several publications including
\citet{zuckerspeagle+19} and \citet{zucker+20}.

This work joins other recent efforts focused on trying to estimate
distances and other stellar properties from astrometry and/or photometry
alone. \citet{bailerjones+18} published a large catalog of over a billion
distances using only astrometric data from the \textit{Gaia} DR2 
and a data-driven model. More recently, \citet{bailerjones+21}
has done the same for \textit{Gaia} EDR3 using a similar data-driven approach
to estimate distances that incorporates both astrometric and photometric \textit{Gaia} data.
An example of work that is most similar to what is presented here is the
{\starhorse} code \citep{santiago+16,queiroz+18,anders+19}, 
which similar to {\brutus} attempts
to estimate stellar parameters, reddenings, \textit{and} distances from
photometric and astrometric data using theoretical stellar models.
Differences between the two approaches will be discussed in more detail
in \S\ref{sec:implementation}.

The outline of the paper is as follows.
In \S\ref{sec:stats}, we describe the underlying Bayesian
statistical framework and modeling, 
including the initial set of Galactic priors over the 3-D distribution of
stars, dust, and related properties. 
In \S\ref{sec:implementation}, we describe the strategy {\brutus} uses
for fast exploration and characterization the probabilistic uncertainties for
a given source. 
In \S\ref{sec:models}, we describe the initial set of
empirical and theoretical stellar models used to infer stellar properties. 
In \S\ref{sec:calib}, we describe how we use both cluster and field stars
to empirically calibrate our theoretical isochrones to improve
inference, particularly at lower masses.
In \S\ref{sec:tests}, we describe a series of tests on mock and real data
used to validate the performance of the models and the code.
We conclude in \S\ref{sec:conc}.

Throughout the paper, individual parameters are denoted using standard
italicized math fonts ($\theta$) while vectors and matrices are denoted using
boldface ($\boldsymbol{\theta}$). Collections of parameters are denoted using sets
($\boldsymbol{\theta} = \{ \theta_i \}_{i=1}^{i=n}$). Vectors should be assumed to
be in column form (i.e. of shape $n \times 1$) unless explicitly stated otherwise.
We will use ``hat'' notation ($\hat{\theta}$) to define
noisy measurements of a particular quantity $\theta$.

\begin{figure*}
\begin{center}
\includegraphics[width=\textwidth]{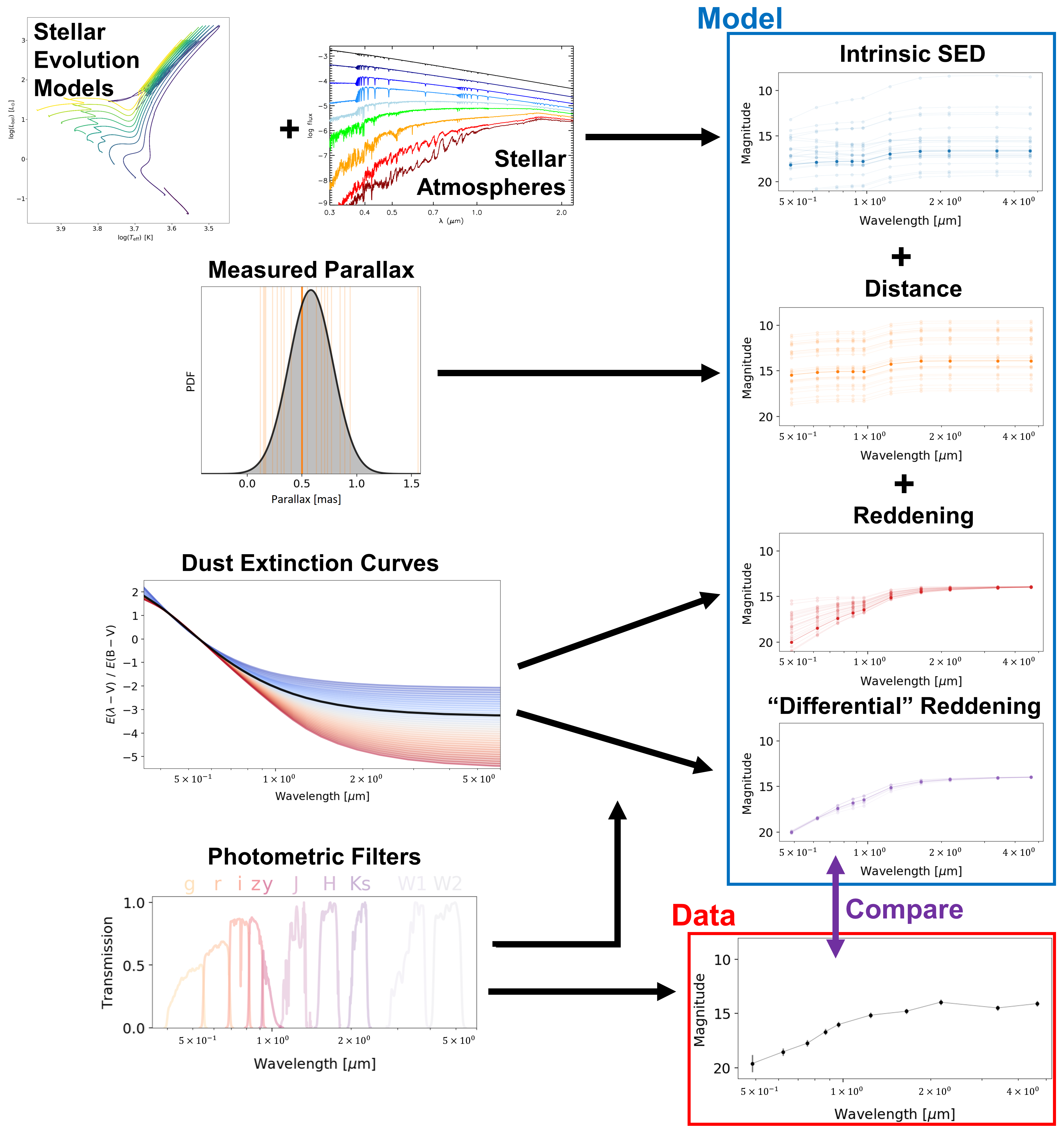}
\end{center}
\caption{An illustration of the components that go into creating
stellar spectral energy distributions (SEDs) in {\brutus}.
The intrinsic (i.e. absolute, unreddened) magnitudes 
$\absmag_{\params}$ (far top) are constructed
using a combination of intrinsic parameters $\params$
that incorporate stellar evolutionary models and
stellar atmospheric models combined with a given set of
photometric filters. These are modified by extrinsic
parameters $\eparams$ (middle) including the distance ($d$), which can
be compared to astrometric parallax measurements, and the
visual extinction ($A_V$) and ``differential'' reddening ($R_V$),
which are based on empirical dust extinction models.
These combine to give predicted magnitudes $\mags_{\params, \eparams}$ 
that can be compared to the noisy observed data $\hat{\mags}$ (bottom).
See \S\ref{subsec:noiseless_model} for additional details.
}\label{fig:overview}
\end{figure*}

\section{Statistical Framework} \label{sec:stats}

Our statistical framework is divided into four parts.
In \S\ref{subsec:noiseless_model}, 
we describe the noiseless (ideal) model
for a given source as a function of stellar parameters,
dust extinction, and distance. In \S\ref{subsec:noisy_data},
we describe our assumptions regarding 
the photometric and astrometric data and the corresponding
likelihoods. In \S\ref{subsec:posterior},
we outline the basis for combining these pieces of information
into a Bayesian posterior probability using relevant priors. 
We discuss these priors in \S\ref{subsec:prior}.

In brief, we assume that:
\begin{itemize}
    \item Our noiseless model can be described as a linear combination 
    (in magnitudes) of intrinsic stellar, dust, and distance components.
    \item The measured flux densities (i.e. photometry)
    and the parallaxes have independently and identically distributed (iid)
    Normal (i.e. Gaussian) uncertainties.
    \item The priors for a wide range of parameters
    can be separated into components involving the stellar
    initial mass function (IMF), 3-D stellar number density, 3-D metallicity
    distribution, 3-D age distribution, and 3-D dust extinction.
    \item The variation in the underlying dust extinction curve
    can be described with a linear one-parameter model.
\end{itemize}
See the subsections below for additional details.

\subsection{Noiseless Model} \label{subsec:noiseless_model}

We assume that the observed magnitudes $\mags \equiv \{ m_i \}_{i=1}^{i=b}$
over a set of $b$ photometric bands can be modeled as
\begin{equation} \label{eq:model_mag}
    \boxed{
    \mags_{\params, \eparams} 
    \equiv \absmag_{\params} + \mu + A_V \times 
    \left(\rvec_{\params} + R_V \times \rvec'_{\params}\right)
    }
\end{equation}
This contains several components:
\begin{itemize}
    \item The \textit{intrinsic absolute magnitude} $\absmag_{\params}$
    of the star as a function of its \textit{intrinsic} 
    (stellar) parameters $\params$.
    \item The \textit{distance modulus} $\mu \equiv 5 \log(d/10)$ where
    $d$ is the distance to the object in pc.
    \item The \textit{dust extinction} $A_V \equiv V_{\rm obs} - V_{\rm true}$
    in magnitudes, measured using the difference between the observed
    $V_{\rm obs}$ and true $V_{\rm true}$ magnitudes in the $V$-band.
    \item The \textit{reddening vector} $\rvec_{\params}$ 
    that determines the wavelength-dependence of extinction 
    across the $b$ filters.\footnote{The dependence on $\params$ is
    due to the changing shape of the stellar spectrum across each filter.}
    \item The \textit{differential extinction} 
    $R_V \equiv A_V / (A_B - A_V) \equiv A_V / {\rm E}(B-V)$
    in the $V$-band versus the $B$-band, where ${\rm E}(B-V)$ is often referred
    to as the color excess.
    \item The \textit{differential reddening vector} $\rvec'_{\params}$ 
    that modifies the shape of the underlying reddening vector $\rvec_{\params}$.
\end{itemize}
For compactness, we define
$\eparams$ to be the combined set of 
all \textit{extrinsic} parameters (here $\mu$, $A_V$, and $R_V$)
that modify modify the observed magnitudes to be different from $\absmag_{\params}$.
The combined effect of $\params$ and $\eparams$ then generate the observed
magnitudes $\mags_{\params, \eparams}$. We will discuss 
the logic behind separating $\params$ and $\eparams$ into these two separate
categories in \S\ref{sec:implementation}.

\begin{figure*}
\begin{center}
\includegraphics[width=\textwidth]{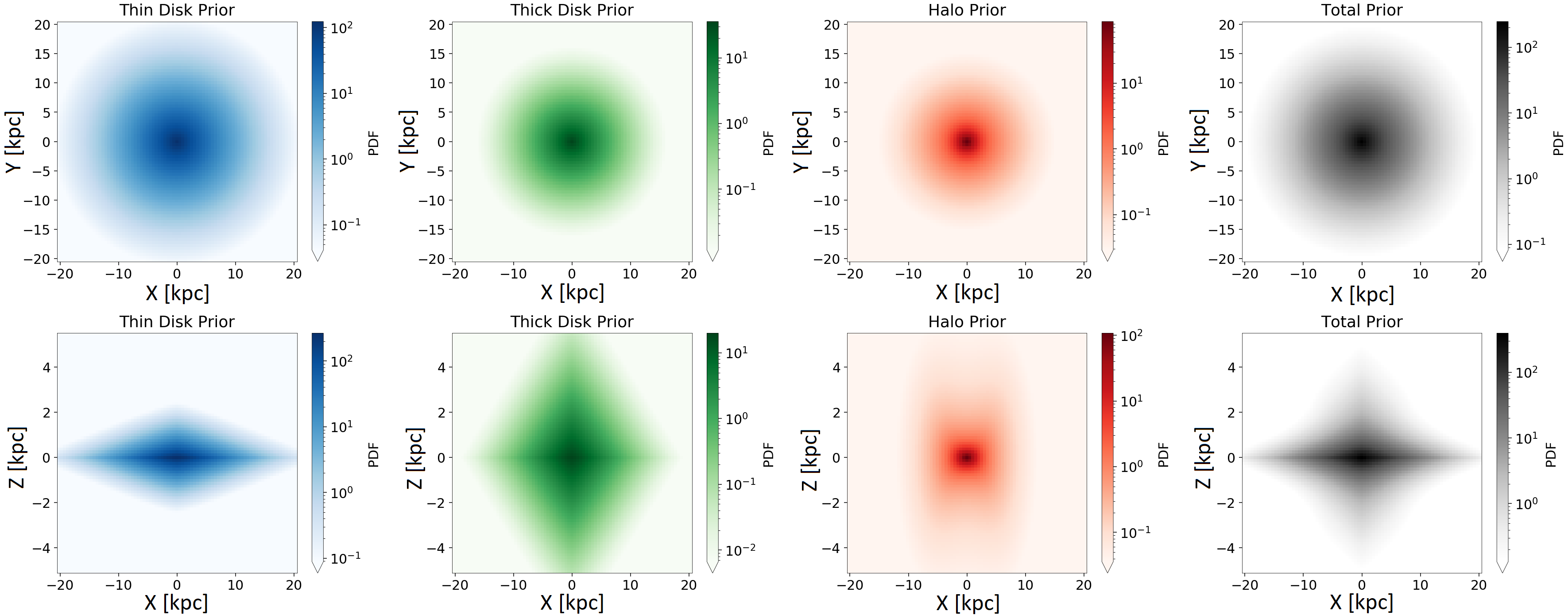}
\end{center}
\caption{The 3-D stellar number density prior used in {\brutus},
marginalized and projected into Galactocentric Cartesian 
$X$-$Y$ (top) and $X$-$Z$ (bottom) coordinates. This
is divided into thin disk (blue, far left), thick disk (green, center left),
and halo (red, center right) components along with their combined contributions
(black, far right). The effective prior (not shown here)
includes a ${\rm d}V \propto d^2$ component centered on
the Sun at $(X, Y, Z) \approx (-8, 0, 0)$ kpc 
to account for changes in the volume element,
leading to suppression/enhancement of the prior relative to the true underlying
number density for nearby/faraway sources.
See Table \ref{tab:prior} and \S\ref{subsec:prior_stars} for more details.
}\label{fig:prior_density}
\end{figure*}

\subsubsection{Intrinsic Parameters} \label{subsubec:noiseless_intrinsic}

For the majority of this paper, we define 
intrinsic (stellar) parameters $\params$ as
\begin{equation}
    \params \equiv 
    \begin{bmatrix}
    M_{\rm init} \\
    \feh_{\rm init} \\
    t_{\rm age}
    \end{bmatrix}
\end{equation}
where $M_{\rm init}$ is the initial mass (in $M_\odot$), 
$\feh_{\rm init}$ is the initial metallicity (relative to solar),
and $t_{\rm age}$ is the current age (in Gyr).
This is a vast oversimplification of stellar physics and evolution,
ignoring the contributions of stellar rotation \citep{gossage+19},
$\alpha$-process element abundance variations \citep{thomas+03},
binarity \citep{eldridge+17}, and more. However, given the limited
resolution of current, publicly-available photometric data
and the difficulty in modeling all of these processes simultaneously,
we follow recent work such as \citet{anders+19}
by approximating stellar evolution using only these three parameters.
We hope to improve on this in future work.

Theoretical stellar evolutionary tracks can relate these
intrinsic parameters $\params$ to surface-level stellar parameters
\begin{equation}
    \params_\star \equiv
    \begin{bmatrix}
    \log g \\
    \log T_{\rm eff} \\
    \log L_{\rm bol} \\
    \log R_\star \\
    \feh_{\rm surf}
    \end{bmatrix}
\end{equation}
where $g$ is the surface gravity (in cgs), $T_{\rm eff}$ is the
effective temperature (in K), $L_{\rm bol}$ is the bolometric luminosity
(in $L_\odot$), $R_\star$ is the radius (in $R_\odot$),
and $\feh_{\rm surf}$ is the surface metallicity (relative to
solar).\footnote{The present-day abundances $\feh_{\rm surf}$
differ from the \textit{initial} abundances at birth
$\feh_{\rm init}$.} $\params_\star$ can then be connected to the 
more direct observables through the use
of stellar atmospheric models \citep{kurucz70,gustafsson+08} and associated
line lists \citep{piskunov+95,kupka+00,ryabchikova+15}
that relate these parameters to corresponding spectral flux densities
$F_\nu(\lambda|\params)$ as a function of wavelength $\lambda$.

The corresponding intrinsic absolute magnitude $M_i(\params)$
in a given filter $i$ with filter transmission curve $T_i(\lambda)$ is then
\begin{equation}
    M_i(\params) \equiv -2.5 
    \log\left(\frac{\int_{0}^{\infty} F_\nu(\lambda|\params) T_i(\lambda) \lambda^{-1} d\lambda}
    {\int_{0}^{\infty} S_\nu(\lambda) T_i(\lambda) \lambda^{-1} d\lambda} \right)
\end{equation}
where $S_\nu(\lambda)$ is the spectral flux density used to normalize
the observations. In the Vega magnitude system, $S_\nu(\lambda)$ is
the spectrum of Vega, while in the AB system 
$S_\nu(\lambda) = 3631\,{\rm Jy}$ is a constant.

Combining these ingredients together then gives
us a way to generate the intrinsic magnitudes $\absmag_{\params}$:
\begin{equation*}
    \quad\:\:\: \params 
    \:\:\quad\quad \xrightarrow[]{{\rm isochrones}} \quad\quad\:
    \params_\star 
    \:\:\quad\quad \xrightarrow[{\rm filters}]{{\rm atmospheres}} \:\: 
    \absmag_{\params}
\end{equation*}
\begin{equation*}
    \begin{bmatrix}
    M_{\rm init} \\
    \feh_{\rm init} \\
    t_{\rm age}
    \end{bmatrix}
    \xrightarrow[]{{\rm isochrones}}
    \begin{bmatrix}
    \log g \\
    \log T_{\rm eff} \\
    \log L_{\rm bol} \\
    \log R_\star \\
    \feh_{\rm surf}
    \end{bmatrix}
    \xrightarrow[{\rm filters}]{{\rm atmospheres}}
    \begin{bmatrix}
    M_1 \\
    \vdots \\
    M_b
    \end{bmatrix}
\end{equation*}

\subsubsection{Extrinsic Parameters} \label{subsubec:noiseless_extrinsic}

Our intrinsic magnitudes $\absmag_{\params} = \{ M_i(\params) \}_{i=1}^{i=b}$ 
over our $b$ photometric bands are extinguished by foreground
dust with some wavelength-dependent optical depth per $A_V$, $\tau_\lambda$.
The integrated effect for a given scaling factor $A$ at a distance
$d$ is then
\begin{equation}
    m_i(\params | A, d) = -2.5 
    \log\left(\frac{d^{-2}\int_{0}^{\infty} e^{-A\tau_\lambda} 
    F_\nu(\lambda|\params) T_i(\lambda) \lambda^{-1} d\lambda}
    {\int_{0}^{\infty} S_\nu(\lambda) T_i(\lambda) \lambda^{-1} d\lambda} \right).
\end{equation}
For small $A$, we can approximate this expression as
\begin{equation}
    m_i(\params|A,d) \approx M_i(\params) + A \times R_i(\params) + \mu(d)
\end{equation}
where the reddening $R_i(\params)$ in the $i$th band is
\begin{equation}
    R_i(\params) \equiv \frac{2.5}{\ln 10} \frac{\int_{0}^{\infty} \tau_\lambda 
    F_\nu(\lambda|\params) T_i(\lambda) \lambda^{-1} d\lambda}
    {\int_{0}^{\infty} F_\nu(\lambda|\params) T_i(\lambda) \lambda^{-1} d\lambda}
\end{equation}
and we have added the distance modulus $\mu = 5 \log(d/10)$ to explicitly account
for the impact of the distance $d$ of the object.
For $A = A_V$ and $\rvec_{\params} = \{ R_i(\params) \}_{i=1}^{i=b}$,
this then becomes a simplified version of equation \eqref{eq:model_mag}.

While the above approximation
only strictly holds true for small $A_V$, it
still generally serves as a good model for
observed extinction and reddening for $A_V \gg 1$ \citep{green+14}.
In addition, while the reddening vector
$\rvec_{\params}$ clearly depends on the
underlying spectrum $F_{\nu}(\lambda|\params)$ and 
therefore will be different for each star,
many approaches further approximate the reddening vector as being
independent of the underlying spectrum
(i.e. $\rvec_{\params} = \rvec$) \citep{green+15,green+18,green+19}. 
As our ability to model the observed extinction is most often limited 
by our imperfect knowledge of $\tau_\lambda$ and its variation within the Galaxy,
this approximation is often reasonable \citep{schlafly+16, green+21}.
We will discuss this further in \S\ref{sec:models}.

It is important to note that $\tau_\lambda$ is not
universal for foreground dust. It instead depends
on a variety of intrinsic properties such as the dust grain size distribution
along a given line of sight and the relative composition of dust grains
(e.g., silicate versus carbonaceous grains). While the underlying physics
are complex \citep{draine03}, in practice it has been shown
\citep{fitzpatrick99,schlaflyfinkbeiner11,schlafly+16}
that most of this variability in the optical and 
near-infrared (NIR) can be modeled with a single parameter $R_V$ such that
\begin{equation}
    \tau_{\lambda, \rm eff}(R_V) \approx \tau_\lambda + R_V \times \tau_\lambda'
\end{equation}
where $\tau_\lambda'$ characterizes the wavelength-dependence of
``differential extinction'' as a function of $R_V$.
Adding in this term then gives the corresponding differential
reddening in the $i$th band as
\begin{equation}
    R_i'(\params) \equiv \frac{2.5}{\ln 10} \frac{\int_{0}^{\infty} \tau_\lambda' 
    F_\nu(\lambda|\params) T_i(\lambda) \lambda^{-1} d\lambda}
    {\int_{0}^{\infty} F_\nu(\lambda|\params) T_i(\lambda) \lambda^{-1} d\lambda}
\end{equation}
Letting $\rvec'_{\params} = \{ R_i'(\params) \}_{i=1}^{i=b}$ then gives
the final component of the model in equation \eqref{eq:model_mag}.
See \S\ref{subsubsec:mist_dust} for further discussion on $R_V$.
This gives a fiducial set of extrinsic parameters $\eparams$ as
\begin{equation}
    \eparams \equiv 
    \begin{bmatrix}
    d \\
    A_V \\
    R_V
    \end{bmatrix}
\end{equation}

A schematic illustration of each of the various components of
our basic model is shown in Figure \ref{fig:overview}.

\begin{figure*}
\begin{center}
\includegraphics[width=0.6\textwidth]{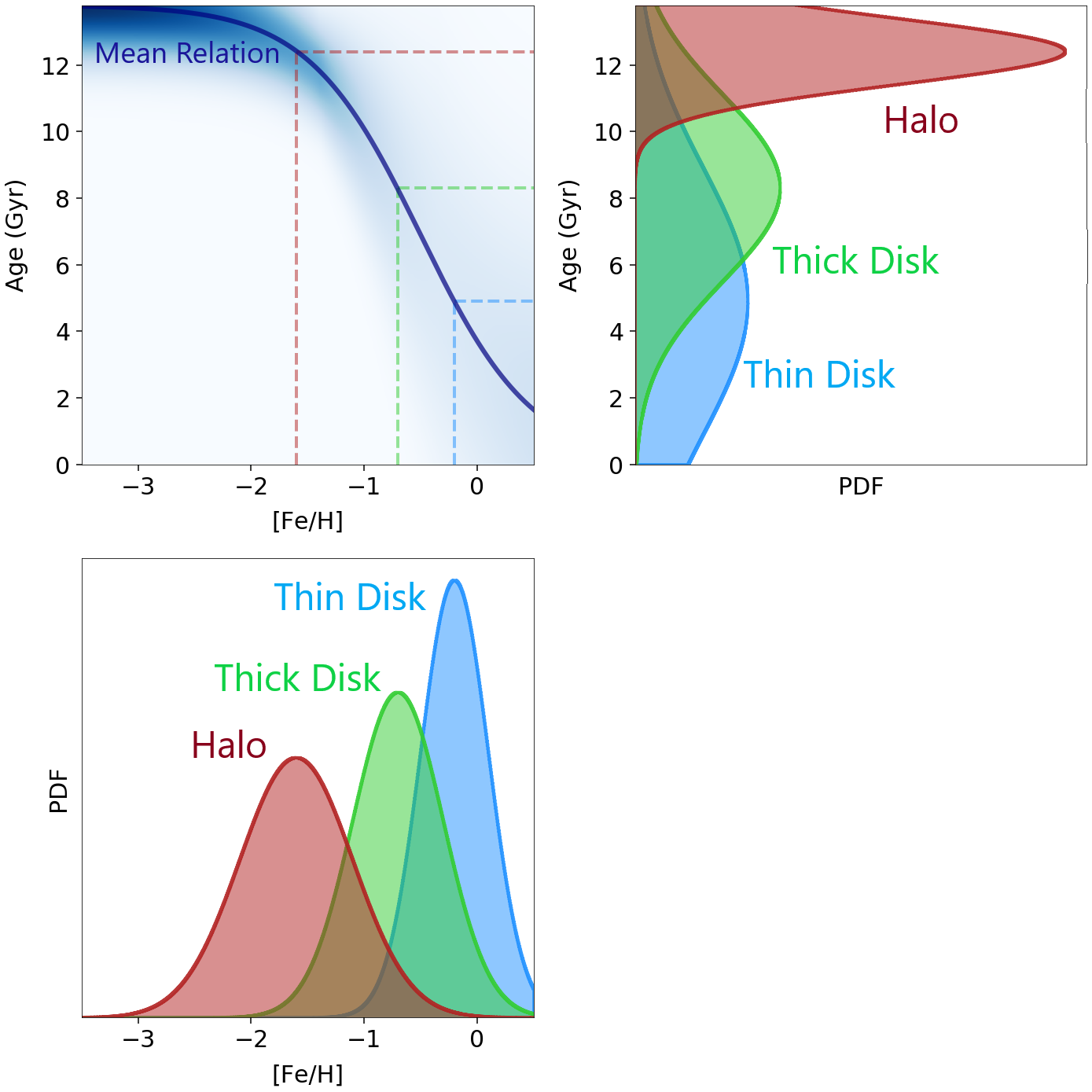}
\end{center}
\caption{The age-metallicity relation (shaded blue density; top left) used to set the
thin disk (blue), thick disk (green), and halo (red) 1-D metallicity
(bottom-left) and 1-D age (top-right) priors used in {\brutus}. The mean relationship
is highlighted as the solid dark blue line. The associated mean values for the 
ages and metallicities are indicated by the dashed red, green, and blue lines.
See Table \ref{tab:prior} and 
\S\ref{subsec:prior_feh} and \S\ref{subsec:prior_age} for more details.
}\label{fig:prior_agefeh}
\end{figure*}

\subsection{Noisy Data} \label{subsec:noisy_data}

\subsubsection{Photometry} \label{subsubsec:data_phot}

We assume that our data contain a set of noisy 
flux densities $\hat{\flux} = \{ \hat{F}_i \}_{i=1}^{i=b}$
in $b$ photometric bands that are distributed
following a Normal distribution
around the true flux densities $\flux = \{ F_i \}_{i=1}^{i=b}$
with corresponding uncertainties 
$\errors_{\flux} = \{ \sigma_{F,i} \}_{i=1}^{i=b}$. More formally,
\begin{equation}
    \hat{\flux} \sim \Normal{\flux}{\cov_{\flux}}
\end{equation}
where $\data \sim \Normal{\meanvec}{\cov}$ indicates 
that the data $\data = \hat{\flux}$
is drawn from a Normal probability density function (PDF) 
with mean vector $\meanvec = \flux$ 
and covariance matrix $\cov = \cov_{\flux} = \mathrm{diag}(\errors_{\flux})$, where
$\mathrm{diag}(\errors_{\flux})$ indicates a diagonal matrix with the $i$th value
of $\errors_{\flux}$ located in the $(i, i)$ matrix position 
and zeros everywhere else.

\begin{figure*}
\begin{center}
\includegraphics[width=\textwidth]{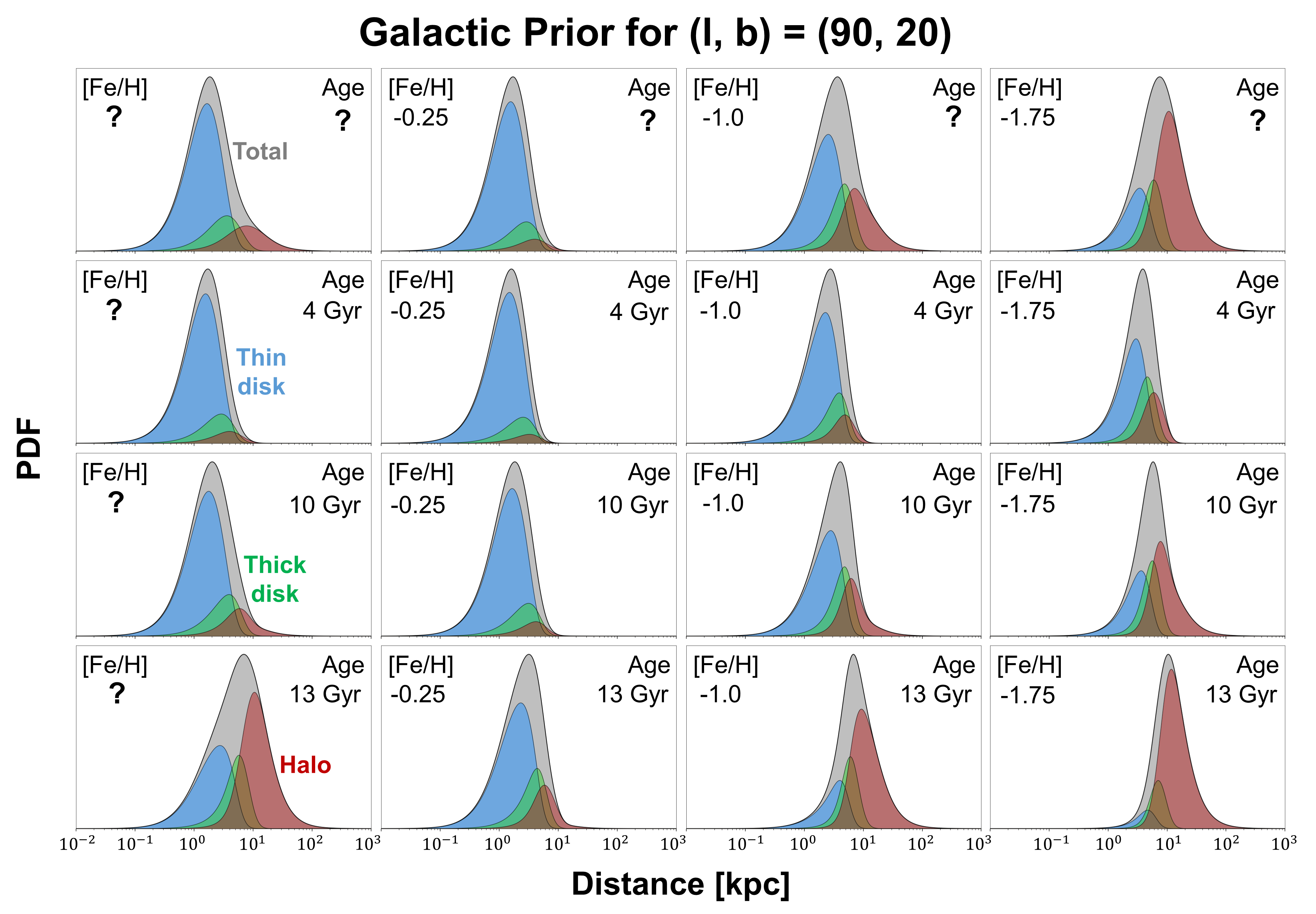}
\end{center}
\caption{An example of the Galactic prior over distance,
metallicity, and age used in {\brutus} evaluated
for a given sightline with Galactic coordinates $(\ell, b) = (90^\circ, 20^\circ)$. 
Each panel is broken into the total
probability density (gray) along with the contribution from the thin disk (blue),
thick disk (green), and halo (red). The top-left corner highlights
the prior $\prior(d|\ell, b)$ marginalized over (i.e. assuming unknown values of)
stellar metallicity $\feh$ and age $t_{\rm age}$, as indicated by question marks.
The left-most column shows the prior $\prior(d|\ell, b, t_{\rm age})$ 
conditioning only on $t_{\rm age}$, while the top-most row shows the prior
$\prior(d|\ell, b, \feh)$ but conditioning on $\feh$ instead. Each sub-panel
shows the prior $\prior(d|\ell, b, \feh, t_{\rm age})$ conditioning on both
$\feh$ and $t_{\rm age}$. Marginalized over $\feh$ and $t_{\rm age}$,
the prior prefers a source to be in the thin disk with small but non-negligible
contributions from the thick disk and halo. Conditioning on
low $\feh$ or high $t_{\rm age}$, however, begins to favor a source being in the halo.
Only after conditioning on both low $\feh$ and high $t_{\rm age}$
is a source strongly favored to be in the halo. As expected, at high $\feh$ and low
$t_{\rm age}$ a source is almost entirely associated with the thin disk, although
small contributions from the thick disk and halo remain due to their much larger 
number densities (relative to the thin disk) at larger distances.
See Table \ref{tab:prior}, \S\ref{subsec:prior}, 
and \S\ref{ap:priors} for more details.
}\label{fig:prior_sightline}
\end{figure*}

\begin{figure*}
\begin{center}
\includegraphics[width=\textwidth]{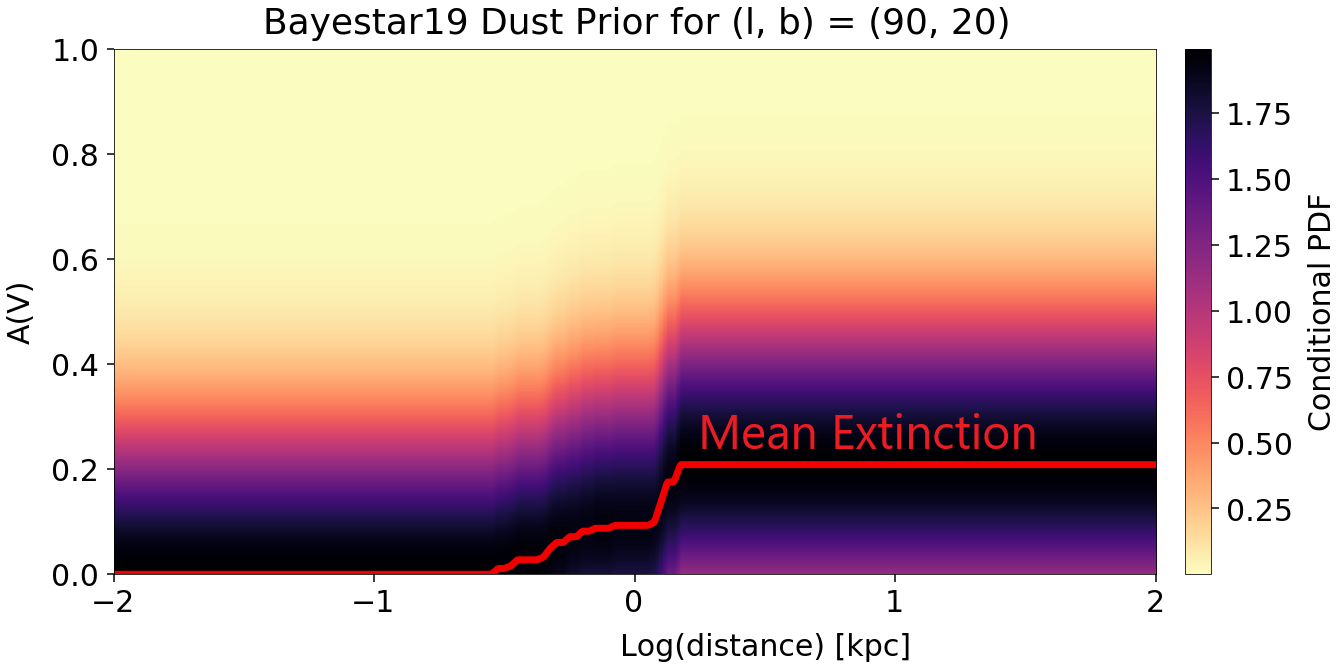}
\end{center}
\caption{An illustration of the \citet{green+19} {\bayestarmap}
3-D dust prior $\prior(A_V|\ell, b, d)$ used in {\brutus}
along the same $(\ell, b)=(90^\circ, 20^\circ)$ sightline as Figure \ref{fig:prior_sightline}.
The mean relationship is highlighted as the solid red line. 
There is evidence for diffuse dust between 0.3\,kpc and 1.0\,kpc and a concentrated
dust feature between 1.1\,kpc to 1.3\,kpc. This increases the preferred $A_V$ value
for a particular source at larger distances, although there is still a substantial
amount of variation allowed by the prior.
See Table \ref{tab:prior} and \S\ref{subsec:prior_av} for more details.
}\label{fig:prior_dust}
\end{figure*}

The log-likelihood of the observed flux density as a function of 
$\params$ and $\eparams$ follows the PDF
of a multivariate Normal distribution:
\begin{align}
    -2 \ln\likelihood_{\rm phot}(\params, \eparams) =
    \,&(\hat{\flux} - \flux_{\params, \eparams})^{\T} \cov_{\flux}^{-1} 
    (\hat{\flux} - \flux_{\params, \eparams}) \nonumber \\
    &+ \ln\left[{\rm det}\left(2\pi\cov_{\flux}\right)\right]
\end{align}
where $\T$ is the transpose operator, $\cov^{-1}$ is the
matrix inverse of $\cov$ (i.e. the precision matrix),
and ${\rm det}(\cdot)$ is the determinant of the matrix.
In our case where we assume $\cov_{\flux}$ is diagonal, this reduces to
\begin{equation}
    \boxed{
    -2 \ln\likelihood_{\rm phot}(\params, \eparams) = 
    \mathlarger{\mathlarger{\sum}}_{i=1}^{b}
    \frac{(\hat{F}_i - F_{\params,\eparams, i})^2}{\sigma_{\flux,i}^2}
    + \ln(2\pi\sigma_{\flux,i}^2)
    }
\end{equation}

Throughout the rest of the paper, we will assume that flux densities are defined
in units of ``maggies'' (i.e. in units relative to the standard reference
used to define the magnitude system; \citealt{finkbeiner+04})
such that we can convert from flux density to magnitude via
\begin{equation}
    \mags_{\params, \eparams} \equiv 
    \left\{ -2.5 \log \left( F_{\params, \eparams, i} \right) \right\}_{i=1}^{i=b}
\end{equation}
Note that while this assumption simplifies the 
majority of the subsequent derivations, it does not impact our results in any way.
We will return to this mismatch between our model (in magnitudes) and
our data (in flux densities) in \S\ref{sec:implementation}.

\subsubsection{Astrometry} \label{subsubsec:data_astr}

In addition to flux densities, we may also have astrometric measurements
for many of the sources from \textit{Gaia} \citep{lindegren+18}. While these include both
noisy parallax and proper motion measurements, in this work
we will only utilize the parallaxes for our inference and leave incorporating
proper motions to future work. We assume that
the noisy parallax $\hat{\varpi}$ is Normally distributed
about the true parallax $\varpi \equiv 1/d$ (in mas) 
for a given distance to the object $d$ (in kpc) with some
scatter $\sigma_{\varpi}$ such that
\begin{equation}
    \hat{\varpi} \sim \Normal{\varpi}{\sigma_{\varpi}}.
\end{equation}
The corresponding log-likelihood is then
\begin{equation}
    \boxed{
    -2\ln\likelihood_{\rm astr}(\eparams) = 
    \frac{(\hat{\varpi}-\varpi(\eparams))^2}{\sigma_{\varpi}^2} + 
    \ln(2\pi\sigma_{\varpi}^2)
    }
\end{equation}
since $\varpi(\eparams) = \varpi(\mu) = \varpi(d)$ is one of the extrinsic 
parameters we are interested in.


\subsection{Posterior Probability} \label{subsec:posterior}

The probability $P(\params, \eparams | \hat{\flux}, \hat{\varpi})$ 
for a particular set of intrinsic parameters $\params$ and extrinsic parameters
$\eparams$ given the observed $b$ flux densities $\hat{\flux}$, parallax
$\hat{\varpi}$, and prior knowledge $P(\params, \eparams)$ about $\params$
and $\eparams$ can be derived using Bayes Theorem:
\begin{align}
    P(\params, \eparams | \hat{\flux}, \hat{\varpi}) &\propto
    P(\hat{\flux}, \hat{\varpi} | \params, \eparams) P(\params, \eparams)
    \nonumber \\
    &\boxed{\equiv 
    \likelihood_{\rm phot}(\params, \eparams) 
    \likelihood_{\rm astr}(\eparams)
    \prior(\params, \eparams)
    }
\end{align}
where $P(\params, \eparams | \hat{\flux}, \hat{\varpi})$
is the \textit{posterior} probability for $\params$ and $\eparams$,
$P(\hat{\flux}, \hat{\varpi} | \params, \eparams)
\equiv \likelihood_{\rm phot}(\params, \eparams) \likelihood_{\rm astr}(\eparams)$
is the \textit{likelihood}, which we have split into
photometric $\likelihood_{\rm phot}(\params, \eparams)$ 
and astrometric $\likelihood_{\rm astr}(\eparams)$ terms, 
and $P(\params, \eparams) \equiv \prior(\params, \eparams)$
is the \textit{prior}.

Combined, this allows us to translate from a set of observed
flux densities $\hat{\flux}$ and parallax $\hat{\varpi}$, along with their
corresponding errors $\errors_{\flux}$ and $\sigma_{\varpi}$,
into constraints on the distance ($\mu$),
dust extinction ($A_V$, $R_V$), and intrinsic stellar properties
($\params$) for each source.

\subsection{Priors} \label{subsec:prior}

Our prior $\prior(\params, \eparams)$ 
over $\params$ and $\eparams$ represents a Galactic model describing
the 3-D distribution of stars, dust, and their associated properties 
throughout the Milky Way. Within the stable implementation of 
{\brutus} in use at the time of 
writing\footnote{\texttt{v0.7.5}: \url{http://doi.org/10.5281/zenodo.3711493}},
the prior is divided up into a few independent components
describing several different processes:
\begin{align}
    \pi(\params, \eparams) \propto 
    &\:\:\underbrace{\pi(M_{\rm init})}_{\rm IMF} \nonumber \\
    &\times \underbrace{\pi(d | \ell, b)}_{\rm 3D\:number} \nonumber \\
    &\times \underbrace{\pi(\feh_{\rm init} | d, \ell, b)}_{\rm 3D\:metallicity}
    \nonumber \\
    &\times \underbrace{\pi(t_{\rm age} | d, \ell, b)}_{\rm 3D\:age} \\
    &\times \underbrace{\pi(A_V | d, \ell, b)}_{\rm 3D\:extinction} \nonumber \\
    &\times \underbrace{\pi(R_V)}_{\rm Dust\:curve} \nonumber
\end{align}
where $d$ is the heliocentric distance of a source and $(\ell, b)$ 
are the Galactic longitude and latitude, respectively.
While this assumption makes the problem more straightforward,
it enforces a couple of assumptions regarding, e.g.,
the universality of the IMF or the lack of correlations between individual
stellar ages and corresponding metallicities.

Schematic illustrations of our 3-D stellar and extinction
priors are shown in Figures \ref{fig:prior_density},
\ref{fig:prior_agefeh}, \ref{fig:prior_sightline}, and
\ref{fig:prior_dust}.
A summary of the priors and associated constants
are described in Table \ref{tab:prior}. A detailed
description can be found in \S\ref{ap:priors}.
We hope to add in more options for more complex
Galactic stellar priors (number densities,
metallicities, ages, $\alpha$-abundance variations, etc.)
and 3-D dust extinction priors, such as those included in 
{\dustmaps}\footnote{\url{https://github.com/gregreen/dustmaps}}
\citep{green18}, in the future.

\begin{deluxetable*}{lcclcc}
\tablecolumns{6}
\tablecaption{Description of default priors and 
their corresponding hyper-parameters in {\brutus}
based on \citet{blandhawthorngerhard16} and
\cite{xue+15}. See \S\ref{subsec:prior} and
\S\ref{ap:priors} for additional details.
\label{tab:prior}}
\tablehead{
\colhead{Description} & \colhead{Symbol} & {Value} &
\colhead{Description} & \colhead{Symbol} & {Value}
}
\startdata
\cutinhead{\textbf{Initial Mass Function}}
Low-mass power law slope & $\alpha_1$ & $1.3$ &
High-mass power law slope & $\alpha_2$ & $2.3$ \\
\cutinhead{\textbf{3-D Stellar Number Density}}
Solar radius & $R_\odot$ & $8.2\,{\rm kpc}$ 
& Halo smoothing radius & $R_s$ & $1\,{\rm kpc}$ \\
Solar height & $Z_\odot$ & $0.025\,{\rm kpc}$ 
& Halo oblateness at $r=0$ & $q_0$ & $0.2$ \\
Thin disk scale radius & $R_{\rm thin}$ & $2.6\,{\rm kpc}$ 
& Halo oblateness at $r=\infty$ & $q_\infty$ & $0.8$ \\
Thin disk scale height & $Z_{\rm thin}$ & $0.3\,{\rm kpc}$ 
& Halo scale radius & $r_q$ & $6\,{\rm kpc}$ \\
Thick disk scale radius & $R_{\rm thick}$ & $2.0\,{\rm kpc}$ 
& Halo power law slope & $\eta$ & $4.2$ \\
Thick disk scale height & $Z_{\rm thick}$ & $0.9\,{\rm kpc}$ 
& Halo fractional contribution at $R_\odot$ & $f_{\rm halo}$ & $0.005$ \\
Thick disk fractional contribution at $R_\odot$ & $f_{\rm thick}$ & $0.04$ \\
\cutinhead{\textbf{Stellar Metallicity}}
Thin disk mean metallicity  & $\mu_{\feh, {\rm thin}}$ & $-0.2$ 
& Halo mean metallicity & $\mu_{\feh, {\rm halo}}$ & $-1.6$ \\
Thin disk metallicity scatter & $\sigma_{\feh, {\rm thin}}$ & $0.3$
& Halo metallicity scatter & $\sigma_{\feh, {\rm halo}}$ & $0.5$ \\
Thick disk mean metallicity & $\mu_{\feh, {\rm thick}}$ & $-0.7$ \\
Thick disk metallicity scatter & $\sigma_{\feh, {\rm thick}}$ & $0.4$ \\
\cutinhead{\textbf{Stellar Age}}
Maximum age & $t_{\rm max}$ & $13.8\,{\rm Gyr}$
& Thin disk mean age & $\mu_{t, {\rm thin}}$ & $4.9\,{\rm Gyr}$ \\
Minimum age & $t_{\rm min}$ & $0\,{\rm Gyr}$ 
& Thin disk age scatter & $\sigma_{t, {\rm thin}}$ & $4\,{\rm Gyr}$ \\
Maximum age scatter & $\sigma_{\rm max}$ & $4\,{\rm Gyr}$ 
& Thick disk mean age & $\mu_{t, {\rm thick}}$ & $8.3\,{\rm Gyr}$ \\
Minimum age scatter & $\sigma_{\rm min}$ & $1\,{\rm Gyr}$
& Thick disk age scatter & $\sigma_{t, {\rm thick}}$ & $2.8\,{\rm Gyr}$ \\
Age-metallicity relation pivot & $\xi_{\feh}$ & $-0.5$ 
& Halo mean age & $\mu_{t, {\rm halo}}$ & $12.4\,{\rm Gyr}$ \\
Age-metallicity relation scale-length & $\Delta_{\feh}$ & $0.5$
& Halo age scatter & $\sigma_{t, {\rm halo}}$ & $1\,{\rm Gyr}$ \\
Standard deviation from maximum age & $n_\sigma$ & $2$ \\
\cutinhead{\textbf{3-D Dust Extinction}}
3-D $A_V$ mean & $\mu_A(d|\ell, b)$ & {\bayestarmap} mean 
& 3-D $A_V$ scatter & $\Delta_A$ & $0.2\,{\rm mag}$ \\
3-D $A_V$ uncertainty & $\sigma_A(d|\ell, b)$ & {\bayestarmap} scatter \\
\cutinhead{\textbf{Dust Curve Variation}}
$R_V$ mean & $\mu_R$ & $3.32$
& $R_V$ scatter & $\sigma_R$ & $0.18$ \\
\enddata
\end{deluxetable*}

\section{Implementation} \label{sec:implementation}

\begin{figure*}
\begin{center}
\includegraphics[width=\textwidth]{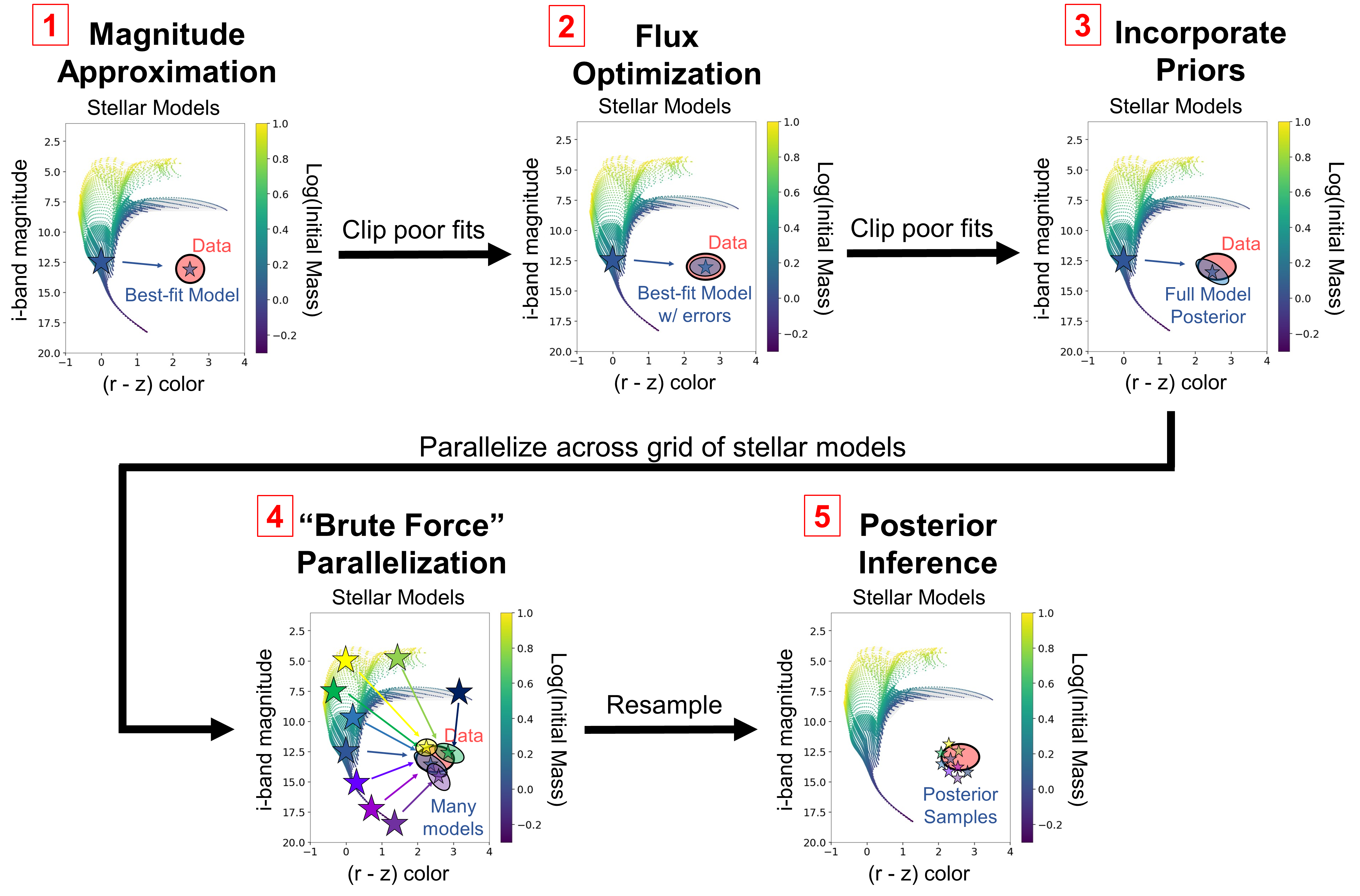}
\end{center}
\caption{A schematic illustration of the approach {\brutus} takes to
estimate stellar parameters across a $b$-band SED 
(only 3 are shown for visual clarity).
First, {\brutus} constructs a ``quick approximation''
of the observed magnitudes (1, top left; see \S\ref{subsubsec:impl_mag}).
After clipping poor fits, {\brutus} transforms the data to the native
flux densities and conducts a limited optimization of the best-fit
parameters (2, top middle; see \S\ref{subsubsec:impl_flux}). 
After clipping poor fits again using more stringent criteria, 
{\brutus} uses importance sampling to incorporate constraints from 
our priors (see \S\ref{subsec:prior}) as well as any measured
parallax (3, top right; see \S\ref{subsubsec:impl_mc}).
These steps are processed in parallel across a grid of
stellar models (4, bottom left) and subsequently resampled
to approximately sample from the underlying posterior
(5, bottom right; see \S\ref{subsubsec:impl_timing}).
The whole process takes only a few seconds for a typical 
SED with $b \sim 10$ bands and parallax measurements
with low-to-moderate signal-to-noise ratios.
See \S\ref{sec:implementation} for more details and Table
\ref{tab:settings} for a summary of important hyper-parameters.
}\label{fig:brutus_algorithm}
\end{figure*}

\begin{figure*}
\begin{center}
\includegraphics[width=\textwidth]{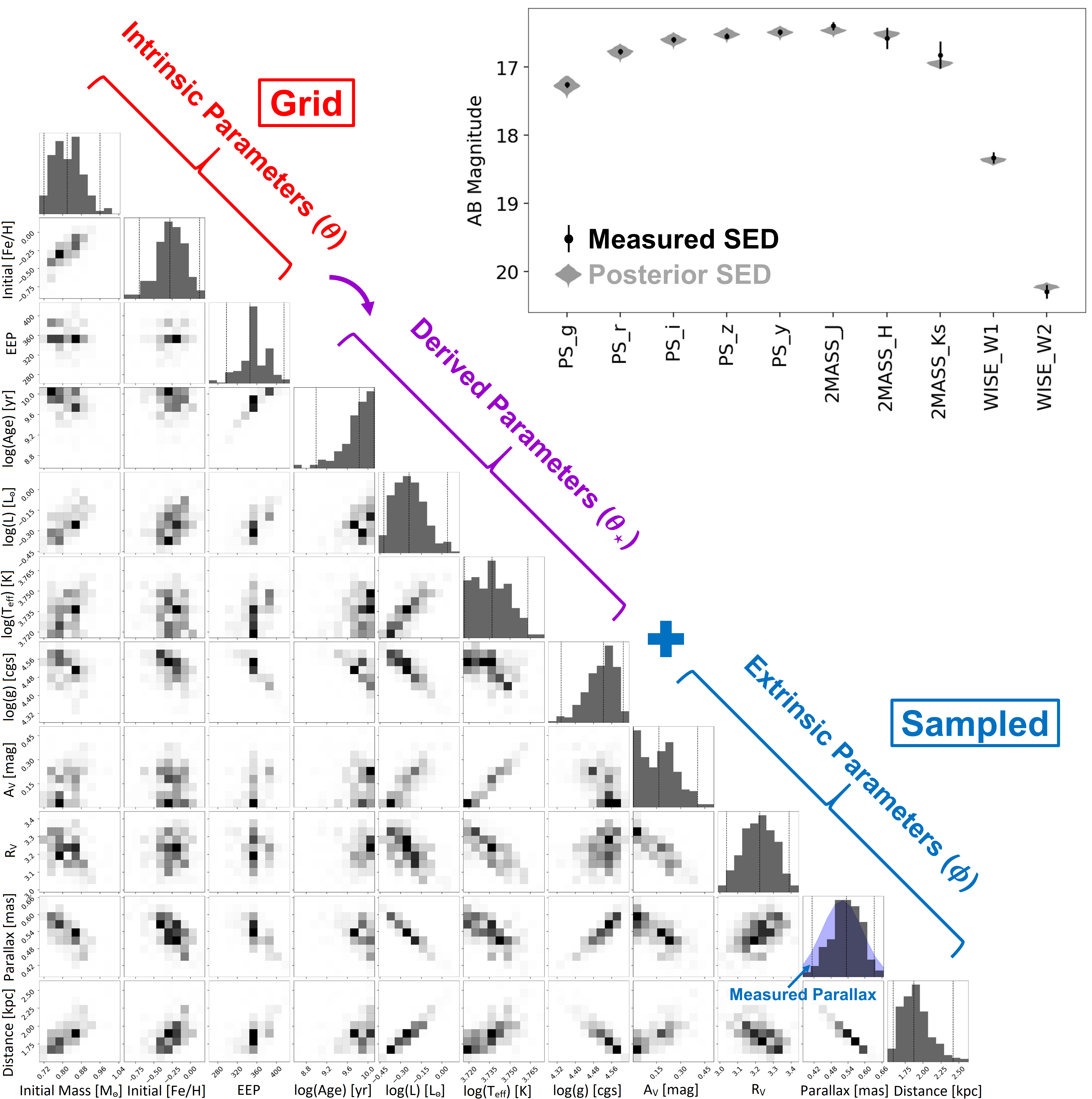}
\end{center}
\caption{An example of parameters estimated from {\brutus}
using the default priors (see \S\ref{subsec:prior}) and
setup (see \S\ref{sec:implementation} and Table \ref{tab:settings})
and the {\mist} models (see \S\ref{sec:models} and \S\ref{sec:calib})
for a real object observed in Pan-STARRS (PS), 2MASS, and WISE
with a \textit{Gaia} parallax measurement.
The upper-right region shows the measured SED (black points)
with the associated 2-sigma errors (black lines) along with
SED realizations from the posterior (gray shaded regions).
The lower-left region highlights the 1-D and 2-D marginalized
posterior distributions estimated using the $n_{\rm post} = 250$
samples saved to disk with 10 evenly-spaced bins in each dimension,
with the grayscale indicating the relative density of samples
across the $10 \times 10$ bins.
The labels, which are read in directly
from the stellar model grid, correspond to the set of 
(gridded) intrinsic stellar parameters $\params$ 
and their derived surface-level stellar parameters $\params_\star$
along with the (sampled) extrinsic parameters $\eparams$.
The astrometric constraints from the measured parallax alone
are highlighted in light blue. Since the fits for each model
$\params_i$ on the grid are done entirely in parallel, 
all correlations that emerge are a result of
the posterior resampling process.
}\label{fig:example}
\end{figure*}

{\brutus} uses a combination of linear regression,
Monte Carlo sampling, and brute force methods to
generate fast but robust approximations to the underlying posterior.
This approach is able to capture strong covariances between parameters, trace
extended structures in this distribution, and characterize multiple possible solutions.
The basic procedure works as follows:
\begin{enumerate}
    \item For all models in a given grid of (intrinsic) stellar parameters,
    use linear regression in magnitude space to solve for the best-fit solution for our
    extrinsic parameters $\eparams$ given the intrinsic stellar parameters $\params$.
    \item After removing models that are poor fits, improve the
    remaining fits by transforming the best-fit solutions from 
    magnitudes into flux densities and using linear regression
    to solve for first-order corrections.
    \item After further removing models with low expected posterior probabilities,
    use Monte Carlo sampling to numerically integrate over the prior.
    \item (Re)sample $\params$ and $\eparams$ from the estimated posterior.
\end{enumerate}

Previous work such as {\starhorse} \citet{santiago+16,queiroz+18,anders+19} also use grids
of parameters over stellar models to estimate $\params$ and $\eparams$. The main
differences between the approach taken in {\brutus} outlined above and those
in many previous approaches are as follows:
\begin{enumerate}
    \item \textit{Parameter grids}: 
    Many past approaches fit grids in both intrinsic $\params$ and extrinsic $\eparams$
    stellar parameters. {\brutus} only requires grids in $\params$, which are then used to
    generate continuous estimates for $\eparams$.
    \item \textit{Variation in dust curves}: While many previous approaches
    can deal with variations in $A_V$, {\brutus} can model both variation in $A_V$ and $R_V$.
    \item \textit{Flux density vs magnitude}: Previous approaches often compare models and data in magnitudes,
    which can cause issues for data with low signal-to-noise ratios (SNRs). 
    {\brutus} fits performs better at lower SNR by comparing models and data in flux densities.
    \item \textit{Error modelling}: Some past approaches do not propagate certain aspects of 
    measurement and/or model uncertainties when evaluating individual models. {\brutus}
    includes methods to explicitly try and model both.
\end{enumerate}

A schematic illustration of our approach is shown in Figure \ref{fig:brutus_algorithm}.
An example of the output stellar parameters and the associated SED
can be seen in Figure \ref{fig:example}.
A detailed description can be found in \S\ref{ap:implementation}.
We find that for a typical source observed in $\sim 8$ 
optical-to-NIR photometric bands with weak parallax
constraints, {\brutus} is able to generate $\sim 250$
samples from the posterior in $\sim 5$ seconds
for a grid of $\sim 7.5 \times 10^5$ models.
See \S\ref{subsec:methods_cuts} for additional discussion.

\section{Stellar and Extinction Models} \label{sec:models}

While {\brutus} can in theory incorporate an arbitrary set of
stellar and extinction models via a corresponding grid in 
$\{ \params_i \}_{i=1}^{i=n} \rightarrow 
\{ \params_{\star, i} \}_{i=1}^{i=n} \rightarrow
\{\absmag_{\params,i}, \rvec_{\params, i}, \rvec'_{\params, i}\}_{i=1}^{i=n}$, 
it is currently designed to work with two models by default:
\begin{itemize}
    \item The theoretical MESA Isochrone and Stellar Tracks
    (\mist) models \citep{choi+16} combined
    with the $R_V$-dependent extinction curve from
    \citep{fitzpatrick04}.
    \item The empirical {\bayestar} models 
    \citep{green+14,green+15,zuckerspeagle+19}
    combined with the empirical $R_V$-dependent extinction
    curve from \citet{schlafly+16}.
\end{itemize}
We hope to incorporate additional models
such as PARSEC \citep{bressan+12} and BPASS \citep{eldridge+17}
in the future.

\begin{figure*}
\begin{center}
\includegraphics[width=\textwidth]{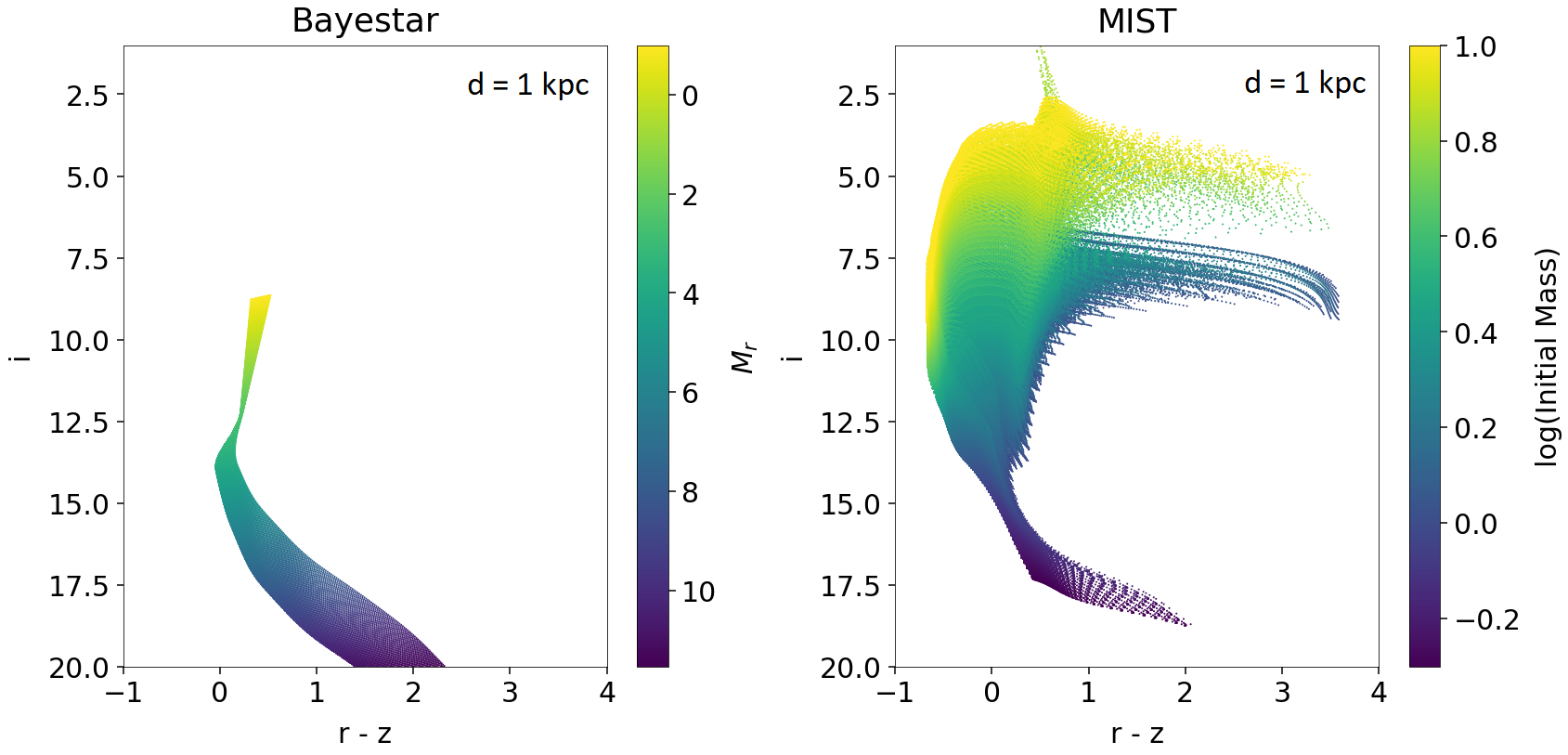}
\end{center}
\caption{Color-magnitude diagrams (CMDs) in $i$ magnitude 
(at $1\,{\rm kpc}$) versus $r-z$ color in the Pan-STARRS filters
for {\bayestar} (\texttt{grid\_bayestar\_v5}, left) 
and {\mist} (\texttt{grid\_mist\_v8}, right) models,
implemented by default in {\brutus}.
The $n \sim 4 \times 10^4$ 
empirical {\bayestar} models (see \S\ref{subsec:bayestar})
are defined over a grid of metallicity
values along with absolute Pan-STARRS $r$-band magnitudes $M_r$
(left color scale), which serve as a rough proxy of intrinsic luminosity
(and hence initial mass). The $n \sim 7.5 \times 10^5$
theoretical {\mist} models (see \S\ref{subsec:mist}), 
including the empirical corrections 
discussed in \S\ref{sec:calib}, are defined over a grid of ages,
metallicites, and initial masses (right color scale).
While the {\bayestar} models extend down to fainter intrinsic luminosity
(lower initial mass) on the Main Sequence (MS; i.e. ``dwarfs''), 
they only contain limited models at higher initial
masses and little-to-no post-MS models (giants). They also are
not calibrated outside a limited set of photometric bands.
By contrast, while the {\mist} models are constrained to be above
$M_{\rm init} \sim 0.5\,M_\odot$ in this work (see \S\ref{sec:calib}),
they include a wide variety of evolutionary phases and 
extend to higher initial masses.
}\label{fig:mist_vs_bayestar}
\end{figure*}

A comparison of the {\mist} and {\bayestar}
models is shown in Figure \ref{fig:mist_vs_bayestar}.
A detailed description of the MIST models and the corresponding
pre-generated photometric grids in {\brutus} 
is provided in \S\ref{subsec:mist}.
A detailed description of the {\bayestar} models and the
corresponding photometric grids in {\brutus} is provided in
\ref{subsec:bayestar}. All of the relevant
data products described in this section are available
online as part of the {\brutus} codebase.

\subsection{{\mist}} \label{subsec:mist}

\subsubsection{Isochrones} \label{subsubsec:mist_iso}

The MESA Isochrone and Stellar Tracks (\mist) models
are a set of theoretical isochrones built off of the
Modules in Stellar Evolution 
\citep[\mesa;][]{paxton+11,paxton+13,paxton+15,paxton+18,paxton+19}
codebase that connect intrinsic stellar evolutionary parameters
$\params$ to physical surface-level parameters $\params_\star$
as described in \S\ref{subsec:noiseless_model}.
A full description of the models can be found in
\citep{choi+16}. We utilize the {\mist}
Version 1.2 \textit{non-rotating} models computed over
a grid of initial masses, initial metallicities, and ages
as described on the {\mist} 
website\footnote{\url{http://waps.cfa.harvard.edu/MIST/}}.

As described in \citet{dotter16}, since the timescale of stellar
evolution is sensitive to the initial mass, different stars will
reach different evolutionary phases at different times and
over different timescales. As a result, it is advantageous
to define a grid in \textit{Equivalent Evolutionary Points} (EEPs)
instead, which interpolate smoothly between particular stellar 
evolutionary states. While these have a monotonic relation with
age, they are defined such that evolutionary phases where there are rapid
changes to the surface or interior stellar properties are adequately
captured in relevant evolutionary tracks and isochrone tables.
EEPs can therefore be converted to ages $t_{\rm age}$ using associated 
lookup tables for a particular $M_{\rm init}$ 
and $\feh_{\rm init}$. To account for the subsequent 
unequal spacing in age, the grid spacing
$\Delta_i$ for each model $\params_i$ for our {\mist} grid
also includes the associated
$\Delta t_{{\rm age}, i}/\Delta {\rm EEP}_i$ estimated at each EEP
using central finite differences (which have errors at second order).

While the {\mist} models have been generally successful at
reproducing stellar behavior across a wide range of
masses, metallicities, and ages, we want to highlight
two particular areas where more work is needed:
\begin{enumerate}
    \item \textit{Rotation:}
    Non-rotating stellar evolutionary tracks are used in this work for simplicity.
    This is a limitation since stars rotate,
    which can alter the relationship between age and EEP as well as
    the associated surface parameters $\params_\star$, especially
    around the Main Sequence turn-off for $M_{\rm init} > 1.2 \, M_\odot$
    \citep[see, e.g.,][]{gossage+18}.
    We note that this inevitably will introduce
    systematic uncertainties in any inferred stellar parameters
    (both $\params$ and $\params_\star$).
    \item \textit{Abundance patterns:} 
    The {\mist} models assume entirely solar-scaled abundance
    patterns. However, substantial populations of stars both in clusters
    and in the field display enhancement/depletion of
    $\alpha$-process elements (Ne, Mg, Si, S, Ar, Ca, and Ti)
    relative to the Sun (i.e. $\afe \neq 0$). $\afe$ variations
    can induce changes in the effective temperature on the order
    of $\pm 100\,{\rm K}$ that are somewhat degenerate with other
    effects such as changes in metallicity and intervening dust
    extinction \citep{dotter+07}. These effects
    will be included in an upcoming suite of newer {\mist} models
    (Dotter et al. in prep.) and are not addressed in this work.
\end{enumerate}

The 3-D grid over $M_{\rm init}$, $\feh_{\rm init}$, and ${\rm EEP}$ we use
for the {\mist} models is described in Table \ref{tab:mist_grid}.
This is adaptively spaced in $M_{\rm init}$ and defined in
EEP to have coarser resolution on the Main Sequence (MS)
(${\rm EEP} = 202-454$) and finer resolution on the post-MS
(${\rm EEP} > 454$). It does not include any models
on the pre-MS (${\rm EEP} < 202$) or 
at or beyond the start of the thermally-pulsing 
asymptotic giant branch (${\rm EEP} > 808$).\footnote{
See the {\mist} website and/or \citet{dotter16} for
additional information on the EEP definitions used in {\mist}.}

We use linear interpolation to generate
the corresponding grid over the surface-level parameters
$\{ \params_i \}_{i=1}^{i=n} 
\rightarrow \{ \params_{\star,i} \}_{i=1}^{i=n}$.
After removing models with unphysical combinations of parameters
(e.g., ages substantially exceeding the estimated 
current age of the Universe), we are left with
a total of $n \sim 7.5 \times 10^5$ models.\footnote{
While {\brutus} allows for modeling additional unresolved
binary components, internal testing found current systematic
uncertainties (see \S\ref{sec:calib}) are too large
to infer meaningful quantities for many
individual stars. This feature is therefore disabled by default.}
Cross-validation and hold-out tests over
the original grid indicates interpolation errors can become significant
($\gtrsim 10\%$) above $2\,M_\odot$, but remain at the few percent level
between $0.5\,M_\odot$ and $2\,M_\odot$.

\begin{deluxetable}{lcc}
\tablecolumns{3}
\tablecaption{Default grid of parameters for the {\mist} models
used in {\brutus} (\texttt{grid\_mist\_v8}). 
See \S\ref{subsec:mist} for additional details.
\label{tab:mist_grid}}
\tablehead{Minimum & Maximum & Spacing}
\startdata
\cutinhead{\textbf{Initial Mass} ($M_{\rm init}$)}
$0.5\,M_\odot$ & $2.8\,M_\odot$ & $0.02\,M_\odot$ \\
$2.8\,M_\odot$ & $3.0\,M_\odot$ & $0.1\,M_\odot$ \\
$3.0\,M_\odot$ & $8.0\,M_\odot$ & $0.25\,M_\odot$ \\
$8.0\,M_\odot$ & $10.0\,M_\odot$ & $0.5\,M_\odot$ \\
\cutinhead{\textbf{Initial Metallicity} ($\feh_{\rm init}$)}
$-4.0$ & $+0.5$ & $0.06$ \\
\cutinhead{\textbf{Equivalent Evolutionary Point} (${\rm EEP}$)}
$202$ & $454$ & $12$ \\
$454$ & $808$ & $6$ \\
\enddata
\end{deluxetable}

\subsubsection{Atmospheric Models} \label{subsubsec:mist_c3k}

As in \citet{cargile+20}, 
we connect each model $\params_\star$ to an underlying
spectrum $F_\nu(\lambda|\params_\star)$
using the {\ctk} atmospheric models (C. Conroy, priv. comm.)
given the predicted value of $\params_\star(\params)$ from
linear interpolation and the relevant
filter transmission curves (see \S\ref{subsec:noiseless_model}).
These synthetic spectra are calculated using the 1-D 
local thermodynamic equilibrium (LTE) plane-parallel atmosphere and
radiative transfer codes {\atlas} and
{\synthe} maintained by R. Kurucz \citep{kurucz70,kuruczavrett81,kurucz93}.
The line list used in the radiative
transfer calculations was provided by R. Kurucz (private communication)
after being empirically tuned to the observed, 
ultra-high resolution spectra of the Sun and Arcturus
(Cargile et al. in prep). The micro-turbulence is assumed
to be constant with a velocity of
$v_{\rm micro} = 1\,{\rm km}\,{\rm s}^{-1}$.
The {\ctk} atmosphere models have been shown to 
reproduce the observed relationship between
color-$T_{\rm eff}$ relations for all but the very
lowest mass stars (i.e. M dwarfs) at the few percent level in $T_{\rm eff}$.

The {\ctk} atmosphere models are originally constructed over a 4-D grid
in effective temperature $T_{\rm eff}$, surface gravity $\log g$,
surface metallicity $\feh_{\rm surf}$, and surface $\alpha$-abundance
enhancement $\afe_{\rm surf}$ over an adaptive grid as outlined in
Table \ref{tab:ctk_grid}. This includes a total of $n=26561$ models 
after removing unphysical combinations of parameters.
Since the current {\mist} models only use solar-scaled abundance patterns, 
when predicting spectra and/or photometry {\brutus} always sets
$\afe_{\rm surf} = 0$ by default.

\begin{deluxetable}{lcc}
\tablecolumns{3}
\tablecaption{Grid of parameters for the {\ctk} 
stellar atmosphere models used in {\brutus}. 
See \S\ref{subsec:mist} for additional details.
Dust extinction is incorporated using the $R_V$-dependent
dust curve from \citet{fitzpatrick04}.
\label{tab:ctk_grid}}
\tablehead{Minimum & Maximum & Spacing}
\startdata
\cutinhead{\textbf{Effective Temperature} ($T_{\rm eff}$)}
$2500\,{\rm K}$ & $50000\,{\rm K}$ & $200\,{\rm K}-10000\,{\rm K}$ \\
\cutinhead{\textbf{Surface Gravity} ($\log g$)}
$-1.0$ & $5.0$ & $0.5$ \\
\cutinhead{\textbf{Surface Metallicity} ($\feh_{\rm surf}$)}
$-4.0$ & $-3.0$ & $0.5$ \\
$-3.0$ & $+0.5$ & $0.25$ \\
\cutinhead{\textbf{Surface $\alpha$ Abundance} ($\afe_{\rm surf}$)}
$-0.2$ & $+0.6$ & $0.2$ \\
\cutinhead{\textbf{Dust Extinction} ($A_V$)}
$0.0$ & $5.0$ & $1.0$ \\
\cutinhead{\textbf{Dust Curve Variation} ($R_V$)}
$2.0$ & $5.0$ & $1.0$ \\
\enddata
\end{deluxetable}

\subsubsection{Dust Extinction Curve} \label{subsubsec:mist_dust}

To incorporate the impact of dust extinction, we add two
dimensions to the {\ctk} atmospheric model grid in both
$A_V$ and $R_V$ based on the $R_V$-dependent dust extinction curve 
from \citet{fitzpatrick04}. This dust curve has been
shown to accurately reproduce detailed observations 
from spectra and photometry in the optical and NIR
\citep{fitzpatrick04,schlafly+16}. The associated grid
in $A_V$ and $R_V$ is shown in Table \ref{tab:ctk_grid}.
The final 6-D grid contains $n \sim \sn{6.4}{5}$ models.

\subsubsection{Photometry} \label{subsubsec:mist_phot}

Photometry is computed from each synthetic spectrum 
for a large set of photometric systems derived from various
imaging surveys. This currently includes:
\begin{itemize}
    \item Pan-STARRS: $g$, $r$, $i$, $z$, 
    $y$, $w$, and $w_{\rm open}$.
    \item DECam: $u$, $g$, $r$, $i$, $z$, and $Y$.
    \item Bessel: $U$, $B$, $V$, $R$, and $I$.
    \item 2MASS: $J$, $H$, and $K_s$.
    \item UKIDSS: $Z$, $Y$, $J$, $H$, and $K$.
    \item WISE: $W_1$, $W_2$, $W_3$, and $W_4$.
    \item \textit{Gaia}: $G$, $BP$, and $RP$.
    \item \textit{Tycho}: $B$ and $V$.
    \item \textit{Hipparcos}: $H_p$.
    \item \textit{Kepler}: $D_{51}$ and $K_p$.
    \item \textit{TESS}: $w_{\rm TESS}$.
\end{itemize}
Note that the \textit{Gaia} filter curves and zeropoints
are computed using the DR2 photometric calibrations from
\citet{evans+18} and \citet{maizapellanizweiler18}. For
the $BP$ filter, where the behavior on the bright and faint end
substantially differ, we have opted to utilize the ``faint''
version of the filter curve. See \citet{choi+16} and the {\mist}
website for additional discussion on the filters, 
photometric systems, and corresponding ``zero-points''. 
We will return to some of these in \S\ref{sec:calib}.

To generate photometry in a given band $M_i(\params_\star, A_V, R_V)$,
we need to both integrate each model over the corresponding
filter transmission curve $T_b(\lambda)$
and interpolate over the corresponding 6-D grid.
While the former operation is straightforward, the latter
is slightly more difficult due to the larger dimensionality
of our grid and smooth (but non-linear) behavior in the synthetic 
spectra as a function of our parameters.

As discussed in \citet{ting+19}, the use of 
artificial neural networks (NNs) to predict synthetic spectra (and in our
case photometry) has advantages over other interpolation methods.
We train a multi-layered feed-forward NN using {\pytorch}
over the 6-D grid of normalized photometry after integrating
the synthetic spectra over each filter transmission curve.
We use the same architecture and training
procedure described in \citet{cargile+20}. Briefly,
our network has 4 layers (2 hidden layers) with 64 neurons
per layer and sigmoid activation functions. Training is performed using
adaptive cross-validation procedures over regular epochs
and where the density of models are increased where the
network’s predictions are the least accurate. The
predicted magnitudes have a mean square error (MSE)
of $\sim 0.01$\,mag over both the training data and a subset of
hold-out test data. This is below the level of known systematic 
uncertainties in the {\mist} models \citep{choi+16}
and seen in \S\ref{sec:calib}.

\subsubsection{Linear Reddening Approximation} \label{subsubsec:mist_linear}

As described in \S\ref{subsec:noiseless_model},
we need to approximate the impact of dust extinction 
through the use of a linear reddening vector $\rvec_{\params}$
and differential reddening vector $\rvec'_{\params}$ over
our filter curves. As expected, we find that
for large changes in $A_V$ and $R_V$ the corresponding
change in magnitudes is not fully described by a single
linear vector but requires additional polynomial terms in
both parameters. While this will impact the ``absolute''
inferred $A_V$ and $R_V$, it is consistent both with the definition
of $E(B-V) \rightarrow A_V$ from the 3-D dust map of
\citet{green+19} (henceforth {\bayestarmap}) 
and of $R_V$ from \citet{schlafly+16} and \citet{schlafly+17}.

The details of our approximation are described in \S\ref{ap:linear_dust}.
Overall, we find agreement at the few percent level across a wide range
of $A_V$ and $R_V$ values.

\subsection{{\bayestar}} \label{subsec:bayestar}

\subsubsection{Stellar Models} \label{subsubsec:bayestar_star}

As described in \citet{green+14,green+15}, 
the {\bayestar} stellar models are obtained by fitting
a stellar locus in 7-D color space in the 
Panoramic Survey Telescope and Rapid Response System
\citep[Pan-STARRS;][]{chambers+16} $grizy$
and Two Micron All Sky Survey \citep[2MASS;][]{skrutskie+06} $JHK_s$ 
bands following the procedure described
in \citet{newbergyanny97}.
These are derived from $\sim 1$
million stars with detections in all bands,
estimated magnitude errors of $< 0.5\,{\rm mag}$, and
estimated dust extinction $E(B-V) < 0.01$
based on the 2-D integrated dust map from
\citet{sfd98}.
The photometry is then ``de-reddened''
by assuming all stars have the same reddening vector
(i.e. $\rvec_{\params} = \rvec$), where $\rvec$
is estimated from a $T_{\rm eff} = 7000\,K$ solar-metallicity
source spectrum at $E(B-V)=0.4\,{\rm mag}$ using the
extinction curve from \citet{cardelli+89} with
$R_V = 3.1$.

Once this 7-D locus has been derived, metallicity-dependent
absolute magnitudes $\absmag_{\params}$ are obtained
using the metallicity-dependent photometric parallax
relation given in \citet{ivezic+08}.
This makes the {\bayestar} stellar models primarily dependent
on only two parameters: the absolute magnitude $M_r$ in
the Pan-STARRS $r$-band and the ``metallicity''
$\feh$ of the star. The locus is extended
artificially to lower stellar masses (fainter $M_r$)
using the prescription outlined in \citet{green+14}.

Since this fitting is done exclusively in color space,
it is not sensitive to evolved giants with nearly-identical colors
as MS stars (i.e. ``dwarfs''). 
As a result, giant templates from \citet{ivezic+08},
derived from linear fits to globular cluster color-magnitude
diagrams (CMDs), are ``grafted'' onto the stellar locus based on their
corresponding $r - i$ color. While these templates are
quite limited relative to the diverse evolutionary states
included in the {\mist} models (see Figure \ref{fig:mist_vs_bayestar}),
they still capture basic features of the dwarf-giant degeneracy.

Since these templates are only defined based on specific photometric
bands and are not connected to any underlying stellar spectra,
they cannot be easily extended to other bands. 
Currently, they are only supported in the DECam, Pan-STARRS, and
2MASS system \citep{green+14,green+15,zuckerspeagle+19}.

\subsubsection{Dust Extinction Curve} \label{subsubsec:bayestar_red}

For the {\bayestar} models, we follow the approach taken in
\citet{green+14} and subsequent work by assuming
the reddening vector is again both linear and constant for all models
such that $\rvec_{\params} = \rvec$. As in \citet{green+19},
we take this reddening vector to be the empirical
reddening vector derived from \citet{schlafly+16}
re-normalized to scale with $A_V$ as described in \citet{zuckerspeagle+19}.
Similarly, we also assume that the differential reddening vector
is the same for all models such that $\rvec'_{\params} = \rvec'$,
where $\rvec'$ is again derived from \citet{schlafly+16}.

As with the empirical stellar models, it is also difficult
to extend these results to other photometric systems without
constraints on the underlying dust extinction curve.
\citet{schlafly+16}, however, do
provide a simple functional form for the dust curve that
reproduces the observational constraints 
for a particular stellar spectrum. This was used in
\citet{zuckerspeagle+19} to (re-)derive $\rvec$ and $\rvec'$
in the DECam, Pan-STARRS, and 2MASS photometric system.

\begin{figure*}
\begin{center}
\includegraphics[width=0.9\textwidth]{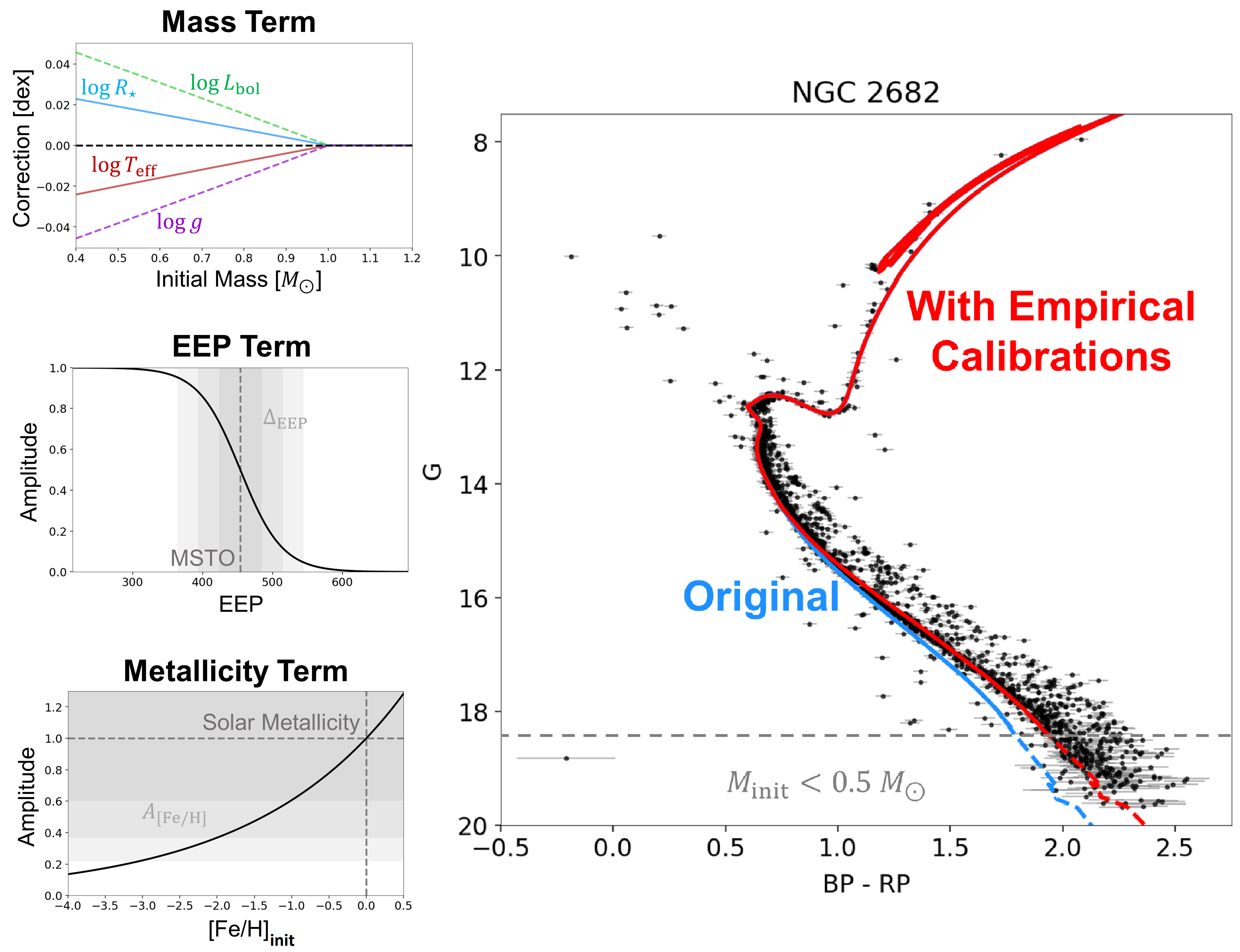}
\end{center}
\caption{An illustration of the components used in our
empirical {\mist} isochrone corrections. The top left panel shows
the underlying corrections, which we take to be linear adjustments
in $\log R_\star$ (solid blue) and $\log T_{\rm eff}$ (solid red)
as a function of initial mass, with no corrections above $M_{\rm init} = 1\,M_\odot$.
These then propagate to corrections in $\log g$ (dashed purple) 
and $\log L_{\rm bol}$ (dashed green) \textit{at fixed $T_{\rm eff}$}.
The middle left panel shows the function used to suppress the overall amplitude
of these corrections near the Main Sequence turn-off (MSTO) as a function of EEP, 
with intervals of the suppression scale $\Delta_{\rm EEP}$ highlighted with shaded grey
regions. The bottom left panel shows the function used to
modify the overall amplitude as a function of initial metallicity
relative to $\feh_{\rm init} = 0$, with intervals of the
$e$-folding amplitude $e^{A_{\feh}}$ highlighted with shaded gray regions.
The combined effect these have on a specific isochrone
is shown in the \textit{Gaia} $G$ versus $BP-RP$
color-magnitude diagram of NGC 2682 (i.e. M67) on the right,
with the data shown in black, the original isochrone in blue, and the
isochrone after applying these corrections in red. The approximate
location where $M_{\rm init} < 0.5\,M_\odot$ is indicated
with a dashed grey line (see Figure \ref{fig:ngc2682}).
The overall shape and behavior of the 
isochrone displays much better agreement with the data,
especially at lower masses, while the behavior near and beyond the MSTO
remains relatively unchanged.
See \S\ref{subsec:calib_terms} and \S\ref{subsec:cluster_eiso} for additional details.
}\label{fig:iso_corr}
\end{figure*}

\subsubsection{Priors} \label{subsubsec:bayestar_priors}

While we have used the same notation $\params$
to refer to the intrinsic parameters for both the {\mist} and {\bayestar}
models, the {\bayestar} models do not have age $t_{\rm age}$ 
or initial mass $M_{\rm init}$ estimates. As a result, inference
over these models does not incorporate our age prior
$\prior(t_{\rm age} | d, \ell, b)$ or IMF prior $\prior(M_{\rm init})$.
Instead, the {\bayestar} models are subject to a luminosity
function prior $\prior(M_r)$ taken from \citet{green+14}.
This is derived by computing the luminosity function 
$\prior(M_r | \feh, t_{\rm age})$ for a particular metallicity $\feh$
and age $t_{\rm age}$ based on the
PARSEC models \citep{bressan+12} assuming a
\citet{chabrier03} IMF and then marginalizing over
$\feh$ and $t_{\rm age}$ such that
\begin{align}
    \prior(M_r) = \int \prior(M_r | \feh, t_{\rm age}) 
    \prior(\feh, t_{\rm age}) \,{\rm d}\feh \,{\rm d}t_{\rm age}
\end{align}
where the prior over $\feh$ and $t_{\rm age}$ is the product
of two independent Normal distributions
\begin{equation}
    \prior(\feh, t_{\rm age}) = \Normal{\mu_{\feh}}{\sigma^2_{\feh}}
    \times \Normal{\mu_{t}}{\sigma_t^2}
\end{equation}
given $\mu_{\feh} = -0.5$, $\sigma_{\feh} = 0.5$, 
$\mu_t = 7\,{\rm Gyr}$, and $\sigma_t = 2\,{\rm Gyr}$.
The resulting $\pi(M_r)$ prior is tabulated over
a large grid of $M_r$ values that
is provided as part of the {\brutus} package.

\section{Empirical Calibration of MIST Isochrones} \label{sec:calib}

Unlike the {\bayestar} models, which are derived directly from
data, the {\mist} models (both the underlying {\mist} V1.2
isochrones and the atmospheric {\ctk} models) have known
systematic offsets in predicted stellar properties $\params_\star$
and observed photometric flux densities $\flux$
over particular sets of intrinsic parameters $\params$ \citep{choi+16}. 
This is particularly acute for, e.g., the very broad \textit{Gaia} filters.
These systematics can also become pronounced 
at lower stellar masses ($M_{\rm init} \lesssim 0.75$),
where the impact of stellar activity, magnetic fields,
convection, and molecular absorption features, among
others, become increasingly important.

To ameliorate some of these differences, we
develop limited empirical calibrations
for the {\mist} models described above.
The corrections and models are available as part of the
{\brutus} package.

We perform these calibrations in two steps. First,
we utilize a set of ``benchmark'' open clusters to
calibrate mass-dependent and metallicity-dependent
offsets in predicted surface properties from
the {\mist} isochrones as well as a series of
preliminary photometric offsets. Afterwards, we validate and
refine our photometric offsets using a large population
of nearby, high-latitude, low-reddening field stars
with good parallax measurements.

\begin{figure*}
\begin{center}
\includegraphics[width=\textwidth]{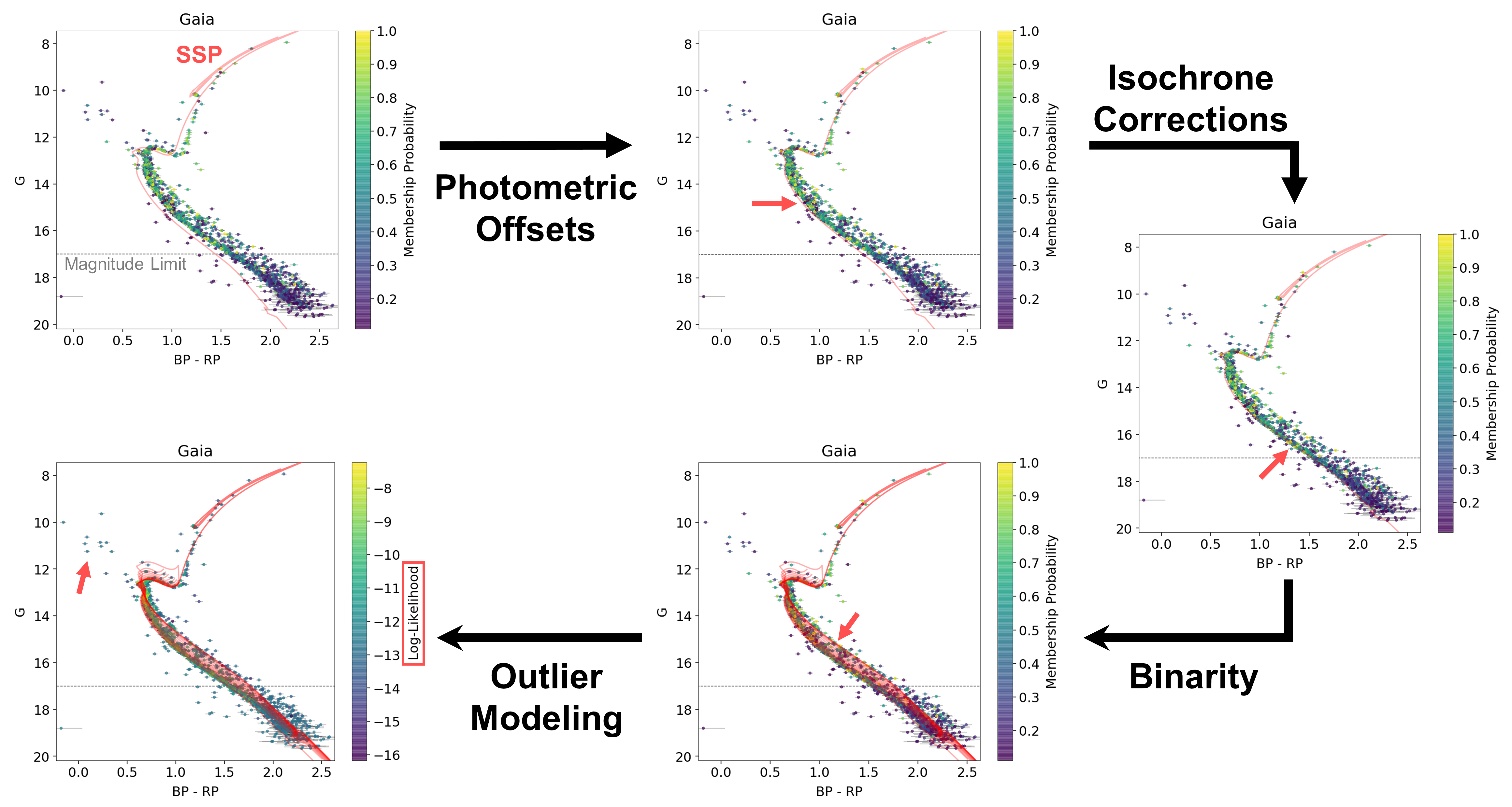}
\end{center}
\caption{An illustration of the components in the cluster model we use
as part of our empirical calibration of the {\mist} models, 
highlighting data from NGC 2682 (i.e. M67; see Figure \ref{fig:ngc2682}).
A total of six clusters were used.
The baseline component is a simple stellar population (SSP; top left),
which we integrate over for every single source for a given
set of cluster parameters $\params_{\rm cluster}$ such as age and metallicity.
We then add in a set of empirical
``corrections'' including overall photometric offsets
(top middle) and mass-dependent luminosity and temperature shifts (right)
that adjust the SSP model to better match the data 
(see Figure \ref{fig:iso_corr}).
We then integrate over possible contributions from unresolved binaries
(bottom middle) with unknown secondary companion masses. Finally,
we use information on cluster membership from astrometric measurements
from \textit{Gaia} DR2 along with mixture modeling
to compute robust likelihoods (bottom left).
This is done simultaneously over all sources and in all bands,
allowing us to leverage the full SED for each source along with
corresponding parallax measurements.
See \S\ref{ap:calib_clusters} for additional details.
}\label{fig:iso_fit}
\end{figure*}

We provide an overview of our empirical corrections
in \S\ref{subsec:calib_terms} and illustrate them 
in Figure \ref{fig:iso_corr}.
Our calibration using over a set of ``benchmark'' open
clusters is described in \S\ref{subsec:calib_benchmark}
and illustrated in Figure \ref{fig:iso_fit}.
Our calibration with field stars is described
in \S\ref{subsec:calib_field}.

Additional details describing our cluster model 
can be found in \S\ref{ap:calib_clusters}.

\subsection{Overview of Empirical Corrections} \label{subsec:calib_terms}

We use a series of \textit{empirically-motivated corrections} to address
systematic modeling issues derived from the use of
theoretical isochrones such as {\mist} and synthetic
spectra such as {\ctk}. These change the output surface-level parameters
 of a star to a ``corrected'' version 
$\params_\star(\params) \rightarrow \params_\star'(\params_\star, \params)$
as a function of the original predicted surface-level parameters
$\params_\star$ as well as the underlying stellar evolution
parameters $\params$ from the {\mist} isochrones. In particular,
we opt to modify the stellar radius $\log R_\star$ and the
effective temperature $\log T_{\rm eff}$ (and by proxy the
surface gravity $\log g$ and bolometric luminosity $\log L_{\rm bol}$)
such that
\begin{equation*}
    \begin{bmatrix}
    M_{\rm init} \\
    \feh_{\rm init} \\
    t_{\rm age}
    \end{bmatrix}
    \xrightarrow[]{}
    \begin{bmatrix}
    \log g \\
    \log T_{\rm eff} \\
    \log L_{\rm bol} \\
    \log R_\star \\
    \feh_{\rm surf}
    \end{bmatrix}
    \xrightarrow[]{}
    \begin{bmatrix}
    \log g' \\
    \log T_{\rm eff}' \\
    \log L_{\rm bol}' \\
    \log R_\star' \\
    \feh_{\rm surf}
    \end{bmatrix}
    \xrightarrow[]{}
    \begin{bmatrix}
    M_1 \\
    \vdots \\
    M_b
    \end{bmatrix}
\end{equation*}

We choose to apply empirical corrections to $\log R_\star$ and
$\log T_{\rm eff}$ for two reasons:
\begin{enumerate}
    \item We expect both to be strongly affected by magnetic fields
    (which are not included in {\mist}), 
    which appear to ``puff up'' stars (making them larger) and contribute to
    sunspot activity (making them cooler), especially at lower masses
    \citep{berdyugina05,somerspinsonneault15,somers+20}. 
    \item Detailed modeling of binaries already suggests
    that the {\mist} models deviate slightly from the observations
    in these two parameters \citep{choi+16}.
\end{enumerate}

To keep our empirical corrections as simple as possible,
we only introduce corrections for masses below $M_{\rm init} = 1\,M_\odot$
and ``suppress'' the effects of our derived corrections after
stars evolve off the MS (i.e. after stars have ${\rm EEP} > 454$)
and for sub-solar metallicities (where the fits are relatively unconstrained;
see \S\ref{subsec:calib_benchmark}). We further assume
that our corrections only involve a single parameter, $M_{\rm init}$,
and that they are fully linear\footnote{While we experimented with more
complex functional forms, we found that there was not
enough data to warrant using them.} with slopes $c_R$ and $c_T$ for 
$\log R_\star$ and $\log T_{\rm eff}$, respectively. We suppress
effects for evolved stars\footnote{While there are known 
disagreements between the {\mist} models
and observations for post-MS stellar evolutionary phases 
\citep{choi+16}, investigating them is beyond the scope of this work.}
over a scale of $\Delta_{\rm EEP}$
and for sub-solar metallicities with an amplitude of $A_{\feh}$.
Effects are then propagated to other parameters such as $\log g$ and
$\log L_{\rm bol}$ in a self-consistent manner.

\begin{deluxetable}{lcc}
\tablecolumns{3}
\tablecaption{A summary of the parameters used to apply the
empirical {\mist} isochrone corrections used in {\brutus}. 
See \S\ref{ap:calib_clusters} for additional details.
\label{tab:iso_corr}}
\tablehead{Description & Symbol & Value}
\startdata
Slope of $M_{\rm init}$-dependent $T_{\rm eff}$ correction & $c_T$ & $+0.09$ \\
Slope of $M_{\rm init}$-dependent $R_\star$ correction & $c_R$ & $-0.09$ \\
Scale of EEP suppression & $\Delta_{\rm EEP}$ & 30 \\
Amplitude of $\feh$ suppression & $A_{\feh}$ & 0.5 \\
\enddata
\end{deluxetable}

\begin{figure*}
\begin{center}
\includegraphics[width=\textwidth]{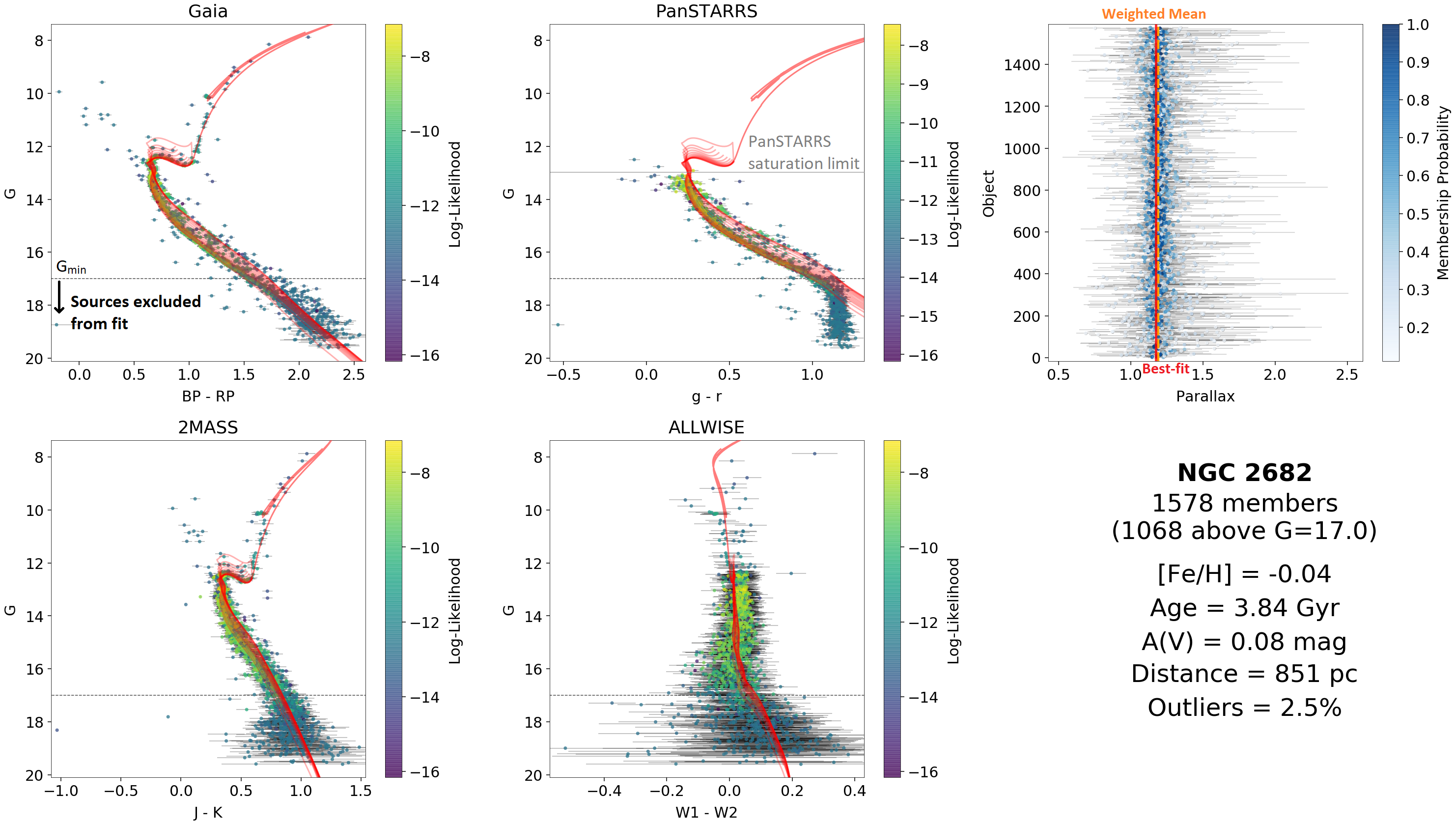}
\end{center}
\caption{Our best-fit cluster model (see \S\ref{ap:calib_clusters})
to \textit{Gaia} DR2, Pan-STARRS, 2MASS, and AllWISE data 
for NGC 2862 (i.e. M67) for all stars with $G > 17.0$
(horizontal dashed black line). Color-magnitude diagrams (CMDs) for
the sources associated with the cluster are shown in the left four
panels. Points are colored according to their log-likelihoods 
with respect to our best-fit model, whose associated isochrone 
tracks are overplotted in red. The parallax measurements for the individual
sources, colored according to their membership probability, are shown in the top right
along with the weighted mean (dashed orange line) and best-fit distance
(solid red). A summary of the best-fit parameters are shown on the bottom right.
We see that the overall model is a good fit to the SED overall,
capturing the behavior across the MS and post-MS down to
$G_{\rm min} = 17$ ($M_{\rm init} \sim 0.5\,M_\odot$),
and that the outlier model does a good job
of reducing the contribution of the small ($\approx 2.5\%$)
percentage of outlying sources such as
blue stragglers (top left area of CMDs). The values for
for the age, metallicity, and $A(V)$ are
in excellent overall agreement with measurements from
the literature.
}\label{fig:ngc2682}
\end{figure*}

An illustration of the functional forms used for the corrections
themselves is shown in Figure \ref{fig:iso_corr}.
See Table \ref{tab:iso_corr} for a summary of the parameters used to model
empirical corrections used in this work and their final set of values.
We find that the overall empirical corrections substantially improve behavior down to
$M_{\rm init} \sim 0.5\,M_\odot$, which can be seen more clearly in
Figure \ref{fig:ngc2682}.

In addition to these isochrone-oriented corrections, we also
fit for a set of \textit{photometric offsets} to account for
slightly different photometric calibrations between the synthetic photometry
computed from models versus the real photometry from surveys
or issues with the {\ctk} stellar atmosphere models. 
We model these offsets explicitly by introducing a set of scale-factors
$\mathbf{s}_{\rm em} = \{ s_{{\rm em}, i} \}_{i=1}^{i=b}$ 
that simply rescale the \textit{data} such that
the new flux density $\hat{F}_{i,j}'$ for a given star $i$ in band $j$ is
\begin{equation}
    \hat{F}_{i,j}' = s_{{\rm em}, j} \times \hat{F}_{i, j}
\end{equation}
We are able to do so thanks to
the large number of available parallax measurements from
\textit{Gaia} DR2 that give independent constraints on the distance,
thereby fixing not just offsets in color but offsets in absolute
magnitude.

The impact these two sets of empirical corrections have on 
our cluster model is shown in the fourth and fifth panels
of Figure \ref{fig:iso_fit}.

\subsection{Calibration with ``Benchmark'' Clusters} \label{subsec:calib_benchmark}

The full cluster model (\S\ref{ap:calib_clusters}) includes both
cluster-level parameters $\params_{\rm cluster}$ and the empirical correction
terms described in \S\ref{subsec:calib_terms}. While typically
we would be interested in inferring $\params_{\rm cluster}$,
in this work we are instead interested in constraining 
the empirical correction parameters
$c_T$, $c_R$, and $\mathbf{s}_{\rm em}$ that modify the
{\mist} models. We estimate these parameters by marginalizing over
$\params_{\rm cluster}$ based on fits to a series of ``benchmark''
clusters. To do this, we use the three-step approach:
\begin{enumerate}
    \item Use optimization methods to compute the 
    maximum-likelihood estimate (MLE)
    for $\params_{\rm cluster}$, $c_T$, $c_R$, and 
    $\mathbf{s}_{\rm em}$ for each cluster, allowing
    all parameters to vary.
    \item Fix $c_T$ and $c_R$ to be roughly the median
    value across all of the benchmark clusters, then re-optimze
    the results starting from the previous MLE.
    \item Fix $\mathbf{s}_{\rm em}$ to be the roughly
    the median value across all of the benchmark clusters,
    the re-optimize the results starting from the previous
    MLE.
\end{enumerate}
We perform this iterative conditional optimization using a combination
of Powell's method \citep{powell64} and Nelder-Mead \citep{neldermead65}
implemented within the \textsc{optimize.minimize} routine in {\scipy} 
\citep{virtanen+20}. To avoid being biased by difficulties modeling 
low-mass sources, for each cluster
we impose a \textit{Gaia} DR2 $G$-band cutoff $G_{\rm min}$ below which
no sources are included in the fit.

The data we use to perform this calibration come from six open clusters:
NGC 2548 (M48), NGC 752, NGC 188, NGC 2632 (Praesepe), 
NGC 2682 (M67), and NGC 3532. 
The reasons for choosing these particular open clusters are as follows:
\begin{itemize}
    \item \textit{Well-studied}:
    Each cluster has been the subject of numerous studies and
    has reasonably well-known literature values
    of the overall metallicity, age, and distance,
    often validated using multiple methods.
    \item \textit{Wavelength coverage}: With the exception of NGC 3532
    (which does not have Pan-STARRS coverage),
    all clusters have photometry and astrometry measurements from
    \textit{Gaia} DR2 and photometry from Pan-STARRS, 2MASS, and AllWISE.
    The long wavelength range is crucial for disentangling the relative
    impacts of dust compared to those expected from our empirical corrections.
    \item \textit{Metal-rich}: Because models of evolved stars on the
    horizontal giant branch at lower metallicities are unreliable, all clusters are required
    to have $\feh_{\rm init} \sim 0$. We hope to extend our sample down to
    lower metallicities and improve upon our outlier modelling in future work.
\end{itemize}

Cluster membership probabilities $p_{{\rm mem}, i}$ were assigned using the
Hierarchical Density-Based Spatial Clustering of Applications with Noise (HDBSCAN)
algorithm \citep{campello+13,mcinnes+17a,mcinnes+17b} using proper motion
and parallax measurements from \textit{Gaia} DR2. Photometry was taken from
Pan-STARRS, 2MASS, and AllWISE \citep{cutri+13} surveys and cross-matched
using Pan-STARRS as the primary catalogue within a 0.5 arcsecond radius.
All photometric errors had 0.02 mags added in quadrature to account
for possible systematic effects (see also \S\ref{subsec:calib_field}).

The final fit to NGC 2682 (M67) is shown in Figure \ref{fig:ngc2682}
to illustrate the overall quality of our results. The fits to the remaining
five clusters are shown in \S\ref{ap:cluster}.
The final set of empirical isochrone corrections are listed in
Table \ref{tab:iso_corr} while the final set of
empirical photometric offsets
are listed in Table \ref{tab:phot_off}. We find the
resulting temperature and radius offsets agree well
with observations from binary systems
presented in \cite{choi+16}. When we subsequently
discuss the {\mist} models, the assumption should be that
these isochrone-level corrections have already been applied 
unless explicitly stated otherwise.

\subsection{Calibration with ``Benchmark'' Field Stars} \label{subsec:calib_field}

While the overall behavior of the empirical isochrone
corrections was somewhat stable across clusters, we found
there was substantially more variation in the fitted photometric
offsets $\mathbf{s}_{\rm em}$ from cluster to cluster.
This led us to develop an independent
pipeline to derive and validate photometric offsets
between the {\mist} models and the data using
a sample of nearby, low-reddening stars with high
signal-to-noise parallax measurements.

\begin{figure*}
\begin{center}
\includegraphics[width=\textwidth]{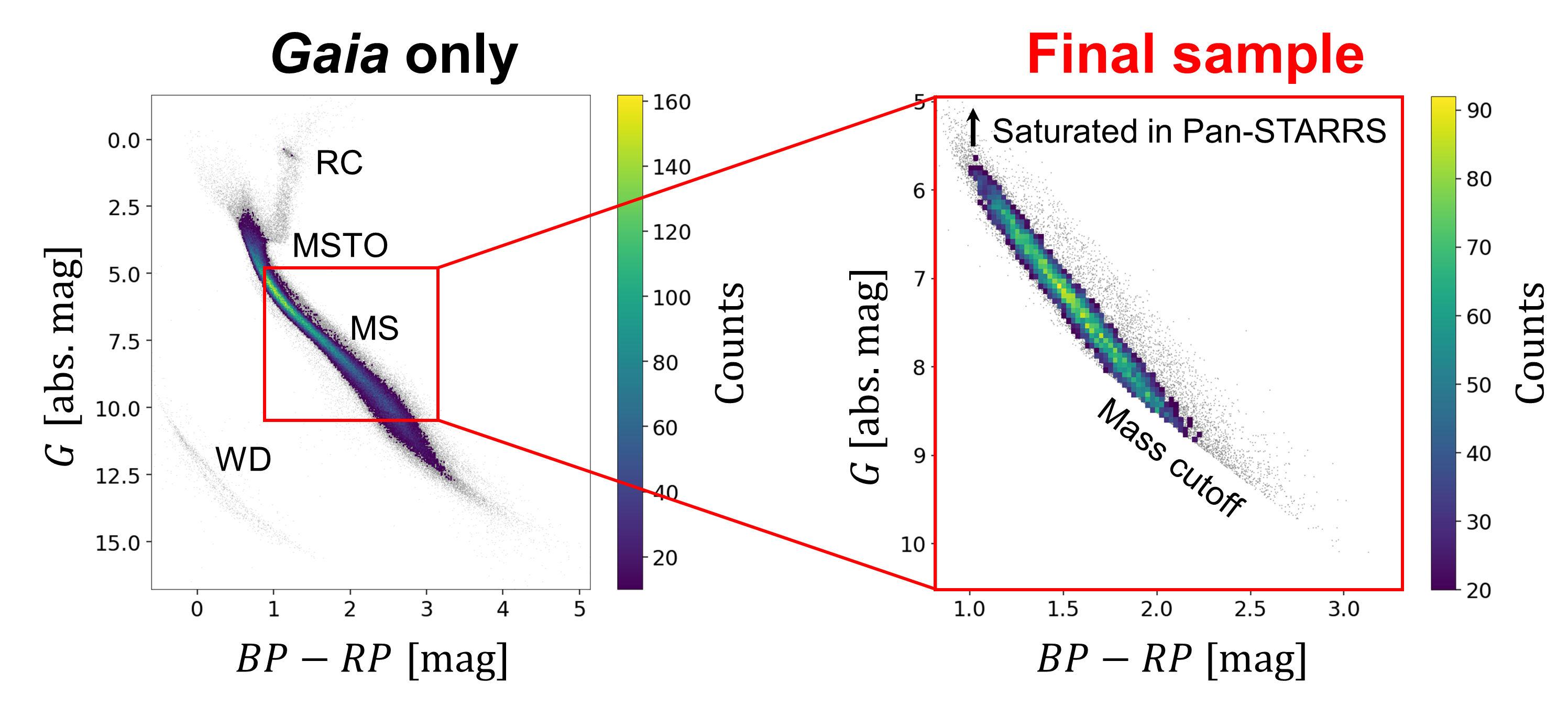}
\end{center}
\caption{The number density of sources across the
\textit{Gaia} $G$ versus $BP - RP$ CMD
for the set of ``benchmark'' field stars used to calibrate 
empirical photometric offsets for the {\mist} and
{\bayestar} models assuming a distance of
$1\,{\rm kpc}=1/(1\,{\rm mas})$. Above a given
threshold colored histograms are used to highlight the number density; below
this threshold individual sources are plotted are gray points. The left panel
shows the set of stars with high Galactic latitude ($|b| > 60$),
low reddening ($E(B-V) < 0.2$), and well-measured parallax
($\hat{\varpi}/\sigma_\varpi > 50$), with a few stellar populations labeled 
(WD = white dwarf, MS = main sequence, MSTO = main sequence turn-off,
RC = red clump). The final collection of $n=22933$ stars after requiring
full photometric coverage in \textit{Gaia} DR2, Pan-STARRS, 2MASS, and AllWISE
is highlighted in the red box and shown on the right. The impact of the
Pan-STARRS $r < 14$ saturation limit and the $M_{\rm init} > 0.55\,M_\odot$
initial mass cutoff are labeled.
}\label{fig:gaia_cmd}
\end{figure*}

\begin{deluxetable}{lccc}
\tablecolumns{4}
\tablecaption{A summary of the photometric offsets
(see \S\ref{subsec:cluster_ephot}) derived for
the empirical {\mist} isochrone corrections used in {\brutus}
along with the estimated systematic uncertainties in the models.
See \S\ref{subsec:calib_benchmark} and \S\ref{subsec:calib_field}
for additional details on the values derived using open clusters
and high-latitude field stars, respectively. 
Note that the field values (highlighted in bold)
are the ones implemented by default. 
We find offsets/uncertainties of roughly 2\% uncertainties in the optical,
3\% in the NIR, and 4\% in the IR.
\label{tab:phot_off}}
\tablehead{Filter & Value (Cluster) & \textbf{Value (Field)} & Uncertainty}
\startdata
\cutinhead{\textit{Gaia} DR2\tablenotemark{a}}
$G$ & 1.03 & \textbf{1.01} & 0.02 \\
$BP$ & 1.05 & \textbf{1.02} & 0.02 \\
$RP$ & 1.00 & \textbf{0.97} & 0.02 \\
\cutinhead{Pan-STARRS}
$g$ & 1.03 & \textbf{1.01} & 0.02 \\
$r$ & 0.96 & \textbf{0.97} & 0.02 \\
$i$ & 0.98 & \textbf{0.97} & 0.02 \\
$z$ & 0.99 & \textbf{0.96} & 0.02 \\
$y$ & 1.00 & \textbf{0.97} & 0.02 \\
\cutinhead{2MASS}
$J$ & 1.00 & \textbf{0.99} & 0.03 \\
$H$ & 1.04 & \textbf{1.04} & 0.03 \\
$K_s$ & 1.02 & \textbf{1.04} & 0.03 \\
\cutinhead{WISE}
$W1$ & 1.02 & \textbf{1.02} & 0.04 \\
$W2$ & 1.07 & \textbf{1.03} & 0.04 \\
\enddata
\tablenotetext{a}{Offsets were derived using
the latest \textit{Gaia} DR2 filter curves from
\citet{maizapellanizweiler18}. They shift by
$\sim 0.03$ mag when using previously 
published filter curves.}
\end{deluxetable}

The idea behind this method is straightforward.
If a star has very precise parallax measurements,
there is not much allowed variation in the overall
distance. The overall lack of dust also guarantees a fairly
accurate measure of the underlying luminosity.
Taken together, we thus expect that any photometric offsets
between the model and the data must be primarily due to
$\textbf{s}_{\rm em}$. While this measurement will be noisy for
any individual source and likely vary as a function of $\params$,
averaging over a large enough number of
sources allows us to estimate $\textbf{s}_{\rm em}$ robustly.
Details on this procedure can be found in \S\ref{ap:phot_offsets}.

We apply this overall procedure to a set of field stars 
selected using the following criteria:
\begin{itemize}
    \item $|b| > 60^\circ$: High Galactic latitude to avoid crowding and dust.
    \item $E(B-V) < 0.2$ mag: Low total line-of-sight reddening, as estimated
    from the \citet{sfd98} dust map.
    \item $\hat{\varpi}/\sigma_\varpi > 50$: Extremely high signal-to-noise
    parallax measurements to ensure absolute distances are essentially fixed.
    \item \textit{Full photometric coverage}: We require all sources have
    detections in \textit{Gaia}, Pan-STARRS, 2MASS, and AllWISE.
    \item $M_{\rm init} > 0.55\,M_\odot$: We require all sources to
    have initial masses that are sufficiently away from the
    low edge of our mass grid ($M_{\rm init} = 0.5\,M_\odot$).
    Using the {\mist} models, we approximate this to be the case
    when $G < -0.052 \times (BP-RP)^2 + 1.88 \times (BP-RP) + 4.97$, where
    $G$ is the \textit{absolute} \textit{Gaia} $G$-band magnitude estimated using
    the measured parallax.
\end{itemize}
This leaves us with a sample of $n=22933$ stars. Their 
distribution in the \textit{Gaia} $G$ versus $BP-RP$ CMD 
is shown in Figure \ref{fig:gaia_cmd}.
The resulting photometric offsets are listed in Table \ref{tab:phot_off}.

Overall, we find that the offsets for the empirical {\bayestar} models
are on the order of $\lesssim 1\%$, which is expected since 
the {\bayestar} models were originally constructed 
using these exact datasets. The resulting offsets
for the {\mist} models (with empirical isochrone corrections)
are on the order of $\sim 3\%$, in agreement with
the results from \S\ref{subsec:calib_benchmark}, although the exact
values differ somewhat between bands. We find that applying these offsets
substantially improves the quality of the fits for the {\mist}
models, increasing the number of objects with $\chi^2/b \lesssim 1$
by $\sim 20\%$. In total, we find that $\sim 95\%$ of the sample
is well-fit by both sets of models.

The behavior of these offsets for the {\bayestar} and {\mist}
models as a function of magnitude and position on the \textit{Gaia}
CMD can be found in \S\ref{ap:cmd_resid}.

\subsection{Additional Remarks} \label{subsec:calib_remarks}

We emphasize that the corrections outlined in the previous sections 
should only be seen as ``functional'' rather
than fundamental. They have been calibrated and are designed to be applied
to a particular set of stellar models to improve performance in a particular
context, and are not expected to apply more generally. This is because
these corrections really are a combination of four separate underlying
issues:
\begin{enumerate}
    \item \textit{Shortcomings in stellar evolution models}. 
    The models do not account for magnetic fields, radius inflation, 
    and stellar activity, all of which are important factors for low-mass and evolved stars.
    \item \textit{Shortcomings in the stellar atmospheric models}. While
    the {\ctk} models reproduce color-$T_{\rm eff}$ relations for most stars
    except those at the lowest masses, limited molecular line lists and treatment
    of convection might lead to $T_{\rm eff}$-dependent systematics
    at $< 4000\,{\rm K}$ (where molecules become important) and for giants
    (where 3-D effects become more important).
    \item \textit{Shortcomings in the dust extinction curve}. The dust extinction curve
    is known to vary substantially throughout the Galaxy. While the \citet{fitzpatrick04}
    curve used here has been shown to reproduce the overall reddening for giants,
    modeling errors would lead to systematics as a function of $A_V$ and $R_V$ that
    are almost entirely degenerate with changes in $T_{\rm eff}$.
    \item \textit{Shortcomings in the flux calibration in the data}. Based on
    many recent surveys and work done in 
    \citet{portillospeaglefinkbeiner20} investigating magnitude-dependent biases, we expect these
    to be at the $\sim 2\%$ level, although it may be higher in some cases. This should
    lead to a single, global offset for each band (up to the non-linearity of the detector).
\end{enumerate}

Using internal testing, we have verified that the offsets derived here are broadly consistent
across similar photometric systems in both the optical and NIR, which suggests that the bulk of the
offsets in Table \ref{tab:phot_off} are \textit{not} simply absolute zero-point issues. Investigating
the exact causes for these effects, however, is beyond the scope of this paper. 
We hope to investigate these in more detail in future work.

\section{Validation} \label{sec:tests}

To validate and examine the performance of {\brutus}, we consider two classes of tests:
\begin{enumerate}
    \item \textit{Tests on mock data}: Tests on mock (i.e. simulated) data
    are important to understand in detail
    how {\brutus} performs in various scenarios
    and how well it can recover stellar parameters.
    \item \textit{Independent comparisons}: Comparing results against
    independent methods allows the characterization of systematics
    when applying {\brutus} to real data.
\end{enumerate}

We describe each class of tests in the following subsections.
In \S\ref{subsec:test_mock}, we discuss 
a suite of mock tests to illustrate how well {\brutus} is able
to recover intrinsic stellar pameters $\params$ under various conditions. 
In \S\ref{subsec:test_h3}, we examine how well
{\brutus} is able to estimate distances compared to
\textit{Gaia} DR2 parallax measurements and spectrophotometric
distances from \citet{cargile+20}.

A third class of tests (``A/B tests'') where various features of the
code were changed one-at-a-time to examine the impact of underlying
model choices is discussed in \S\ref{ap:test_cases}.

\subsection{Tests on Mock Data} \label{subsec:test_mock}

\begin{figure*}
\begin{center}
\includegraphics[width=\textwidth]{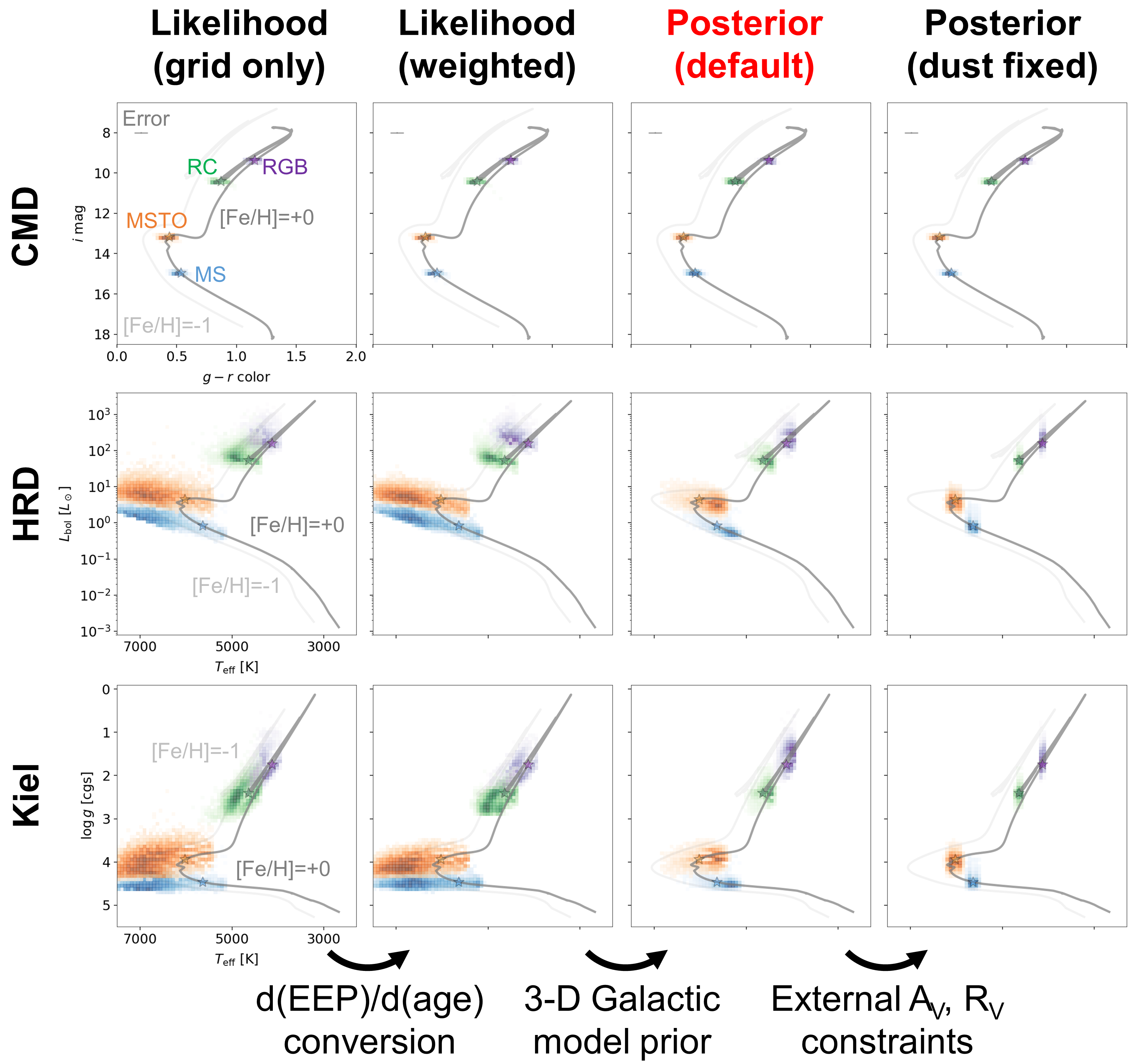}
\end{center}
\caption{Parameter recovery for a mock 10-band SED (Pan-STARRS, 2MASS, and WISE)
with a noisy parallax measurement ($\hat{\varpi}/\sigma_\varpi = 5$),
$A_V=0.4$, $R_V=3.3$, and $d=1\,{\rm kpc}$ located along the same sightline 
$(\ell, b) = (90^\circ, 20^\circ)$ shown in 
Figures \ref{fig:prior_sightline} and \ref{fig:prior_dust}.
Four cases are highlighted: a star on the
main sequence (MS; blue), at the MS turn-off (MSTO; orange), 
on the red giant branch (RGB; purple), and at the red clump (RC; green). 
The input values on the $i$ versus $g-r$ color-magnitude diagram (CMD; top),
Hertzsprung-Russel diagram (HRD; middle), and Kiel diagram (bottom)
are indicated with a star. $\feh_{\rm init} = 0$
and $\feh_{\rm init}=-1$ isochrones are also 
shown in dark gray and light gray, respectively.
The 2-sigma measurement errors are highlighted in the top left corner of the CMD.
Samples from the resulting probability density functions (PDFs)
derived from {\brutus} are shown as shaded regions taking into account
constraints from the SED and measured parallax (far left), 
uneven model sampling over our grid (middle left),
the full 3-D Galactic prior (middle right), and additional (tight)
constraints on $A_V$ and $R_V$ (far right).
Axes labels are only included in the leftmost panels for visual clarity.
The initial mass function (IMF) prior, combined with the 3-D dust prior,
leads {\brutus} to prefer intrinsically redder, cooler sources on/near the MS,
although the true solution is clearly still captured. 
Strong constraints on $A_V$ and $R_V$ give unbiased parameter recovery.
Note that the CMD PDFs show the quality of the fits in all
cases are essentially identical, highlighting the strong impact our 
prior has on the inferred parameters.
}\label{fig:mock_pars_1}
\end{figure*}

\begin{figure*}
\begin{center}
\includegraphics[width=\textwidth]{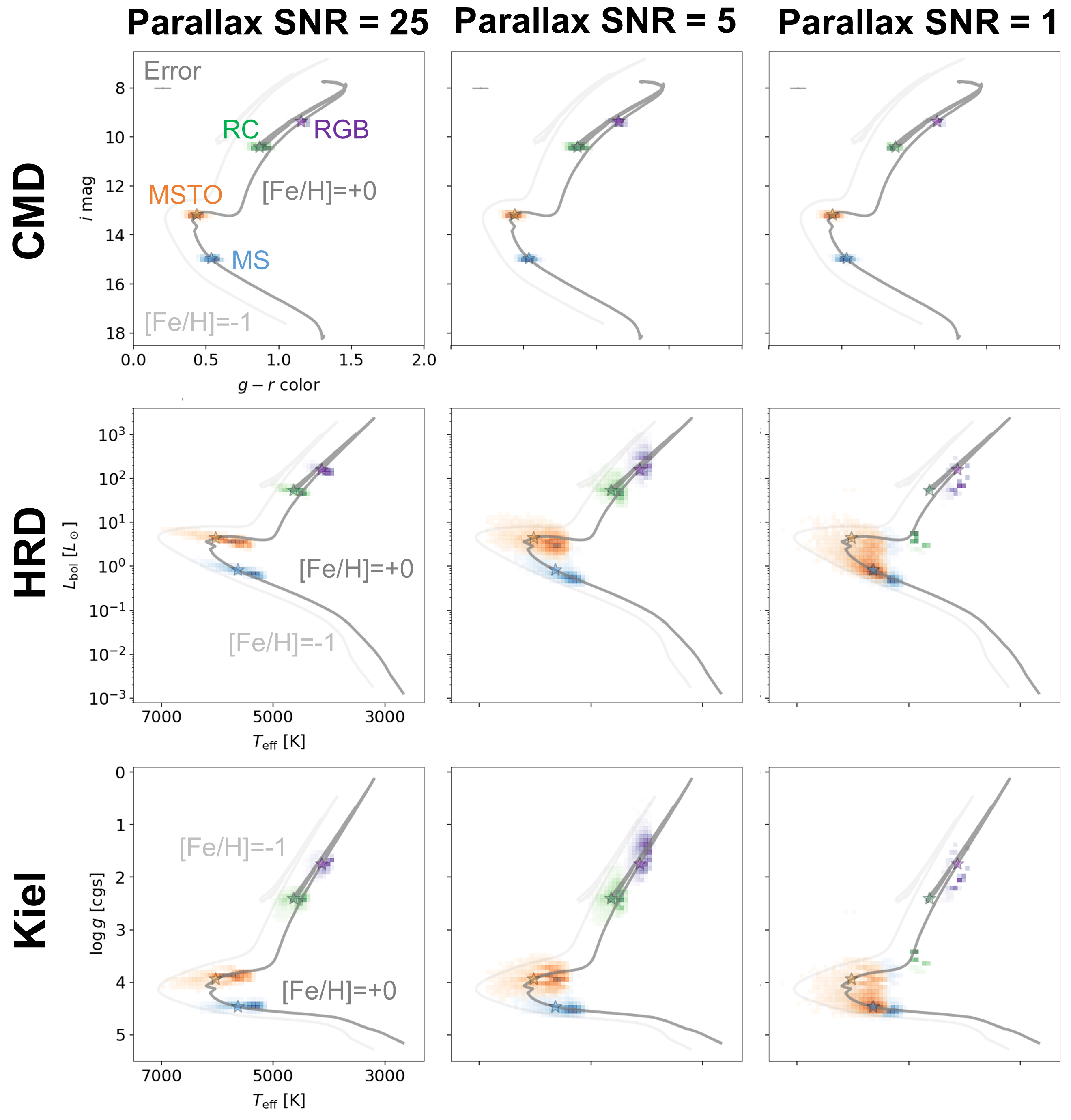}
\end{center}
\caption{As Figure \ref{fig:mock_pars_1}, but now showing
the posterior as a function of the the measured parallax
signal-to-noise (SNR) ratio ${\rm SNR}=\hat{\varpi}/\sigma_\varpi$ 
for ${\rm SNR}=25$ (left), ${\rm SNR}=5$ (middle), and ${\rm SNR}=1$.
As in Figure \ref{fig:mock_pars_1}, we see that 
even for objects with well-constrained distances 
uncertainty on the dust extinction
leads to specific degeneracies in the HR and Kiel diagrams 
along with preferences for cooler, lower-mass sources.
These uncertainties are broadened in $\log L_{\rm bol}$
and $\log g$ for parallaxes with lower SNR due to larger
allowed variation in the estimated distance. At extremely low
SNR, the impact of the prior leads to a strong preference
for lower masses and longer-lasting phases of stellar evolution,
leading to lower estimates on the MS, a preference for the MS over
the MSTO, and a preference for the MSTO over the RC. For models that
have evolved off the post-MS, the preference for lower-mass solutions
and lower dust extinction leads to a preference of higher metallicities
and therefore intrinsicly redder and cooler SEDs, shifting solutions 
down and to the right on the HR and Kiel diagrams.
}\label{fig:mock_pars_2}
\end{figure*}

\begin{figure*}
\begin{center}
\includegraphics[width=\textwidth]{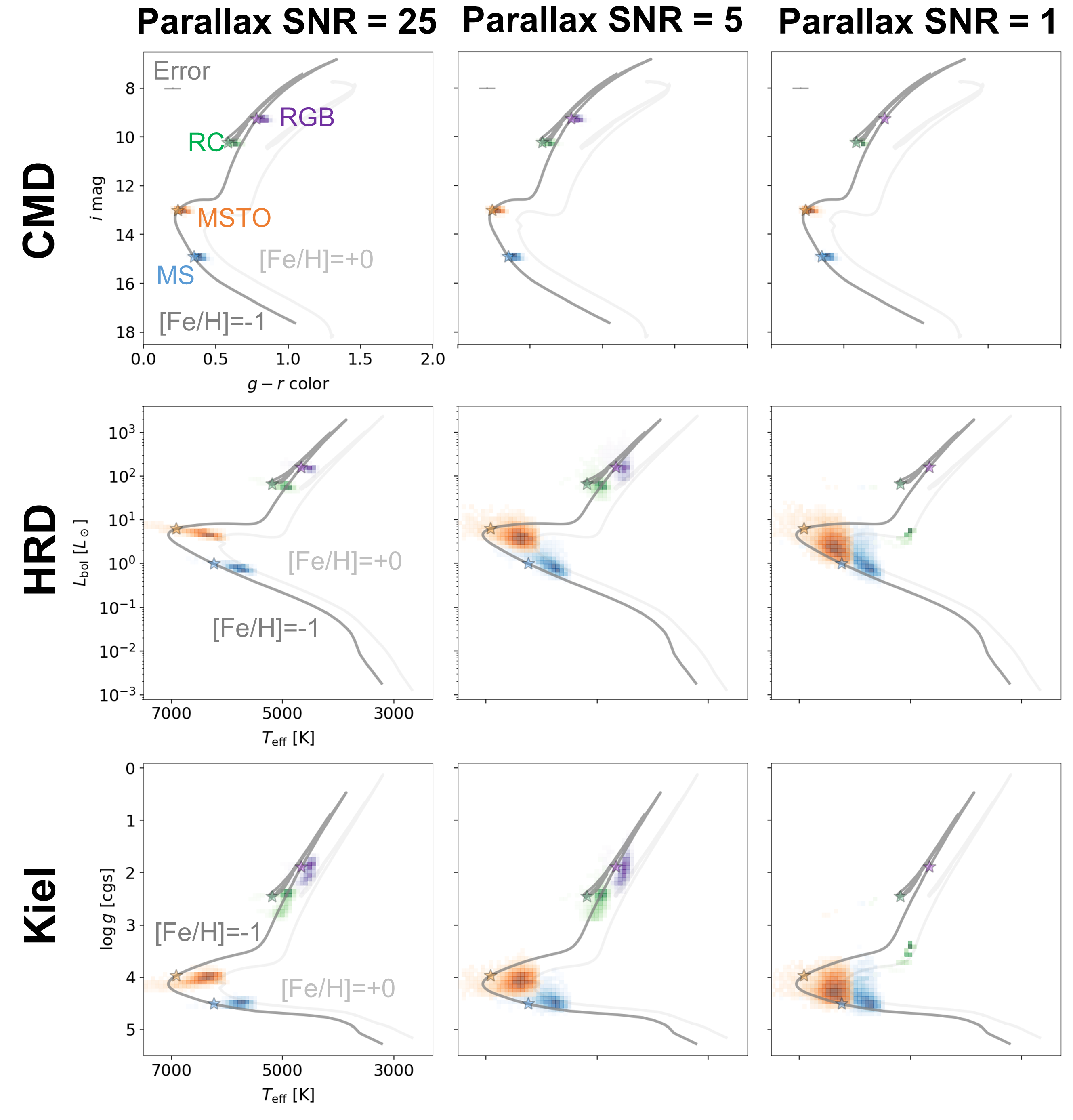}
\end{center}
\caption{As Figure \ref{fig:mock_pars_2}, but now for
sources with $\feh_{\rm init} = -1$ rather than $\feh_{\rm init} = 0$.
At this location, our prior strongly prefers metallicities that
are consistent with the thin disk, leading to a bias
towards redder observed colors in the observed CMD. This remains
true even at high parallax SNR (although
the true solution remains within the uncertainties). 
For the RGB solution at low parallax SNR, the prior belief against
the correct stellar parameters is so strong that all relevant fits are
prematurely removed before the optimization and Monte Carlo sampling
steps (see \ref{subsec:methods_cuts}), causing {\brutus} to fail to produce
a reliable posterior PDF.
}\label{fig:mock_pars_3}
\end{figure*}

Methods that utilize both spectra and photometry are able to get independent
constraints on stellar parameters from the spectra (e.g., $\log g$, $\feh$)
and photometry (e.g., $T_{\rm eff}$). 
This enables complementary constraints on distance 
(via $\log g \rightarrow \log R_\star \rightarrow \log L_{\rm bol}$)
as well as on intrinsic color (via $\feh_{\rm surf}$).
Photometry alone, however, does not provide these additional
constraints and therefore can't break these degeneracies. This
leads to a much stronger reliance on both the underlying 
Galactic prior and on independent distance constraints from
parallax measurements $\hat{\varpi}$.

In order to examine how well {\brutus} is able to recover intrinsic stellar 
parameters $\params$ from photometry and astrometry alone,
we generate mock data for stars at several evolutionary states:
\begin{itemize}
    \item ${\rm EEP} = 350$: On the Main Sequence (MS).
    \item ${\rm EEP} = 450$: On the MS turn-off (MSTO).
    \item ${\rm EEP} = 550$: On the first ascent up the Red Giant branch (RGB).
    \item ${\rm EEP} = 650$: On the Red Clump (RC).
\end{itemize}
As in \citet{cargile+20}, we choose to generate models from a
$t_{\rm age} = 5\,{\rm Gyr}$ {\mist} isochrone. This spans
a wide range of evolutionary states (with differing evolutionary timescales)
and a moderate range in initial mass $M_{\rm init}$, leading to a diverse
set of SEDs.

To highlight the impact of the prior, we choose to place these sources
at a distance of $d = 1\,{\rm kpc}$ along the line-of-sight located
at Galactic coordinates of $(\ell, b) = (90^\circ, 20^\circ)$. As highlighted
in Figure \ref{fig:prior_sightline}, this sightline intersects
all major components of our prior, thereby providing an
illustrative example of how {\brutus} performs. At
$d = 1{\rm kpc}$, in particular, an object is strongly
favored to be a member of the thin disk. As a result,
we generate two models: one with $\feh_{\rm init} = 0$, 
which is consistent with the expected metallicity of our prior, and
one at $\feh_{\rm init} = -1$, which is more consistent with
the thick disk and halo and in tension with our prior.

As shown in Figure \ref{fig:prior_dust}, this sightline also
displays a small but non-negligible amount of dust extinction. 
To examine systematic deviations from the prior, we set the
extinction $A_V=0.4\,{\rm mag}$ to be 
somewhat above the mean expected from the 
3-D dust extinction prior at $d=1\,{\rm kpc}$. We opt
to leave $R_V = 3.3$ to keep it in line with prior expectations.

In Figure \ref{fig:mock_pars_1}, we illustrate
the impact of the prior for a 10-band SED (Pan-STARRS,
2MASS, and WISE) for a source with a measured parallax
$\hat{\varpi} = 1.0\,{\rm mas}$ with a SNR of 5.
As shown in the top panels highlighting the CMDs,
the fitted SEDs from {\brutus} are nearly identical.
However, the inferred parameters differ substantially
between the initial fitted likelihoods and the final inferred posteriors.
In particular, we see that the IMF prior, combined with the 3-D dust prior,
leads {\brutus} to prefer \textit{intrinsically redder} sources
on/near the MS. This leads to a bias towards cooler 
inferred $T_{\rm eff}$ values, 
although the true solution is clearly still captured. 
To illustrate that this is the ultimate cause, we add in strong
constraints on $A_V$ and $R_V$ so that there is no additional
degeneracy between $T_{\rm eff}$ and $A_V$. As expected, in this case
we achieve unbiased parameter recovery. In general, we find
we are able to recover parameters at each EEP quite well, with the
uncertainties decreasing for more evolved stars with more distinctive SEDs.

In Figure \ref{fig:mock_pars_2}, we illustrate how the inferred
parameters vary as a function of parallax SNR. At high SNR
($\hat{\varpi}/\sigma_\varpi=25$), the distance is tightly constrained,
leading to the majority of the variation captured in the inferred $T_{\rm eff}$
and $A_V$. As expected, due to the IMF prior and the slightly higher
$A_V$ value compared to what is expected from the prior, the
posterior prefers somewhat intrinsically redder sources, leading to
cooler inferred $T_{\rm eff}$ (although the true solution 
is again still captured). At $\hat{\varpi}/\sigma_\varpi=5$,
the distance is still constrained to be around $d \sim 1\,{\rm kpc}$
but there is more flexibility to shift the star around, leading
to larger uncertainties in $\log L_{\rm bol}$ and $\log g$.
At $\hat{\varpi}/\sigma_\varpi=1$, however,
the constraint on the distance is much weaker. At this point,
the influence of the prior dominates, preferring stars
to be at longer-lasting evolutionary stages and lower masses.
This shifts the MS star to lower masses, the MSTO start to be
on the MS, and the RC star to instead be at the end of the MSTO
before beginning the ascent up the RGB. The RGB star, which 
is uniquely constrained by the SED, instead simply
prefers a lower-mass, higher-metallicity solution.

Finally, in Figure \ref{fig:mock_pars_3} we illustrate
the $\feh_{\rm init}=-1$ case where the metallicity deviates
substantially from the expected value based on the Galactic prior. 
In this case, the prior is strong enough that
we actually observe it ``pulling'' the SED to be intrinsically
redder in color, as shown on the CMDs. This also leads the
prior to infer metallicities around $\feh_{\rm init} = -0.5$
rather than $-1$, making the inferred temperatures substantially
cooler (and the intrinsic SED redder) compared to their actual
values, although again the true solution is clearly 
still captured. When the parallax SNR becomes $\sim 1$,
we further see solutions being pulled towards the
$\feh_{\rm init} = 0$ isochrone and shifted in intrinsic colors
appropriately as a result, mirroring the behavior from 
Figure \ref{fig:mock_pars_2}. We also
find the RGB solution behaves differently than the other three
cases. Due to its rapid evolution, we find that
with the default settings the prior belief against the RGB
solution is so strong the {\brutus} actually clips the
true solution before the final optimization/sampling step
and fails to produce a reasonable posterior estimate.
This highlights one of the risks with the implementation
outlined in \S\ref{subsec:methods_cuts}, which can
eliminate extremely rare (but possible) solutions from
the fitting process. We confirm that true solution can be
properly recovered when the thresholds used for clipping
fits are substantially relaxed.\footnote{Some of these
changes have been implemented in more recent versions of the code, 
but are not included in \texttt{v0.7.5}.}

\begin{figure}
\begin{center}
\includegraphics[width=0.45\textwidth]{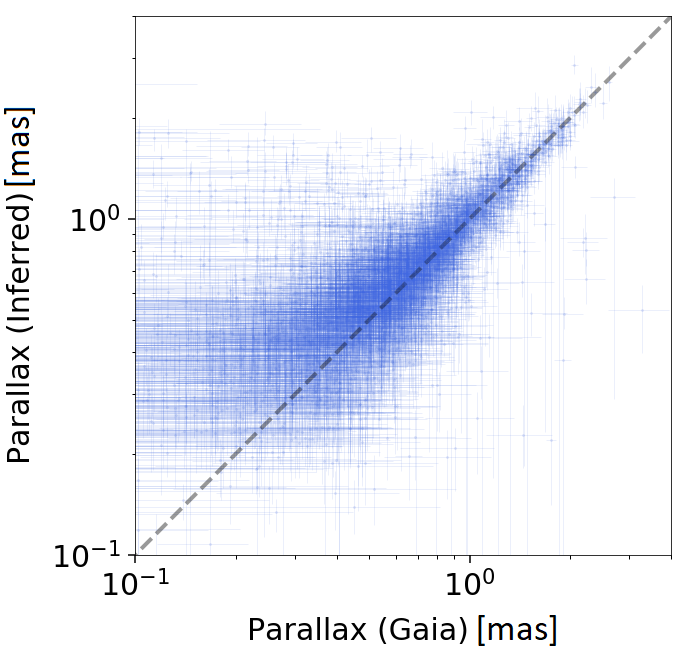}
\end{center}
\caption{The inferred parallax for $n \sim 5100$ representative
objects with (only) a 10-band Pan-STARRS, 2MASS, and WISE SED
from {\brutus} versus the estimated parallax from 
\textit{Gaia} DR2. 1-sigma errors from both sources are plotted
for each point and the one-to-one relation
is shown with a dashed gray line. The inferred
parallaxes from {\brutus} agree well with the measurements
down to low parallax SNR. For smaller parallaxes, we observe
a bias towards larger inferred values due to the
3-D stellar density prior, as well as objects with much larger
inferred values caused by giants being incorrectly modeled as
dwarfs. Both of these effects are expected
(see Figures \ref{fig:mock_pars_1}, 
\ref{fig:mock_pars_2}, and \ref{fig:mock_pars_3}) and largely
mitigated once parallax measurements are incorporated
into the modelling (see Figure \ref{fig:h3}).
}\label{fig:parallax}
\end{figure}

\begin{figure*}
\begin{center}
\includegraphics[width=\textwidth]{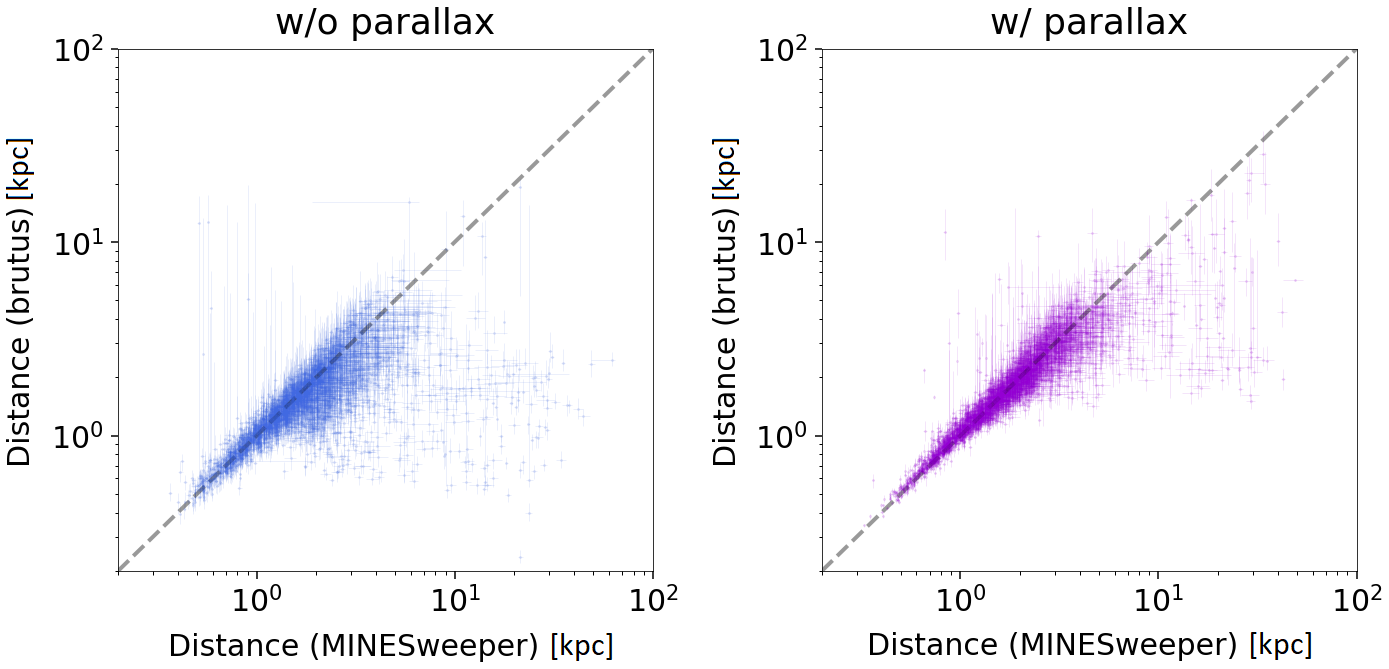}
\end{center}
\caption{A comparison of the distances derived from
{\brutus} (photometry only) and {\ms} (photometry and spectroscopy)
for the same sample of $n \sim 5100$
objects shown in Figure \ref{fig:parallax} without (left, blue)
and with (right, purple) \textit{Gaia} parallax constraints.
1-sigma errors from both sources are plotted
for each point and the one-to-one relation is shown with a dashed gray line.
In both cases, we find excellent agreement between the two
methods for the majority of objects. Without parallax constraints,
however, {\brutus} overwhelmingly prefers to model sources as
being MS dwarfs compared to {\ms} (which has access to $\log g$
information from the spectrum), leading to a substantial fraction 
of post-MS giants being incorrectly classified as MS dwarfs. 
Many of these sources are correctly classified once
the parallax information is included in the fits since
\textit{even an extremely low SNR parallax still rules out small distances}.
In other words, while a parallax of $\hat{\varpi} \sim 0$ gives almost no
constraint on the distance, it \textit{does} imply that $d \gtrsim 2\,{\rm kpc}$
or so, which can be enough to substantially disfavor modeling faraway giants
sources as nearby dwarfs.
}\label{fig:h3}
\end{figure*}

\subsection{Independent Comparison with H3} \label{subsec:test_h3}

As a final series of tests, we want to examine how well
{\brutus} can recover distances to real stars. To make
this comparison, we utilize a subset of $n \sim 5100$
stars from the H3 survey \citep{conroy+19a}.
In brief, the H3 Survey is a high-latitude
($|b| > 30^\circ$), high-resolution ($R=32,000$) spectroscopic
survey of the distant ($d \gtrsim 2\,{\rm kpc}$ Galaxy. 
Targets are selected purely on their Gaia parallax 
($\varpi < 0.4-0.5\,{\rm mas}$), brightness ($15 < r_{\rm PS1} < 18$), 
and accessibility to the 6.5\,m MMT in Arizona, USA (${\rm dec} > -20^\circ$). 
The survey measures radial velocities to $0.1$\,km/s precision, surface abundances
($\feh_{\rm surf}$ and $\afe_{\rm surf}$) to $0.1$\,dex precision, and
spectrophotometric distances to $10\%$ precision using {\ms} \citep{cargile+20}.

For internal testing purposes, spectrophotometric distances for a few thousand 
objects were derived from {\ms} both with and without
\textit{Gaia} parallax measurements. In addition to
being a representative, low-reddening subsample of sources,
all of these measurements were estimated using the same
underlying {\mist} isochrones, excluding the empirical corrections
and photometric offsets derived in this work. This makes the
comparisons both independent (estimated using different
codebases and with/without spectra) while still remaining
internally consistent (similar photometry and underlying stellar models).

In Figure \ref{fig:parallax}, we show the inferred
parallaxes for this subsample of stars from {\brutus}
\textit{without} including the \textit{Gaia} DR2 parallax
measurements using up to 10 bands of photometry (Pan-STARRS,
2MASS, and AllWISE). The results show that {\brutus} is able to
recover accurate parallaxes \textit{exclusively} from photometry
to many nearby sources and that the results are mostly unbiased
down to objects with smaller parallaxes. 

For sources at $\gtrsim 2\,{\rm kpc}$, we observe
a bias towards larger inferred values. This is expected and
arises for two reasons. The first is the 3-D Galactic prior,
which prefers objects being a few kpc away 
(see Figure \ref{fig:prior_sightline}), disfavoring larger
distances (and therefore smaller parallaxes). This ``bias'' is physically-motivated
and expected, as discussed in \citet{bailerjones+18}.
The second is the mis-classification of giants as dwarfs,
as illustrated in Figures \ref{fig:mock_pars_2} and \ref{fig:mock_pars_3},
which leads to anomalously large parallax estimates compared to the
measured values from \textit{Gaia}.

We now investigate how the distance estimates from {\brutus}
compare with those from {\ms} for the same sample of stars.
In Figure \ref{fig:h3}, we show the resulting distances
estimated with and without \textit{Gaia} DR2 parallaxes.
In the left panel of the figure, we see that indeed
without parallax constraints {\brutus} systematically mis-classifies
giants as dwarfs relative to the spectroscopic classifications
obtained by {\ms}. Outside of these sources at 
large distances, however, the results
are in excellent agreement between the two codes and datasets.

Once the parallaxes are included, the overall distance agreement improves,
as expected. We also observe that the rate of
mis-classification between {\brutus} and {\ms} decreases substantially,
even for objects with extremely small parallaxes and correspondingly low
parallax SNRs. This result is due to one fundamental feature of parallaxes:
\textit{even an extremely low SNR parallax still rules out small distances}.
In other words, while a parallax of $\hat{\varpi} \sim 0$ gives almost no
constraint on the distance, it \textit{does} imply that $d \gtrsim 2\,{\rm kpc}$
or so, which is enough to substantially disfavor modeling faraway giants
sources as nearby dwarfs. This further highlights just how crucial parallax 
measurements \textit{at all SNR} are for accurately recovering 
not only intrinsic stellar
parameters but also accurate distances to many sources.

\section{Conclusion} \label{sec:conc}

One of the main challenges of studying the Milky Way is transforming
observations of the projected 2-D positions of sources
on the sky into full 3-D maps. This has become particularly important
in recent years as large-area surveys such as SDSS \citep{york+00},
Pan-STARRS \citep{chambers+16}, and \textit{Gaia} \citep{gaia+16}
provide measurements to billions of stars.
These datasets have been crucial to driving new discoveries
related to the structure and evolution of the Milky Way
\citep{koppelman+18,belokurov+18,helmi+18,antoja+18}.

To build the reliable 3-D maps much of these discoveries
are based on, the raw observations (photometry, astrometry,
spectroscopy, etc.) from these surveys must be converted into
physical quantities such as 3-D positions and velocities,
metallicities, masses, ages, and more. Part of this
process requires developing an algorithm that
can quickly yet robustly derive many of these
parameters from photometry and astrometry alone,
which characterizes the vast majority ($\sim 99\%$)
of the data currently available in the Milky Way.

To assist with this process, in this paper we presented
{\brutus}, a public, open-source {\python}
package that uses a combination of statistical
approaches to infer stellar properties, distances,
and extinctions for sources using photometry and astrometry.
We described the statistical framework behind
the code (\S\ref{sec:stats}), the techniques used to
perform inference (\S\ref{sec:implementation}), and the
stellar models that are currently implemented (\S\ref{sec:models}).
All of these portions of the code are designed to be highly
modular so that each component can be adjusted and varied
through various input files and command-line arguments.

We then outlined a procedure for ``calibrating''
the underlying stellar models in {\brutus} using a novel approach
to fit stellar clusters (\S\ref{ap:calib_clusters}).
We applied this procedure to a series of ``benchmark''
open clusters, finding excellent overall fits
(\S\ref{subsec:calib_benchmark}). We then
used a sample of $\sim 20,000$ nearby field stars to
further investigate photometric offsets, further
demonstrating the underlying models match the data well
(\S\ref{subsec:calib_field}).

Finally, we validated the performance of the code
using a series of validation tests over both mock data
and real observations (\S\ref{sec:tests}). These illustrated
that {\brutus} performs well overall while also highlighting
the limitations of the code and the sensitivity of
parameter recovery to the Galactic prior and parallax observations.

Overall, we hope that the paper serves as a useful
overview of the challenges associated with astrophotometric
stellar parameter estimation and the various parts of {\brutus}
that attempt to solve this problem. The application of the
code to a sample of $>10^8$ stars are described
in a companion paper \citep{speagle+24}.

\acknowledgements

\noindent \textit{Contributions:}

The author list is divided up into 3 groups: 
\begin{enumerate}
    \item a list of primary authors who made direct 
    contributions to the {\brutus} codebase and
    were actively involved in development and testing (JSS, CZ, and ABe),
    \item an alphabetized secondary list of authors who made direct contributions
    to data products, indirect contributions to the code, and/or were
    involved in detailed discussions surrounding its development (PAC to EFS), and
    \item an alphabetized tertiary list of authors who provided
    useful feedback during the development process and/or on the paper itself
    (ABo to IAZ).
\end{enumerate}

\noindent \textit{Personal:}

JSS is immensely grateful to Rebecca Bleich for being an amazing and
supportive partner during these difficult times.
JSS would also like to thank Jan Rybizki, Stefan Meingast, 
Jo{\~a}o Alves, Seth Gossage, and Nayantara Mudur
for insightful discussions that improved the quality of this work. \\

\noindent \textit{Funding:}

JSS and CZ were partially supported by the Harvard Data Science Initiative.
HMK  acknowledges  support  from  the  DOE  CSGF  under  grant  number DE-FG02-97ER25308. 
AD received support from the National Aeronautics and 
Space Administration (NASA) under Contract No. NNG16PJ26C issued 
through the WFIRST Science Investigation Teams Program.  
CZ and DPF acknowledge support from NSF grant AST-1614941, 
“Exploring the Galaxy: 3-Dimensional Structure and Stellar Streams.”
AKS gratefully acknowledges support by a National Science 
Foundation Graduate Research Fellowship (DGE-1745303). 
YST acknowledges financial support from the Australian Research Council
through DECRA Fellowship DE220101520. \\

\noindent \textit{Code}:

This work has benefited from the following packages:
\begin{itemize}
    \item {\astropy} \citep{astropy+13,astropy+18}
    \item {\numpy} \citep{vanderwalt+11}
    \item {\scipy} \citep{virtanen+20}
    \item {\matplotlib} \citep{hunter07}
    \item {\healpy} \citep{gorski+05,zonca+19}
    \item {\corner} \citep{foremanmackey16}
    \item {\hdbscan} \citep{mcinnes+17a,mcinnes+17b}
\end{itemize}

\bibliography{ref}

\appendix

\section{Detailed Description of Priors} \label{ap:priors}

\subsection{Initial Mass Function} \label{subsec:prior_imf}

We assume that stars are all born with initial masses $M_{\rm init}$
independently sampled from a universal stellar initial mass function (IMF).
We assume that this follows a broken power law such that
\begin{equation}
    \pi(M_{\rm init}) \propto
    \begin{cases}
    0 & M_{\rm init} < 0.08 \\
    M_{\rm init}^{-\alpha_1} & 0.08 \leq M_{\rm init} < 0.5 \\
    M_{\rm init}^{-\alpha_2} & M_{\rm init} \geq 0.5
    \end{cases}
\end{equation}
where we exclude initial masses below the deuterium-burning limit
of $\sim 0.08 M_\odot$ and we set $\alpha_1 = -1.3$ and $\alpha_2=-2.3$
following \citet{kroupa01}.
As we currently do not recommend modeling stars below $0.5 M_\odot$
(see \S\ref{sec:calib}), this is functionally equivalent
to a \citet{salpeter55} IMF. We hope to extend our models and priors 
down to lower stellar masses in future work.

\subsection{3-D Stellar Number Density} \label{subsec:prior_stars}

For a given number density distribution
$n(d | \ell, b)$ along a particular
line of sight (LOS) specified by Galactic coordinate $\ell$ and $b$,
the probability of observing stars at a given distance $d$
involves accounting for the increasing differential volume from the associated
shell ${\rm d}V/{\rm d} d = 4 \pi d^2$ as a function of distance so that
\begin{equation}
    \pi(d | \ell, b) \propto n(d | \ell, b) \times d^2
\end{equation}

We assume that the number density of stars in the Galaxy come from three components:
\begin{itemize}
    \item A \textit{thin disk} of younger stars with higher metallicities.
    \item A \textit{thick disk} of slightly older stars with lower metallicities.
    \item A \textit{halo} of substantially older stars with low metallicities.
\end{itemize}
This implies our number density is
\begin{equation}
    n(d | \ell, b) = \sum_{x} n_x(d | \ell, b) 
\end{equation}
where the sum over $x$ is taken over each component
$n_{\rm thin}(d | \ell, b)$, $n_{\rm thick}(d | \ell, b)$, 
and $n_{\rm halo}(d | \ell, b)$.
Unlike other recent work \citep[e.g.,][]{anders+19}, 
we do not currently include any components corresponding to the Galactic
bulge and/or bar. We hope to include these in future work.
Note that while we use terms like ``thick disk'' to refer to these functions,
they are intended to be purely \textit{operational} terms that
serve as shorthand to describe the underlying components and are not meant to
imply any particular formation history.

Since measuring absolute number densities can be difficult,
observations often can better constrain the \textit{relative} number density
$n'(d | \ell, b)$ such that
\begin{equation}
    n_x(d | \ell, b) = f_x \times n_{\odot, \rm thin} 
    \times n'_x(d | \ell, b)
\end{equation}
where $n'_x(d=0 | \ell, b) \equiv 1$ at the position of the Sun ($d=0$), 
$n_{\odot, \rm thin}$ is the number density of the thin disk 
at $d=0$, and $f_x$ is a scale factor
specifying the relative contribution for the $x$th component 
(where $f_{\rm thin} = 1$ by definition). Since $n_{\odot, \rm thin}$
is the same for all terms, we can ignore its contribution to the
prior and work directly with $f_x$ and $n'_x(d | \ell, b)$.

We model the relative number density of stars in the thin
and thick disk with an exponential profile as a function of 
Galactocentric radius $R$ and disk height $Z$
\begin{equation}
    n'_{\rm disk}(R, Z) = \exp\left[\frac{R - R_\odot}{R_{\rm disk}}\right]
    \times \exp\left[\frac{|Z| - |Z_\odot|}{Z_{\rm disk}}\right]
\end{equation}
where $R_{\rm disk}$ is the scale radius, $Z_{\rm disk}$
is the scale height, and $R_\odot = 8.2\,{\rm kpc}$ and 
$Z_\odot=0.025\,{\rm kpc}$ are the corresponding
Solar values taken from \citet{blandhawthorngerhard16}.
It is straightforward to convert from Galactic coordinates to
Galactocentric cylindrical coordinates 
$(d, \ell, b) \rightarrow (R, Z, \phi)$
to evaluate this prior from our observed data using 
{\astropy}\footnote{\url{https://www.astropy.org}} \citep{astropy+13,astropy+18}
or other publicly-available {\python} packages.
The values for constants assumed for the
thin and thick disk are taken from \citet{blandhawthorngerhard16}
and summarized in Table \ref{tab:prior}.

Following \citet{xue+15},
we model the relative number density of stars in the halo
using a power law of the form
\begin{equation}
    n'_{\rm halo}(r_{\rm eff}) 
    = \left(\frac{r_{\rm eff}}{r_{\odot, \rm eff}}\right)^{-\eta}
\end{equation}
as a function of effective radius
\begin{equation}
    r_{\rm eff}(R, Z, q(r), R_s) = \sqrt{R^2 + (Z / q(r))^2 + R_s^2}
\end{equation}
where $\eta=4.2$, $q$ is the oblateness ($q=1$ is spherical, $q=0$ is perfectly flat)
and $R_s=1\,{\rm kpc}$ is a smoothing radius that prevents the power law from diverging
near the Galactic center.\footnote{Note that this is larger than the
$R_s = 0.5\,{\rm kpc}$ adopted in \citet{green+19} to avoid being overly sensitive to density
changes near the Galactic center.} The oblateness $q(r)$ is allowed 
to change as a function of Galactocentric
spherical radius $r = \sqrt{R^2 + Z^2}$ such that:
\begin{equation}
    q(r) = q_{\infty} - (q_{\infty} - q_{0}) \times
    \exp\left[1 - \sqrt{1 + (r/r_q)^2}\right]
\end{equation}
where $q_{\infty} = q(r=\infty) = 0.8$ is the oblateness at large radii,
$q_{0} = q(r=0) = 0.2$ is the oblateness at the Galactic center,
and $r_q = 6\,{\rm kpc}$ is the scale radius over which
the oblateness begins to transition from
$q_{0}$ to $q_{\infty}$. To ensure $n'_{\rm halo}(R_\odot, Z_\odot) = 1$,
the function is normalized to
\begin{equation}
    r_{\odot, \rm eff} \equiv \sqrt{R_\odot^2 + (Z_\odot / q_\odot)^2 + R_s^2} 
    \approx 8.26\,{\rm kpc}
\end{equation}
where $q_\odot = q(r=r_\odot) \approx 0.5$ 
for $r_\odot = \sqrt{R_\odot^2 + Z_\odot^2}$.

A schematic illustration of our combined 3-D number density/distance prior
is shown in Figure \ref{fig:prior_density}.

\subsection{3-D Stellar Metallicity} \label{subsec:prior_feh}

We assume a spatially-invariant prior on the metallicities
of stars within each component (indexed by $x$) based loosely on
\citet{blandhawthorngerhard16} and \citet{anders+19} that follow a Normal
distribution with mean $\mu_{\feh, x}$ and standard deviation
$\sigma_{\feh, x}$:
\begin{align}
    \pi_x(\feh_{\rm init}) &= \frac{1}{\sqrt{2\pi\sigma_{\feh, x}^2}}
    \times \exp\left[-\frac{1}{2}\frac{\left(\feh_{\rm init} - \mu_{\feh, x}\right)^2}
    {\sigma_{\feh, x}^2}\right] \\
    &\equiv\Normal{\mu_{\feh, x}}{\sigma_{\feh, x}^2}
\end{align}
where we slightly abuse notation by letting $\Normal{\mu}{\sigma^2}$
represent the PDF of a Normal distribution with
mean $\mu$ and standard deviation $\sigma$.
The metallicity prior at a particular 3-D position 
is then a number-density weighted combination of each component
\begin{equation}
    \pi(\feh_{\rm init} | d, \ell, b) = 
    \sum_{x} \frac{n_x(d|\ell, b)}{n(d|\ell, b)}
    \times \pi_x(\feh_{\rm init})
\end{equation}
The values for the mean and standard deviation of each component
of our prior (thin disk, thick disk, and halo) 
are summarized in Table \ref{tab:prior} and highlighted in Figure \ref{fig:prior_agefeh}.

\subsection{3-D Stellar Age} \label{subsec:prior_age}

Similar to our metallicity prior, we also assume a spatially-invariant prior 
on the ages of stars within each component
loosely following \citet{blandhawthorngerhard16},
\citet{xue+15}, and \citet{anders+19} that follows a truncated Normal distribution
between minimum age $t_{\rm min}$ and maximum age $t_{\rm max}$ with mean
$\mu_{t, x}$ and standard deviation $\sigma_{t, x}$:
\begin{equation}
    \pi_x(t_{\rm age}) = 
    \begin{cases}\frac{\Normal{\mu_{t, x}}{\sigma_{t, x}^2}}
    {\int_{t_{\rm min}}^{t_{\rm max}} \Normal{\mu_{t, x}}{\sigma_{t, x}^2} {\rm d}t_{\rm age}}
    & t_{\rm min} \leq t_{\rm age} \leq t_{\rm max} \\
    0 & {\rm otherwise}
    \end{cases}
\end{equation}
where we choose $t_{\rm min} = 0$ and $t_{\rm max} = 13.8$ (roughly
the current age of the Universe; \citealt{planck16}).
The age prior at a particular 3-D position is 
again a number-density weighted combination of each component
\begin{align}
    \pi(&t_{\rm age} | d, \ell, b) = 
    \sum_{x} \frac{n_x(d|\ell, b)}{n(d|\ell, b)} 
    \times \pi_x(t_{\rm age})
\end{align}

Although we do not account for explicit covariances 
between the metallicities and ages of stars within/across components in this work,
we still want to connect the overall mean
$\mu_{t,x}$ and scatter $\sigma_{t, x}$ in the ages of stars for each component 
to the mean metallicity $\mu_{\feh, x}$ of that component.
We choose a smooth logistic function
\begin{equation}
    \mu_{t, x}(\mu_{\feh, x}) = \frac{t_{\rm max} - t_{\rm min}}
    {1 + \exp\left[\frac{\mu_{\feh, x} - \xi_{\feh}}{\Delta_\feh}\right]} 
    + t_{\rm min}
\end{equation}
where $\xi_{\feh}=-0.5$ is the ``pivot'' metallicity halfway between $t_{\rm min}$ and
$t_{\rm max}$ and $\Delta_{\feh}=0.5$ is the metallicity scale-length over which the
age transitions from $t_{\rm min}$ to $t_{\rm max}$.

The scatter $\sigma_{t, x}$ is subsequently set based on the 
mean age of the population $\mu_{t, x}$ as:
\begin{equation}
    \sigma_{t, x}(\mu_{t, x}) =
    \begin{cases}
    \sigma_{t, {\rm min}} & \frac{t_{\rm max} - \mu_{t, x}}
    {n_\sigma} < \sigma_{t, {\rm min}} \\
    \frac{t_{\rm max} - \mu_{t, x}}{n_\sigma} & 
    \sigma_{t, {\rm min}} \leq \frac{t_{\rm max} - \mu_{t, x}}{n_\sigma} 
    \leq \sigma_{t, {\rm max}} \\
    \sigma_{t, {\rm max}} & \frac{t_{\rm max} - \mu_{t, x}}
    {n_\sigma} > \sigma_{t, {\rm min}}
    \end{cases}
\end{equation}
where $\sigma_{t, {\rm min}}$ is the minimum allowed scatter, 
$\sigma_{t, {\rm max}}$ is the maximum allowed scatter, and $n_\sigma$
scales $\sigma_{t, x}$ relative to the difference between the mean age
and the maximum age $t_{\rm max} - \mu_{t, x}$. This function smoothly varies the
scatter in age between a maximum value of $\sigma_{t, {\rm max}}$ and a minimum
value of $\sigma_{t, {\rm min}}$ as the age increases relative to the maximum allowed
age $t_{\rm max}$. This ensures that younger populations (such as in the disk)
have a higher allowed age dispersion arising from more extended star formation histories,
while older populations (such as in the halo) have a lower allowed age dispersion
due to the shorter timescale over which the stars could have formed.

The values for this ``age-metallicity relation'' and the corresponding
the mean and standard deviation of the age for each component (thin disk, thick disk,
and halo) are summarized in Table \ref{tab:prior} 
and highlighted in Figure \ref{fig:prior_agefeh}. The mean values and scatter
are in broad agreement with similar work such as \citet{anders+19}.
An illustration of the combined prior on number density, metallicity, and age
for a specific line of sight is shown in Figure \ref{fig:prior_sightline}.

\subsection{3-D Dust Extinction} \label{subsec:prior_av}

We assume the distribution of dust is independent of the distribution of
stars and has a density of $\rho_{\rm dust}(d, \ell, b)$. 
The impact of dust on the observed photometry
for any particular star depends on the cumulative dust along a given LOS:
\begin{align}
    A_V(d | \ell, b) &=
    f_{\params}\left(\int_{0}^{d} \rho_{\rm dust}(d', \ell, b) \, {\rm d} d'\right) \nonumber \\
    &\equiv f_{\params}\left(N_{\rm dust}(d | \ell, b)\right)
\end{align}
where $N_{\rm dust}(d | \ell, b)$ is the column density out to distance $d$ along the
LOS defined by $(\ell, b)$ and $f_{\params}(\cdot)$ is an unknown function that translates
$N_{\rm dust}(d | \ell, b) \rightarrow A_V(d|\ell, b)$ into $V$-band extinction
for a given stellar spectrum $F_\nu(\lambda|\params)$.
Typically $f_{\params}(N_{\rm dust})$ is assumed to be 
roughly linear in $N_{\rm dust}$ and independent of stellar parameters.
While this is unimportant for our purposes
since we are placing a prior directly on $A_V$ rather than $\rho_{\rm dust}$,
we hope to try to infer $\rho_{\rm dust}$ directly in future work.

We assume that the $A_V$ for stars at roughly the same $(d, \ell, b)$ are Normally
distributed with mean $\mu_A(d|\ell, b)$ and standard deviation
$\sigma_A(d|\ell, b)$:
\begin{equation}
    \pi(A_V | d, \ell, b) = \Normal{\mu_A(d|\ell, b)}{\sigma_A(d|\ell, b)}
\end{equation}
We take the mean
\begin{equation}
    \mu_A(d|\ell, b) = \frac{1}{N} \sum_{i=1}^{N} A_{V,i}^{\rm B19}(d | \ell, b)
\end{equation}
to be the sample mean computed from extinction realizations 
$A_{V,i}^{\rm B19}(d | \ell, b)$ taken from the 3-D dust map 
from \citet{green+19} (i.e. {\bayestarmap}) and the uncertainty
\begin{equation}
    \sigma_A^2(d|\ell, b) = \frac{1}{N}\sum_{i=1}^{N}
    \left[A_{V,i}^{\rm B19}(d | \ell, b) - \mu_A(d|\ell, b)\right]^2 
    + \Delta_A^2
\end{equation}
to be a combination of the sample variance computed from the same set
of {\bayestarmap} extinction realizations along with some intrinsic
spread $\Delta_A=0.2\,{\rm mag}$ based on comparisons from
\citet{green+19}. This helps to account for both possible systematics
in the underlying dust estimates (which were derived using particular stellar
models; see \S\ref{subsec:bayestar}) as well as small-scale dust structure below
the current resolution of the {\bayestarmap} dust map \citep{zuckerspeagle+19}.
Note that since {\bayestarmap} is defined in terms of
$E(B-V) \equiv A_B - A_V$ (i.e. ``reddening'' rather than ``extinction''),
we convert from $E(B-V)$ to $A(V)$ as described in the {\dustmaps} package
using the conversion factor from \citet{schlaflyfinkbeiner11}.

A summary of the parameters involved in
our 3-D dust extinction prior is included in Table \ref{tab:prior}. 
An illustration is shown in Figure \ref{fig:prior_dust}.

\subsection{Variation in the Dust Extinction Curve} \label{subsec:prior_rv}

As described above, we allow the dust curve to vary linearly
as a function of $R_V$. While results from \citet{schlafly+16,schlafly+17}
suggest that $R_V(d, \ell, b)$ varies as a function of 3-D position,
the results are not available with the same fidelity and resolution of the
3-D {\bayestarmap} dust map. As a result, in this work
we consider variation in $R_V$ to be independent of position.
Following \citet{schlafly+16},
we approximate this variation using a Normal distribution
\begin{equation}
    \pi(R_V) = \Normal{\mu_R}{\sigma_R}
\end{equation}
with a mean $R_V$ of $\mu_R=3.32$ and standard deviation of $\sigma_R=0.18$.

In practice, it is difficult to infer $R_V$ variation from individual sources
observed only in a handful of bands. Even at high signal-to-noise across
$\sim 8$ bands of optical and near-infrared (NIR) photometry, we find
that $R_V$ constraints improve over our prior by at most $\sim 30-50\%$.
As such, we consider this component of our model mostly a means to
accommodate systematic offsets between the models and the data
and increase the uncertainty in the derived posteriors.
We hope to explore modeling coherent 3-D $R_V$ variation 
in future work \citep[see also][]{schlafly+17}.

The corresponding mean $\mu_R$ and standard deviation $\sigma_R$
are included in Table \ref{tab:prior}.

\section{Implementation Details} \label{ap:implementation}

We describe the motivation behind pursuing
the overall approach described in \S\ref{sec:implementation}
in \S\ref{subsec:impl_motivation}.
The strategy for performing linear regression over magnitude is described
in \S\ref{subsec:methods_mags}. The subsequent expansion 
and linear regression in flux density is described in \S\ref{subsec:methods_flux}.
The use of Monte Carlo sampling to incorporate
priors is described in \S\ref{subsec:methods_mc}.
Computing estimates and (re)sampling from the posterior 
over a given stellar grid is described in \S\ref{subsec:methods_grids}.
Additional implementation details are discussed in \S\ref{subsec:methods_cuts}.

\subsection{Motivation} \label{subsec:impl_motivation}

In general, estimating posterior-based quantities
tends to follow two approaches:
\begin{enumerate}
    \item Sampling-based (i.e. Monte Carlo) approaches.
    \item Grid-based (i.e. ``brute force'') approaches.
\end{enumerate}
Both have benefits and drawbacks, which we discuss briefly below.

Monte Carlo methods such as Markov Chain Monte Carlo
\citep[MCMC;][]{brooks+11,sharma17,hoggforemanmackey18,speagle19}
or Nested Sampling \citep{skilling04,skilling06} 
try to characterize the posterior distribution
$P(\params, \eparams | \hat{\flux}, \hat{\varpi})$ by generating a
set of $n$ samples $\{ (\params_i, \eparams_i) \}_{i=1}^{i=n}$
along with a set of corresponding weights $\{ w_i \}_{i=1}^{i=n}$.
Expectation values (i.e. weighted averages)
of functions $f(\params, \eparams)$ taken over the posterior
\begin{equation}
    \meanwrt{f(\params, \eparams)}{P} \equiv
    \int f(\params, \eparams) 
    P(\params, \eparams | \hat{\flux}, \hat{\varpi}) 
    {\rm d}\params\,{\rm d}\eparams
\end{equation}
can then be estimated via:
\begin{equation}
    \meanwrt{f(\params, \eparams)}{P} 
    \approx \frac{\sum_{i=1}^{n} w_i \times f(\params_i,\eparams_i)}
    {\sum_{i=1}^{n} w_i}
\end{equation}

While Monte Carlo methods allow us to in theory derive 
arbitarily precise estimates for 
$\meanwrt{f(\params, \eparams)}{P}$, they suffer from two main
drawbacks when trying to infer stellar properties from photometry:
\begin{enumerate}
    \item Many of the posterior distributions for stars
    have multiple, widely-separated solutions
    due to the ``dwarf-giant degeneracy'' since
    main sequence (MS) stars (``dwarfs'') and post-MS
    stars (``giants'') can often have similar colors
    but very different luminosities.
    Many Monte Carlo methods struggle to characterize 
    these types of multi-modal distributions efficiently.
    \item We expect most posteriors to have extended
    and complex degeneracies since changes in $\params$
    and $\eparams$ (e.g., $T_{\rm eff}$ and $A_V$) 
    can impact observables in similar ways. 
    Accurately characterizing these uncertainties 
    will require more samples and/or longer run-times
    from Monte Carlo methods.
\end{enumerate}

``Brute force'' methods can get around these problem by computing
the posterior probability 
$P(\params_i, \eparams_i | \hat{\flux}, \hat{\varpi})$
over a large grid of $n$ values 
$\{ (\params_i, \eparams_i) \}_{i=1}^{i=n}$.
These can then be used to estimate the expectation
value in a similar way to our initial set of samples via
\begin{equation}
    \meanwrt{f(\params, \eparams)}{P} \approx
    \frac{\sum_{i=1}^{n} \Delta_i \times 
    P(\params_i, \eparams_i | \hat{\flux}, \hat{\varpi})
    \times f(\params_i,\eparams_i)}
    {\sum_{i=1}^{n} \Delta_i \times P(\params_i, \eparams_i | \hat{\flux}, \hat{\varpi})}
\end{equation}
where $\Delta_i = \prod_{j=1}^{m} \Delta_{i,j}$ is the volume of each element
$i$ of the grid with spacing $\Delta_{i,j}$ for the $j$-th parameter (out of $m$ total)
at position $(\params_i, \eparams_i)$. For an evenly-spaced grid in each parameter,
$\Delta_i = \Delta$ is constant.

While grids scale extremely poorly ($\propto e^p$)
as the number of parameters $p$ increases, for $p \lesssim 4$ parameters
grids often only comprise $\sim 10^6$ elements that can be fit extremely
quickly using modern computing architectures that excel at linear algebra
operations. This often makes evaluating grids 1-2 orders of magnitude faster
than sequentially generating samples via Monte Carlo methods. It also
allows grids to be expansive enough to explore
large regions of parameter space at fine enough resolution to
characterize multiple modes robustly.

Unfortunately, when inferring stellar properties from photometry
the number of parameters $p$ is often $\gtrsim 5$. To get around this, 
grid-based approaches often either simplify the problem 
(see \S\ref{subsubec:noiseless_intrinsic})
and/or apply grids to only a few parameters at a time.
This, however, often can underestimate uncertainties
and possibly miss solutions.

To resolve these various difficulties, {\brutus} adopts a hybrid approach
that exploits properties of our statistical model to combine
grids, linear regression, and Monte Carlo methods to approximate
the posterior. This allows us to exploit grids when possible to explore
widely-separated modes while retaining the ability to construct accurate
estimates of particular inferred quantities. These are described
in more detail in the subsequent sections.

\subsection{Linear Regression in Magnitudes} \label{subsec:methods_mags}

The core of our approach centers on the fact that our initial model
is essentially linear in the extrinsic parameters $\eparams$ since
\begin{equation}
    \mu + A_V \times (\rvec_{\params} + R_V \times \rvec'_{\params})
    \equiv \eparams_{\rm mag}^{\T} \dvec_{\params}
\end{equation}
where we have slightly abused notation by defining the 
$3 \times 1$ vector
\begin{equation}
    \eparams_{\rm mag} =
    \begin{bmatrix}
    \mu \\
    A_V \\
    A_V \times R_V
    \end{bmatrix}
\end{equation}
and
\begin{equation}
    \dvec_{\params} = 
    \begin{bmatrix}
    \mathbf{1} \\
    \rvec_{\params} \\
    \rvec'_{\params}
    \end{bmatrix}
\end{equation}
is the $3 \times b$ ``data-generating'' matrix 
where $\mathbf{1} = \{ 1 \}_{i=1}^{i=b}$.

Assuming our errors $\errors_{\mags}$ on the magnitudes $\hat{\mags}$
are approximately Normal and defining
$\Delta\hat{\mags}_{\params} \equiv \hat{\mags} - \absmag_{\params}$, 
the corresponding log-likelihood at fixed $\params$ is then
\begin{align}
    -2 \ln\likelihood_{\rm mag}(\eparams_{\rm mag}|\params) = 
    \left[\Delta\hat{\mags}_{\params} - 
    \eparams_{\rm mag}^{\T} \dvec_{\params}\right]^{\T}
    \cov_{\mags}^{-1} 
    \left[\Delta\hat{\mags}_{\params} - 
    \eparams_{\rm mag}^{\T} \dvec_{\params}\right] 
    + \ln\left[{\rm det}\left(2\pi\cov_{\mags}\right)\right]
\end{align}
If we want to solve for the maximum-likelihood estimate (MLE) for $\eparams_{\rm mag}$,
we set the $3 \times 1$ \textit{Jacobian} vector
$\partial \ln \likelihood_{\rm mag} / \partial \eparams_{\rm mag} = \mathbf{0}$
and solve this linear system for $\eparams_{\rm mag}^{\rm MLE}$. This gives
\begin{equation}
    \eparams_{\rm mag}^{\rm MLE}(\params) = 
    \left(\dvec_{\params} \cov_{\mags}^{-1} \dvec_{\params}^{\T}\right)^{-1}
    \dvec_{\params} \cov_{\mags}^{-1} \Delta\hat{\mags}_{\params}
\end{equation}
Note that this is the standard solution for a 
\textit{weighted least-squares linear regression} problem.

\begin{figure*}
\begin{center}
\includegraphics[width=\textwidth]{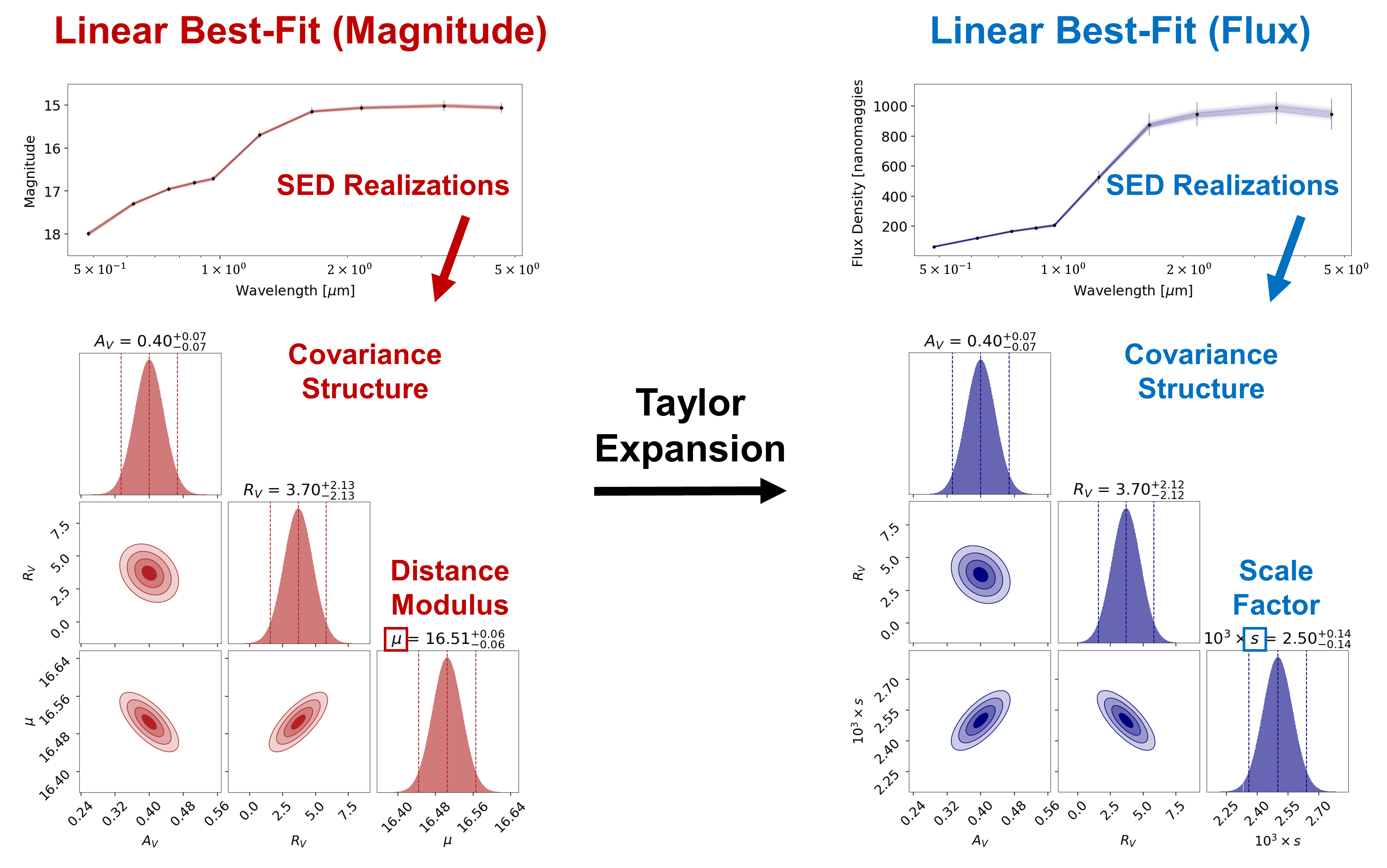}
\end{center}
\caption{An illustration of the best-fit solutions from our
linear model for a particular model SED with intrinsic parameters
$\params$ when fitting for our extrinsic parameters $\eparams$ 
in magnitudes (see \S\ref{subsec:methods_mags}; left)
and after Taylor expanding in flux density 
(see \S\ref{subsec:methods_flux}; right).
The top panels show the measured SED and 2-sigma errors with realizations
around the best-fit solution, while the bottom panels show 1-D and 2-D
marginalized probability densities in dust extinction $A_V$, 
differential extinction $R_V$, and 
distance modulus $\mu$ (left)/scale factor $s$ (right). These
highlight the different covariance structure between the two cases.
Note that the scale factor $s$ is directly related to the parallax $\varpi$
via $s = \varpi^2$.
}\label{fig:linear_approx}
\end{figure*}

This result immediately implies that the conditional 
likelihood of $\eparams$ given $\params$ is
\begin{align}
    \likelihood_{\rm mag}(\eparams_{\rm mag}|\params) 
    = \underbrace{\likelihood_{\rm mag}^{\rm MLE}(\params)
    \times {\rm det}\left[2\pi\cov_{\rm mag}^{\rm MLE}(\params)\right]
    }_{\rm Normalization}
    \times \underbrace{
    \Normal{\eparams_{\rm mag}^{\rm MLE}(\params)}{\cov_{\rm mag}^{\rm MLE}(\params)}
    }_{\rm Distribution}
\end{align}
In other words, $\eparams_{\rm mag}$ is Normally distributed about the 
MLE $\phi_{\rm mag}^{\rm MLE}(\params)$ with a maximum amplitude of
\begin{equation}
    \likelihood_{\rm mag}^{\rm MLE}(\params) \equiv
    \likelihood_{\rm mag}(\params, \eparams_{\rm mag}^{\rm MLE}(\params))
\end{equation}
and a covariance of
\begin{align}
    \cov_{\rm mag}^{\rm MLE}(\params) &= 
    \left(\dvec_{\params} \cov_{\mags}^{-1} \dvec_{\params}^{\T}\right)^{-1} \\
    &= \left(\frac{\partial^2\ln\likelihood_{\rm mag}(\eparams_{\rm mag}|\params)}
    {\partial \eparams_{\rm mag}^2}\right)^{-1}
\end{align}
where $\partial^2\ln\likelihood_{\rm mag}(\eparams_{\rm mag}|\params)
/ \partial \eparams_{\rm mag}^2$ is the $3 \times 3$ \textit{Hessian} matrix 
whose elements are comprised of all the second-order derivatives.
This correspondence between the covariance and the Hessian is
a generic feature for Normal likelihoods.
An example of this covariance structure in $(\mu, A_V, R_V)$
is shown on the left side of Figure \ref{fig:linear_approx}.

Given a prior/independent constraint
$\pi(\eparams) = \Normal{\eparams_{\pi}(\params)}{\cov_{\pi}(\params)}$
on our extrinsic parameters that are Normally distributed 
around mean $\eparams_{\pi}(\params)$ with
covariance $\cov_{\pi}(\params)$, the corresponding posterior can also
be shown to be Normal with a mean equal to the \textit{maximum a posteriori}
(MAP) estimate by solving
\begin{align}
    \left[\cov_{\rm mag}^{\rm MAP}(\params)\right]^{-1}
    \eparams_{\rm mag}^{\rm MAP}(\params) 
    = [\cov_{\rm mag}^{\rm MLE}(\params)]^{-1} \eparams_{\rm mag}^{\rm MLE}(\params)
    +  [\cov_{\rm mag}^{\pi}(\params)]^{-1} \eparams_{\rm mag}^{\pi}(\params)
\end{align}
where the (inverse) covariance is now
\begin{equation}
    \left[\cov_{\rm mag}^{\rm MAP}(\params)\right]^{-1} 
    = [\cov_{\rm mag}^{\rm MLE}(\params)]^{-1} 
    + [\cov_{\rm mag}^{\pi}(\params)]^{-1}
\end{equation}

Unfortunately, we do not have (Normal) priors/independent constraints 
on $\mu$, $A_V$, and $A_V \times R_V$, but rather on
$\varpi$, $A_V(d|\ell, b)$, and $R_V$ separately. This poses a problem,
since in practice we find that the MLE solution is unstable 
without independent constraints on $R_V$, with the reddening vector able 
to adjust to nonphysical values to better model the data.

We resolve this by breaking our problem into two parts. First,
we neglect the contribution of any constraints on $\mu$ or $A_V$,
which are not Normal here, and only consider 
our prior $\pi(R_V)$.
Then, we exploit the fact that this system of equations is
linear in $(\mu, A_V)$ at fixed $R_V$. Likewise, it is linear in
$(\mu, R_V)$ at fixed $A_V$. To find $\eparams_{\rm mag}^{\rm MLE}$, we therefore
alternate between
\begin{itemize}
    \item solving for $(\mu, A_V)_{\rm MLE}$ at fixed $R_V$,
    including the contribution from $\pi(R_V)$, and
    \item solving for $(\mu, R_V)_{\rm MLE}$ at fixed $A_V$.\footnote{{\brutus} can also 
    incorporate Normal priors on $A_V$ if desired.}
\end{itemize}
We iterate between the two until both $A_V$ and $R_V$
converge to within $\delta_{A_V} = \delta_{R_V} = 0.05$, 
which typically occurs within a handful of iterations.
Our strategy for incorporating parallax measurements
and other priors is described in \S\ref{subsec:methods_mc}
and \S\ref{subsec:methods_cuts}.

Finally, even with these additional constraints, it is still possible
that our MAP estimate for $A_V$ and/or $R_V$ will be unphysical (e.g., $A_V < 0$).
To deal with these edge cases, by default the MAP for $A_V$ is limited to be between
$A_{V, {\rm min}} = 0$ and $A_{V, {\rm max}} = 6$ and the MAP for
for $R_V$ to be between $R_{V, {\rm min}} = 1$ and $R_{V, {\rm max}} = 8$.
While these choices may lead to issues in regions with particularly
high extinction, they can easily be changed through in-line arguments.

\subsection{Linear Regression in Flux Density} \label{subsec:methods_flux}

Above, we have assumed that the photometric errors are
Normally distributed in \textit{magnitudes} (i.e. log-flux).
This is a reasonable approximation when the errors are small ($\lesssim 5\%$),
but lead to estimates and errors for parameters that are slightly different
than when we assume errors are distributed following a Normal distribution
in \textit{flux density}.

For small changes $\Delta A_V$ and $\Delta R_V$ around 
a given value of $\eparams$, 
we can Taylor expand our previous linear model in magnitudes to get
a model that is linear in flux density
\begin{equation}
    \flux_{\params, \eparams}(\Delta \eparams)
    \approx s \flux_{\params, \eparams} \odot\, \left[1 - \frac{\ln 10}{2.5} \Delta A_V
    \left(\rvec_{\params} + \Delta R_V \rvec'_{\params}\right)\right]
\end{equation}
where $\odot$ here indicates the Hadamard product 
(i.e. element-wise multiplication),
$\flux_{\params, \eparams}$ is the ``absolute'' model flux density
derived from the absolute model magnitudes $\absmag_{\params, \eparams}$, and
$s$ is now a scale factor that adjusts the normalization.
Slightly abusing notation again and defining our parameters of interest to be
\begin{equation}
    \Delta \eparams_{\rm flux} \equiv
    \begin{bmatrix}
    s \\
    -s\frac{\ln 10}{2.5}\Delta A_V \\
    -s\frac{\ln 10}{2.5}\Delta A_V \Delta R_V
    \end{bmatrix}
\end{equation}
we can again rewrite the above model in matrix form as
\begin{equation}
    \flux_{\params, \eparams}(\Delta \eparams_{\rm flux}) 
    \approx (\Delta \eparams_{\rm flux})^{\T} \data_{\params, \eparams}
\end{equation}
where
\begin{equation}
    \data_{\params, \eparams} =
    \begin{bmatrix}
    \flux_{\params, \eparams} \\
    \flux_{\params, \eparams} \\
    \flux_{\params, \eparams} \\
    \end{bmatrix}
    \,\odot\,
    \begin{bmatrix}
    \mathbf{1} \\
    \rvec_{\params} \\
    \rvec_{\params}' \\
    \end{bmatrix}
\end{equation}

Our log-likelihood then is
\begin{align}
    -2 \ln\likelihood_{\rm flux}(\Delta \eparams_{\rm flux}|\params, \eparams) 
    = \left[\hat{\flux} - 
    (\Delta \eparams_{\rm flux})^{\T} \dvec_{\params, \eparams}\right]^{\T}
    \cov_{\flux}^{-1} 
    \left[\hat{\flux} - 
    (\Delta\eparams_{\rm flux})^{\T} \dvec_{\params,\eparams}\right]
    + \ln\left[{\rm det}\left(2\pi\cov_{\flux}\right)\right]
\end{align}
As \S\ref{subsec:methods_mags}, 
this implies the conditional likelihood is analytic such that
\begin{align}
    \likelihood_{\rm flux}(\Delta\eparams_{\rm flux}|\params,\eparams) 
    = \underbrace{\likelihood_{\rm flux}^{\rm MLE}(\params, \eparams)
    \times {\rm det}\left[2\pi\cov_{\rm flux}^{\rm MLE}(\params,\eparams)\right]
    }_{\rm Normalization}
    \times \underbrace{
    \Normal{\Delta\eparams_{\rm flux}^{\rm MLE}(\params, \eparams)}
    {\cov_{\rm flux}^{\rm MLE}(\params, \eparams)}
    }_{\rm Distribution}
\end{align}
where $\likelihood_{\rm flux}^{\rm MLE}(\params, \eparams)$ is again evaluated
at the MLE, the mean is
\begin{equation}
    \Delta \eparams_{\rm flux}^{\rm MLE}(\params, \eparams) = 
    \left(\dvec_{\params,\eparams} \cov_{\flux}^{-1} 
    \dvec_{\params, \eparams}^{\T}\right)^{-1}
    \dvec_{\params, \eparams} \cov_{\flux}^{-1} \hat{\flux}
\end{equation}
and the covariance is
\begin{equation}
    \cov_{\rm flux}^{\rm MLE}(\params, \eparams) = \left(\dvec_{\params,\eparams} 
    \cov_{\flux}^{-1} \dvec_{\params, \eparams}^{\T}\right)^{-1}
\end{equation}

As with \S\ref{subsec:methods_mags}, in the presence of 
(Normal) priors/independent constraints,
the MAP estimates $\Delta \eparams_{\rm flux}^{\rm MAP}(\params, \eparams)$
and $\cov_{\rm flux}^{\rm MAP}(\params, \eparams)$ are 
straightforward to derive.
Similarly, due to the constraints on $R_V$ needed for the
solution to be well-behaved, we solve for elements of
$\Delta \eparams_{\rm flux}$ iteratively conditioning 
on $A_V$ and $R_V$ in turn.

As the data $\hat{\flux}$ are fundamentally 
in flux density space, we are ultimately interested
in constraining our parameters relative to our linear model
in flux density $\flux_{\params, \eparams}(\Delta\eparams)$. 
We therefore will redefine our earlier set of extrinsic parameters as
\begin{equation}
    \eparams \equiv 
    \begin{bmatrix}
    s \\
    A_V \\
    R_V
    \end{bmatrix}
\end{equation}
where the only change relative to our original parameters is that now
we are inferring the scale-factor $s$ rather than the distance modulus
$\mu$. Note that it is straightforward to convert from $\eparams_{\rm mag}$
and $\Delta \eparams_{\rm flux}$ to $\eparams_{\rm flux}$ and $\eparams$.
Around the true MAP solution $\eparams_{\rm MAP}(\params)$ where 
$\Delta \eparams = (0, 0, 0)$, we can compute
$\cov_{\rm MAP}(\params)$ by explicitly computing the
corresponding Hessian matrix of second-order derivatives.
An example showing the covariance structure in $(s, A_V, R_V)$ versus
that in $(\mu, A_V, R_V)$ from \S\ref{subsec:methods_mags}
is shown in Figure \ref{fig:linear_approx}.

Altogether, this gives us a straightforward recipe for ``optimizing''
our initial MAP solution, derived from magnitudes, to the corresponding
distribution in flux:
\begin{enumerate}
    \item Starting from an initial guess of $A_V = 1$, $R_V = 3.32$,
    derive $\eparams_{\rm mag}^{\rm MAP}(\params)$. Iterate
    in $(\mu, A_V)$ and $(\mu, R_V)$ until convergence.
    \item Convert from $\eparams_{\rm mag}^{\rm MAP}(\params) \rightarrow 
    \eparams_{\rm MAP}(\params)$ and expand around 
    $\eparams_{\rm MAP}(\params)$ to derive
    $F_{\params, \eparams}(\Delta\eparams)$.
    \item Compute first order corrections $\Delta\eparams_{\rm MAP}
    (\params,\eparams)$ for $s$, $A_V$, and $R_V$.
    \item Compute the new MAP solution
    $\eparams_{\rm MAP}^{\rm new} 
    = \eparams_{\rm MAP} 
    + \Delta\eparams_{\rm MAP}(\params,\eparams)$.
    \item Repeat steps 2-4 starting from $\eparams_{\rm MAP}^{\rm new}$
    until the likelihood converges.
\end{enumerate}
In general, we find the likelihood quickly converges 
after only a handful of iterations.

\subsection{Incorporating Priors with Importance Sampling} \label{subsec:methods_mc}

We are interested in estimating expectation values
(see \S\ref{subsec:impl_motivation}) over the \textit{conditional posterior}
\begin{equation}
    P(\eparams | \params, \hat{\flux}, \hat{\varpi}) = 
    \likelihood_{\rm phot}(\eparams | \params) 
    \likelihood_{\rm astr}(\eparams)
    \prior(\eparams|\params)
\end{equation}
From our above calculations, we now have an analytic
approximation for $\likelihood_{\rm phot}(\eparams|\params)$.
Critically, we now know that, conditional on $\params$,
$\eparams$ is approximately distributed as
\begin{equation}
    \eparams(\params) \sim 
    \Normal{\eparams_{\rm MAP}(\params)}
    {\cov_{\rm MAP}(\params)}    
\end{equation}
and therefore easy to simulate. Given a set of
$m$ independently and identically distributed (iid) samples 
$\{ \eparams_{i} \}_{i=1}^{i=m} \sim 
\Normal{\eparams_{\rm MAP}(\params_i)}{\cov_{\rm MAP}(\params_i)}$
drawn from this Normal distribution at a particular value $\params_i$, we can
then approximate expectation values over
posterior using a Monte Carlo approach through
\textit{Importance Sampling} via
\begin{align}
    \meanwrt{f(\eparams|\params_i)}{P}
    &= \int f(\eparams|\params_i) 
    P(\eparams | \params_i, \hat{\flux}, \hat{\varpi})
    \, {\rm d}\eparams \nonumber \\
    &\approx \frac{1}{m}\sum_{j=1}^{m}
    q_{i,j} \times f(\eparams_j|\params_i)
\end{align}
where the importance weight $q_{i,j}$ for each sample $\eparams_{i,j}$
is defined as
\begin{equation}
    q_{i,j} \equiv 
    \frac{P(\eparams_j | \params_i, \hat{\flux}, \hat{\varpi})}
    {\likelihood_{\rm phot}(\eparams_j|\params_i)}
    = \likelihood_{\rm astr}(\eparams_j)
    \prior(\eparams_j|\params_i)
\end{equation}

Generally, we find that for $n_{\rm prior} \gtrsim 30$ samples our estimates
of the integrated conditional probability for a particular model
\begin{equation}
    q_{i} \equiv 
    \int P(\eparams | \params_i, \hat{\flux}, \hat{\varpi})
    \, {\rm d}\eparams 
    \approx \frac{1}{m}\sum_{j=1}^{m} q_{i,j}
\end{equation}
tend to roughly converge. As a result, we choose $n_{\rm prior} = 50$ as the default
number of samples in our computations. This step, which requires generating
a large number of samples for every un-clipped model in our initial grid 
and subsequently evaluating the astrometric likelihood 
and Galactic prior for each sample,
typically takes up the majority of the computation time.

\subsection{Application over Stellar Parameter Grids} \label{subsec:methods_grids}

Above, we showed that we can analytically
solve for the conditional likelihood 
$\likelihood_{\rm phot}(\eparams|\params)$, which implies that we should
be able to generate estimates for those parameters without the use of grids
over $\eparams$.
This allows us to both substantially improve our resolution 
in both $\params$ (over our grid)
and $\eparams$ (with analytic sampling).
In addition, we simultaneously decrease the required computation 
since we only require a grid of $n$ 
values $\{ \params_i \}_{i=1}^{i=n}$ over the 
$m \lesssim 4$ intrinsic set of stellar parameters $\params$
rather than over the $m+3$ parameters from 
both $\params$ and $\eparams$.

We can exploit such a grid by noting that our original posterior can
be rewritten as
\begin{equation}
    P(\params, \eparams | \hat{\flux}, \hat{\varpi}) 
    = P(\eparams | \params, \hat{\flux}, \hat{\varpi})
    P(\params | \hat{\flux}, \hat{\varpi})
\end{equation}
for a given $\params$.
This then implies that we can generate random samples from our joint likelihood in
two steps. First, we draw a random sample $\params'$ from
$P(\params | \hat{\flux}, \hat{\varpi})$. Then,
we draw a corresponding random sample $\eparams'$ from
$P(\eparams | \params', \hat{\flux}, \hat{\varpi})$. More formally,
\begin{align*}
    \params' &\sim P(\params | \hat{\flux}, \hat{\varpi}) \\
    \eparams' &\sim P(\eparams | \params', \hat{\flux}, \hat{\varpi})
\end{align*}

We exploit the weighted Monte Carlo samples from \S\ref{subsec:methods_mc}
to perform this operation using a \textit{bootstrap approximation}.
For a particular grid point $\params_i$, 
generating a random sample $\eparams'$ from the conditional
posterior $P(\eparams | \params_i, \hat{\flux}, \hat{\varpi})$
is roughly equivalent to picking one of the already-generated set of $m$ samples
$\{ \eparams_{i,j} \}_{j=1}^{j=m}$ with a probability 
proportional to its weight $q_{i,j}$. More formally, this implies
\begin{equation}
    \eparams' \sim {\rm Cat}\left[\{\eparams_{i,j} \}_{j=1}^{j=m}, 
    \{ q_{i,j} / q_i\}_{j=1}^{j=m}\right]
\end{equation}
where ${\rm Cat}[\mathbf{x}, \mathbf{p}]$ is the \textit{Categorical distribution}
over $\mathbf{x}=\{x_j\}_{j=1}^{j=m} = \{\eparams_{i,j} \}_{j=1}^{j=m}$ 
with corresponding probabilities
$\mathbf{p}=\{p_j\}_{j=1}^{j=m} = \{ q_{i,j} / q_i\}_{j=1}^{j=m}$ 
and again $q_i = \sum_j q_{i,j}$
is the total conditional probability.

Likewise, we can approximate drawing a sample $\params'$ from
$P(\params | \hat{\flux}, \hat{\varpi})$ using the same strategy
by taking
\begin{equation}
    \params' \sim {\rm Cat}\left[\{\params_i \}_{i=1}^{i=n}, 
    \{ p_i / p\}_{i=1}^{i=n}\right]
\end{equation}
The prior-weighted probability $p_i$ corresponding
to a given $\params_i$ is related to
our previous $q_i$ via
\begin{equation}
    p_i \equiv q_i \times \prior_i \times \Delta_i 
\end{equation}
where $\prior_i = \prior(\params_i)$ is the prior probability
evaluated at $\params_i$ and $\Delta_i$ is the
associated grid spacing. The normalization
\begin{equation}
    p \equiv \sum_i p_i
    \approx \int P(\eparams, \params | \hat{\flux}, \hat{\varpi}) 
    {\rm d}\eparams {\rm d} \params
\end{equation}
is then the estimated marginal likelihood (i.e. the Bayesian evidence)
for the source, which in theory may be useful for model comparisons
in future applications.

Using this strategy, we post-process our MAP fits and Monte Carlo samples
over our grid points to generate $n_{\rm post} = 250$
$\{ (\params_k, \eparams_k) \}_{k=1}^{k=n_{\rm post}}$ samples
that serve as rough approximations to the underlying posterior
$P(\params, \eparams | \hat{\flux}, \hat{\varpi})$. In particular,
we wish to emphasize that these options only generate
\textit{approximations} to the underlying posterior. 
We expect the largest differences between the estimated posterior
relative to the true distribution will be primarily due to
\begin{itemize}
    \item grid resolution effects in $\params$,
    \item ``posterior noise'' caused by noisy estimates of the $p_i$'s, and
    \item ``resampling noise'' due to the procedure used to
    resample the final set of $n_{\rm post}$ samples.
\end{itemize}
See \S\ref{sec:tests} for examples of the impact these may have on
inferring underlying relevant intrinsic and extrinsic stellar properties.
An example of the output stellar parameters and the associated SED
can be seen in Figure \ref{fig:example}.

\subsection{Additional Implementation Details} \label{subsec:methods_cuts}

At multiple points during the fitting process we apply cuts to decrease
the effective size of the stellar parameter grid we are dealing with in order to
reduce the overall run time. The choices we take are summarized
below. The relevant parameters associated with them
are listed in Table \ref{tab:settings}.
The general procedure is shown in Figure \ref{fig:brutus_algorithm}.

\begin{deluxetable}{lcc}
\tablecolumns{3}
\tablecaption{Default settings for hyper-parameters 
used when deriving posterior estimates with {\brutus}.
\label{tab:settings}}
\tablehead{Description & Symbol & Value}
\startdata
\cutinhead{\textbf{Bounds}}
Minimum allowed $A_V$ & $A_{V, {\rm min}}$ & $0\,{\rm mag}$ \\
Maximum allowed $A_V$ & $A_{V, {\rm max}}$ & $6\,{\rm mag}$ \\
Minimum allowed $R_V$ & $R_{V, {\rm min}}$ & $1$ \\
Maximum allowed $R_V$ & $R_{V, {\rm max}}$ & $8$ \\
\cutinhead{\textbf{Magnitude Step}}
Tolerance in $A_V$ & $\delta_{A_V}$ & $0.05\,{\rm mag}$ \\
Tolerance in $R_V$ & $\delta_{R_V}$ & $0.05$ \\
Relative likelihood tolerance threshold
& $f_{\rm tol}^{\rm mag}$ & $0.005$ \\
\cutinhead{\textbf{Flux Step}}
Relative likelihood threshold
& $f_{\rm init}^{\rm flux}$ & $0.005$ \\
Tolerance in $\likelihood_{\rm phot}$
& $\delta_{\likelihood}$ & $0.03$ \\
Relative likelihood tolerance threshold
& $f_{\rm tol}^{\rm flux}$ & $0.01$ \\
\cutinhead{\textbf{Monte Carlo Step}}
Relative posterior threshold
& $f_{\rm post}$ & $0.005$ \\
Number of samples used for integration
& $n_{\rm prior}$ & $50$ \\
Number of resampled posterior draws to save
& $n_{\rm post}$ & $250$ \\
\enddata
\end{deluxetable}

\subsubsection{Magnitude Step} \label{subsubsec:impl_mag}

While models that are reasonable fits to the data
generally converge quickly, the $A_V$ and $R_V$ values for
models that are poor fits can sometimes be ill-behaved.
As a result, we only worry about convergence for 
objects with likelihoods whose values are greater than
$f_{\rm tol}^{\rm mag} = 0.005$ times the maximum
value after the first linear regression step. 
We consider our fits converged when
the maximum variation in $A_V$ and $R_V$ over these ``reasonably fit''
models falls below $\delta_{A_V} = 0.05\,{\rm mag}$
and $\delta_{R_V} = 0.05$, respectively, between one iteration and the next.
This usually occurs within $\lesssim 5$ iterations.

\subsubsection{Flux Density Step} \label{subsubsec:impl_flux}

After our magnitude fits have converged,
we only want to perform additional optimization 
after expanding in flux density
for a smaller subset of grid points, indexed by $i$,
that give ``good'' likelihoods.
Using the MLE values for $\{ \eparams_{{\rm MLE}, i} \}_{i=1}^{n}$,
we compute the corresponding photometric likelihood
$\likelihood_{\rm phot}(\params_i, \eparams_{{\rm MLE}, i})$ 
and, if it is measured, an astrometric likelihood 
$\likelihood_{\rm astr}(\eparams_{{\rm MLE}, i})$
for each object. We only perform the subsequent
flux density expansion and optimization for objects with combined 
astro-photometric likelihoods
that are within $f_{\rm init}^{\rm flux} = 0.005$ 
times the maximum value. This further screens models that
may provide good fits to the photometry but give inconsistent
parallax estimates.

As with our magnitude step, we only
consider convergence over a set of ``reasonably fit'' models
with likelihoods greater than $f_{\rm tol}^{\rm flux} = 0.01$
the current maximum value in order to avoid being overly-sensitive
to the worst-fit models under consideration (i.e. to decrease
our sensitivity to the likely tails of the distribution). 
We take our fits to be converged after the change in the
photometric log-likelihood
$\ln \likelihood_{\rm phot}(\params_i, \eparams_{{\rm MLE}, i})$
between iterations falls below $\delta_{\likelihood} = 0.03$.
As in \S\ref{subsubsec:impl_mag}, 
this usually occurs within $\lesssim 5$ iterations.

\subsubsection{Monte Carlo Step} \label{subsubsec:impl_mc}

Since generating samples
for all of our models is both time-consuming 
and memory-intensive over large grids, we only want to
perform this step (and the subsequent resampling)
over a small subset of models. We subselect these
based on their ``expected'' posterior probabilities,
which we compute in two steps.

First, we ignore all distance-dependent effects such as
our 3-D priors over stellar properties and dust extinction.
The remaining ``static'' contributions of our prior are then
direct functions of our underlying grid of stellar parameters
$\{ \params_i \}_{i=1}^{i=n}$ and can therefore be pre-computed.
This includes contributions from
the spacing of our grid points $\{ \Delta_i \}_{i=1}^{i=n}$
(see \S\ref{subsec:impl_motivation})
and the associated IMF prior over mass $\pi(M_{\rm init})$.

Second, we again want to incorporate possible
constraints from the measured parallax.
Unlike in \S\ref{subsubsec:impl_flux}, however,
we now want to incorporate uncertainties
$\sigma_s(\params_i)$ in the inferred scale-factor $s(\params_i)$
for a particular model $\params_i$. 
The marginal distribution of $s(\params_i)$ derived from
$\likelihood_{\rm phot}(\eparams | \params_i)$
is Normal:
\begin{equation}
    s(\params_i) \sim \Normal{\mu_s(\params_i)}{\sigma_s^2(\params_i)}
\end{equation}
where $\mu_s(\params_i)$ and $\sigma_s(\params_i)$
are just taken from the appropriate
elements of $\eparams_{\rm MAP}(\params_i)$
and $\cov_{\rm MAP}(\params_i)$.
It is straightforward to show that the inferred
scale-factor $s(\params_i)$ is simply the square
of the associated parallax $\varpi(\params_i)$:
\begin{equation}
   s(\params_i) = \varpi^2(\params_i)
\end{equation}

In the case where the measured parallax
$\hat{\varpi}$ is Normally distributed with
at least a moderate signal-to-noise ratio (SNR)
(i.e. $\hat{\varpi}/\sigma_{\varpi} \gg 1$),
it is straightforward to show that the $\varpi^2(\params_i)$ 
is also expected to be approximately Normally distributed such that
\begin{equation}
    \varpi(\params_i) \sim \Normal{\hat{\varpi}}{\sigma_{\varpi}^2} 
    \quad\Rightarrow\quad
    \varpi^2(\params_i) \sim \Normal{\mu_{\varpi^2}}{\sigma_{\varpi^2}^2}
\end{equation}
where $\mu_{\varpi^2} = \hat{\varpi}^2 + \sigma_{\varpi}^2$
and $\sigma_{\varpi^2} = 2 \sigma_{\varpi}^4 
+ 4 \hat{\varpi}^2 \sigma_{\varpi}^2$.
We take our likelihood $\likelihood_s(\params)$ for $s$
to be the convolution of this distribution
with the MAP uncertainties assuming the parallax SNR is
above a minimum SNR threshold $\varpi_{\rm SNR}^{\rm min}$
and uniform otherwise:
\begin{equation}
    \likelihood_s(\params) =
    \begin{cases}
    1 & \hat{\varpi}/\sigma_{\varpi} < \varpi_{\rm SNR}^{\rm min} \\
    \Normal{\mu_{\varpi^2}}{\sigma_{\varpi^2}^2
    + \sigma_s^2(\params)} & \hat{\varpi}/\sigma_{\varpi} \geq \varpi_{\rm SNR}^{\rm min}
    \end{cases}
\end{equation}
We find $\varpi_{\rm SNR}^{\rm min} = 4$ to be a reasonable
threshold where this approximation remains valid to $\sim 10\%$.

After applying these two terms to the photometric likelihoods
computed from our earlier steps, we then select 
the small subset of models 
(often $\lesssim 1\%$ of the original grid) 
with expected posterior probabilities 
greater than $f_{\rm post} = 0.005$ that of the maximum computed value.
We then perform the Monte Carlo integration and resampling procedure
described in \S\ref{subsec:methods_mc} using $n_{\rm prior} = 50$
samples per selected model and saving a total of
$n_{\rm post} = 250$ posterior samples.

\subsubsection{Runtime} \label{subsubsec:impl_timing}

Overall, we find that for grids with $n \sim 7.5 \times 10^5$ models,
{\brutus} is able to perform full posterior estimation
in $\lesssim 5-10$ seconds when fitting $\sim 5-10$ bands of photometry
with low-to-moderate SNR parallax measurements.
This is comparable to the ``rough'' version of {\starhorse}
described in \citet{anders+19} while giving
continuous resolution in the extrinsic parameters $\eparams$
and reasonable resolution over the intrisic stellar parameters $\params$
(see \S\ref{sec:models}).

Due to the adaptive thresholding, Monte Carlo (re)sampling,
and use of parallax measurements, the scaling tends to be
non-linear in the number of bands and the size of the grid.
For smaller grids of $n \sim 4 \times 10^4$ elements (as in
\S\ref{subsec:bayestar}), the typical runtime is $\sim 1$
second, while for much larger grids of $n \sim 2 \times 10^6$
elements the typical runtime can range from $\sim 20-40$ seconds.
With tight parallax constraints, runtimes can be
up to an order of magnitude faster (down to $\sim 0.1$ seconds).

\section{Linear Reddening Approximation} \label{ap:linear_dust}

\begin{figure*}
\begin{center}
\includegraphics[width=\textwidth]{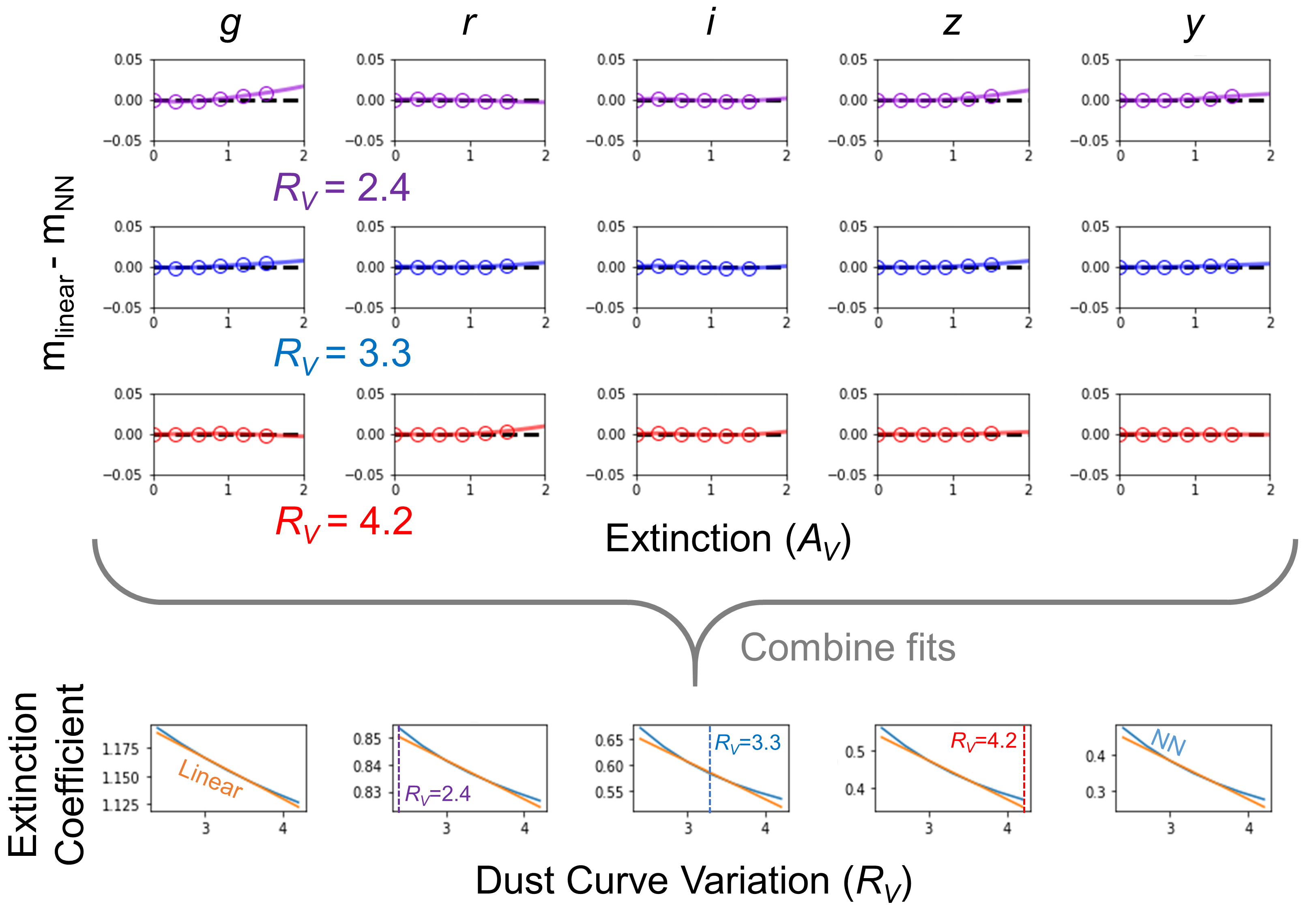}
\end{center}
\caption{An illustration of the linear reddening approximation used in {\brutus}
for the {\mist} models with the \citet{fitzpatrick04} $R_V$-dependent reddening curve
in the Pan-STARRS $g$, $r$, $i$, $z$, and $y$ filters.
First, photometry in each filter is generated using a neural network
(NN) trained on the grid of {\ctk} model atmospheres summarized in Table \ref{tab:ctk_grid}
over a grid of $A_V$ values (open circles) at fixed $R_V$. 
The resulting magnitudes are then fit using weighted linear regression. 
Deviations between the linear predictions ($m_{\rm linear}$)
and the NN baseline ($m_{\rm NN}$) for values of
$R_V=2.4$ (purple), $R_V=3.3$ (blue), and $R_V=4.2$ (red) are highlighted in the
first three rows.
We then perform weighted linear regression over the set of fitted 
linear extinction coefficients.
The resulting linear fits (orange) and the baseline NN predictions
(light blue) are shown in the bottom row.
See \S\ref{subsubsec:mist_dust} for additional details.
}\label{fig:linear_red}
\end{figure*}

We approximate the linear $A_V$ and $R_V$ vector in two steps:
\begin{enumerate}
    \item \textit{$A_V$ step}: At fixed $R_V$, we compute the absolute magnitudes
    $\{ \absmag_{\params}(A_{V,i}|R_V) \}_{i=1}^{m}$ over a grid of $m$
    $A_V$ values from $A_V=0$ to $A_V=1.5$ with a resolution of
    $\Delta A_V = 0.3$. We then perform weighted linear regression
    over the grid points where each point is assigned a weight
    \begin{equation}
    w(A_V) = (10^{-5} + A_V)^{-1}
    \end{equation}
    This ensures the fit essentially goes through $A_V=0$ with
    the $A_V > 0$ grid points assigned a weight inversely
    proportional to the extinction. This then provides
    a set of reddening vectors $\rvec_{\params}(R_V)$
    as a function of $R_V$.
    \item \textit{$R_V$ step}: We compute $\rvec_{\params}(R_V)$ over a grid of $R_V$
    values from $R_V=2.4$ to $4.2$ with a resolution of 
    $\Delta R_V = 0.3$. Then, we perform weighted
    linear regression over the $R_V$ grid points
    where each point is assigned a weight of
    \begin{equation}
    w(R_V) = \exp(-|R_V - 3.3| / 0.5)
    \end{equation}
    This gives maximum weight to the grid point at $R_V=3.3$
    and exponentially declining weight for grid points with
    larger/smaller $R_V$ to ensure we best reproduce the behavior
    for ``typical'' $R_V$ values. This procedure then
    provides a set of $\rvec_{\params}$ and $\rvec_{\params}'$
    associated with each $\absmag_{\params}$.
\end{enumerate}

An example of this overall procedure and the accuracy of
these linear approximations for a solar-like star over several
filters is shown in Figure \ref{fig:linear_red}. Overall, we find
agreement at the few percent level across a wide range
of $A_V$ and $R_V$ values.

\section{Cluster Modeling} \label{ap:calib_clusters}

We build up the basic components of our cluster model below.
In \S\ref{subsec:cluster_ideal}, we outline our baseline model.
In \S\ref{subsec:cluster_binary}, we
introduce our approach for modeling unresolved (non-interacting)
binaries. In \S\ref{subsec:cluster_outlier}, we describe
our approach for dealing with cluster contamination and/or
alternate stellar populations. In \S\ref{subsec:cluster_eiso},
we describe the additional parameters used to add 
empirical corrections to underlying isochrones derived from the {\mist} models.
We finally describe additional empirical photometric corrections
in \S\ref{subsec:cluster_ephot}.

\subsection{Baseline Model} \label{subsec:cluster_ideal}

Assuming that an open cluster can be effectively
approximated by a simple stellar population (SSP),
all stars in the cluster have identical initial
metallicities $\feh_{\rm init}$ and ages $t_{\rm age}$
but have initial masses $M_{\rm init}$
that have been independently sampled from the stellar IMF.
We further assume that stars are located at approximately the same 
distance $d$ and are co-spatial on the sky that they
are behind the same screen of dust (i.e. all have the same
$A_V$ and $R_V$). We will collectively refer to these
``cluster-level'' parameters as 
$\params_{\rm cluster}$.
The log-posterior over $n$ photometric sources
$\{ \hat{\flux}_i\}_{i=1}^{i=n}$ is then proportional to
the product of the likelihoods and priors for each source 
plus the prior over the cluster-level parameters:
\begin{align}
    P(\{M_{{\rm init}, i}\}_{i=1}^{i=n},
    \params_{\rm cluster} | \{ \hat{\flux}_i\}_{i=1}^{i=n})
    &\propto \prior(\params_{\rm cluster})
    \prod_{i=1}^{n} \likelihood_i(M_{{\rm init}, i} | 
    \params_{\rm cluster}) \prior(M_{{\rm init}, i}) \\
    &\equiv \prior_{\rm cluster} \times 
    \prod_{i=1}^{n} \likelihood_i \times \prior_i
\end{align}

We want the likelihoods of each individual source to
incorporate information from various sources when available
without implicitly favoring objects with greater/fewer
measurements. As a result, we consider our
likelihood for each object to follow the $\chi^2$-distribution,
which is substantially less sensitive compared to the
Normal distribution to changes in dimensionality. Our
likelihood is then defined as
\begin{equation}
    \likelihood_i \equiv
	\frac{(\chi^2_i)^{k_i/2-1} \, e^{-\chi^2_i/2}}{2^{k_i/2} \, \Gamma(k_i/2)} 
\end{equation}
where $k_i$ is the effective number of data points ($b_i$ bands
of photometry plus an optional parallax measurement),
$\Gamma(\cdot)$ is the Gamma function, and
\begin{equation}
    \chi^2_i \equiv \sum_{j=1}^{b_i} 
    \frac{(\hat{F}_{i,j} - F_{i,j})^2}{\sigma_{F,i,j}^2}
    + \frac{(\hat{\varpi}_i - 1/d)^2}{\sigma_{\varpi,i}^2}
\end{equation}
is the ``standard'' $\chi^2$-statistic as a function of
the model photometry $\flux_i$ for each object $i$ given
$M_{{\rm init},i}$ and $\params_{\rm cluster}$.

For our baseline model, we assume that our prior
is uniform within some specific range of allowed
$\params_{\rm cluster}$ and $M_{\rm init}$ values such that
$\prior_{\rm cluster}$ and $\{ \prior_i \}_{i=1}^{i=n}$ are constant.
While this assumption is not entirely accurate (e.g., it does not apply
constraints from the IMF), it drastically simplifies the problem
since in practice we do not know $\{ M_{{\rm init}, i} \}_{i=1}^{i=n}$
or other associated parameters for each source.
While proper inference requires utilizing 
a full hierarchical model \citep[e.g.,][]{vonhippel+06,degennaro+09,vandyk+09}
to estimate both $\params_{\rm cluster}$ and $\{ M_{{\rm init}, i} \}_{i=1}^{i=n}$,
we instead simply marginalize over the latter to get
\begin{align}
    P (\params_{\rm cluster} | \{ \hat{\flux}_i\}_{i=1}^{i=n})
    \propto \prod_{i=1}^{n} \int_{M_{\rm min}}^{M_{\rm max}}
    \likelihood_i(M_{\rm init} | \params_{\rm cluster})
    \,{\rm d}M_{\rm init}
\end{align}
where $M_{\rm min} = 0.3\,M_\odot$ (the minimum mass where
the {\mist} models are well-behaved) and $M_{\rm max}=100\,M_\odot$
(the standard upper limit for the IMF).

As described in \citet{dotter16},
isochrones are more evenly sampled in EEP than in $M_{\rm init}$.
As a result, we opt to evaluate this integral using a grid
in ${\rm EEP}$. Similar to \S\ref{subsubsec:mist_iso}, we
account for the unequal spacing in $M_{\rm init}$
using second-order numerical derivatives in
$\Delta M_{\rm init} / \Delta {\rm EEP}$.
By default, we use a grid of $n=2000$ evenly-spaced
values between ${\rm EEP} = 202$ and $808$, 
which spans the beginning of the MS
to the beginning of the thermally-pulsing asymptotic giant branch.

We wish to note that the approach taken here is substantially
different from the analysis of binned ``Hess Diagrams'' 
commonly used in the literature 
\citep[see, e.g.,][]{dolphin97,dolphin02,dejong+08,weisz+11,gouliermis+11,gossage+18}.
While the current approach requires simpler models, it has the
benefit of not being sensitive to any issues related to binning
over the CMD and can jointly model \textit{all} observed
bands rather than just 2-3 at a time.
More complex star formation histories beyond SSPs can also
be explored through the use of, e.g., inhomogeneous Poisson
processes \citep[see, e.g.,][]{leja+19}, although we defer any
such improvements to future work.

This baseline model is shown in the first panel of Figure \ref{fig:iso_fit}.

\subsection{Binaries} \label{subsec:cluster_binary}

As clusters display prominent binary sequences, we
explicitly consider the case where a given source 
could be an unresolved binary with a primary mass of
$M_{\rm init}$ and a binary companion with
secondary mass fraction $q \in [0, 1]$, where $q=0$
is equivalent to a single source and $q=1$ is
an equal-mass binary. Marginalizing 
over both $M_{\rm init}$ and $q$ then gives
\begin{align}
    P(\params_{\rm cluster} | \{ \hat{\flux}_i\}_{i=1}^{i=n})
    \propto \prod_{i=1}^{n} 
    \int_{M_{\rm min}}^{M_{\rm max}} \int_{0}^{1}
    \likelihood_i(M_{\rm init}, q | \params_{\rm cluster})
    \prior(M_{\rm init}, q | \params_{\rm cluster})
    \,{\rm d}M_{\rm init}\,{\rm d}q
\end{align}
We evaluate this integral
in $q$ using an adaptively-spaced grid of
$n=14$ values from $q=0$ to $1$ with a resolution that ranges
from $\Delta q = 0.2$ near the edges to $\Delta q = 0.05$
around $q=0.6$.

Compared to \S\ref{subsec:cluster_ideal},
incorporating binaries involves dealing with a few additional
subtleties. First, we only can model binaries down to $M_{\rm min}$.
This means that while the likelihood for $q=0$
(i.e. a single star) is defined, the likelihood from
$0<q<M_{\rm min}/M_{\rm init}$ is undefined. 
Second, close (unresolved) binaries generally
only exist when the primary (higher-mass)
binary companion has not yet evolved off the MS,
since this process tends to decouple the system. As a result,
our likelihood as a function of $q > 0$ is only defined
for systems with ${\rm EEP} < 480$ 
(i.e. before the first ascent up the giant branch).
Our prior therefore becomes
\begin{align}
    \prior &(M_{\rm init}, q | \params_{\rm cluster})
    \propto 
    \begin{cases}
    1 & \underset{{\rm Single}}{\underbrace{q = 0}} 
    \:{\rm or}\: \underset{{\rm Binary\:System}}{\underbrace{
    \left(\frac{M_{\rm min}}{M_{\rm init}} \leq q \leq 1 
    \:{\rm and}\: {\rm EEP} < 480 \right)}} \\
    0 & {\rm otherwise}
    \end{cases}
\end{align}
where ${\rm EEP}$ again is a function of $M_{\rm init}$ and
$\params_{\rm cluster}$.

A representation of the impact including binaries has on 
our cluster model is shown in the second panel of Figure \ref{fig:iso_fit}.

\subsection{Outliers} \label{subsec:cluster_outlier}

In addition to the $\chi^2$ likelihood defined above, clusters
are subject to additional contamination due 
to background/foreground field stars, which are not associated
with the assumed SSP. In addition, clusters can also have populations
such as blue stragglers \citep{sandage53} whose stellar evolution has been influenced
by interactions with a companion, violating our basic SSP assumption.
To account for these outliers, 
we modify our likelihood for each object to be a weighted
mixture of two components such that
\begin{equation}
    \likelihood_i(f_{\rm out} | p_{{\rm in}, i}) =
    p_{{\rm in}, i} \times \likelihood_{{\rm in}, i} +
    (1 - p_{{\rm in}, i}) \times \likelihood_{{\rm out}, i}
\end{equation}
Here, $p_{{\rm in}, i}$ is the probability that object $i$ is
well-modeled by our SSP (i.e. an ``inlier'') and $1-p_{\rm in}$
is the probability that it is not (i.e. an ``outlier'').

The inlier probability
\begin{equation}
    p_{{\rm in}, i}(p_{{\rm mem}, i}, f_{\rm out}) 
    = p_{{\rm mem}, i} \times (1 - f_{\rm out})
\end{equation}
is defined to be a mixture of two things. The first is
the cluster membership probability $p_{{\rm mem, i}}$ 
for object $i$ based on possible external information such
as spatial position and/or proper motion. The second is 
the baseline outlier fraction $f_{\rm out}$,
which governs the fraction of objects for which our outlier model
$\likelihood_{{\rm out}, i} > \likelihood_{{\rm in}, i}$
serves as a better model than our inlier model (i.e. an SSP).
This is a free parameter that we are interested in modeling in addition
to the cluster parameters $\params_{\rm cluster}$.

We take our outlier model to be an adaptive threshold
such that our likelihood is constant
\begin{equation}
    \likelihood_{{\rm out}, i}(k_i) 
    \equiv \likelihood_{{\rm in}, i}(\chi^2_{\rm max}(k_i), k_i)
\end{equation}
where the value of the likelihood is defined
at the point where the cumulative
probability of our $\chi^2$-distribution with $k_i$
degrees of freedom is less than a particular threshold
$p_{\rm min}$:
\begin{equation}
    \int_{\chi^2_{\rm max}(k_i)}^{\infty}
    \likelihood_{{\rm in}, i}(\chi^2_i, k_i) \, {\rm d}\chi^2_i
    = p_{\rm min}
\end{equation}

This scheme is functionally equivalent to a Bayesian version of ``sigma-clipping''
for a given $p_{\rm min}$ with two main exceptions:
\begin{enumerate}
    \item The threshold used is not a constant value but depends on the
    number of bands $k_i$ observed for each object $i$.
    \item An individual object's contribution to the overall likelihood
    is \textit{de-emphasized} rather than being completely ignored.
\end{enumerate}
We set $p_{\rm min} = 10^{-5}$ by default, a conservative 
value which corresponds to $\sim 4.5$-sigma clipping.

A representation of the impact of outlier modeling on 
our cluster model is shown in the third panel of Figure \ref{fig:iso_fit}.

\subsection{Empirical Isochrone Corrections} \label{subsec:cluster_eiso}

While the above model allows us to incorporate most of the behavior
seen in open clusters (excluding the impact of rotation) such as
binarity (\S\ref{subsec:cluster_binary})
and outliers (\S\ref{subsec:cluster_outlier}), it does not account
for systematic modeling issues in the SSP itself as derived from
theoretical isochrones such as {\mist} and the use of synthetic
spectra such as those derived from the {\ctk} models.
We use a series of \textit{empirically-motivated corrections} 
(see \S\ref{subsec:calib_terms})
to the underlying isochrones in order to address some of these issues.
Note that these are common across all theoretical isochrones, not just {\mist},
with the exception of specialized grids designed to specifically tackle
particular problems.

Our corrections work by adding an additional set of ``corrected''
surface-level parameters $\params_\star'(\params_\star, \params)$
that are a function of the original predicted surface-level parameters
$\params_\star$ as well as the underlying stellar evolution
parameters $\params$ from the {\mist} isochrones. In particular,
we opt to modify the stellar radius $\log R_\star$ and the
effective temperature $\log T_{\rm eff}$ (and by proxy the
surface gravity $\log g$ and bolometric luminosity $\log L_{\rm bol}$)
such that
\begin{equation*}
    \begin{bmatrix}
    M_{\rm init} \\
    \feh_{\rm init} \\
    t_{\rm age}
    \end{bmatrix}
    \xrightarrow[]{}
    \begin{bmatrix}
    \log g \\
    \log T_{\rm eff} \\
    \log L_{\rm bol} \\
    \log R_\star \\
    \feh_{\rm surf}
    \end{bmatrix}
    \xrightarrow[]{}
    \begin{bmatrix}
    \log g' \\
    \log T_{\rm eff}' \\
    \log L_{\rm bol}' \\
    \log R_\star' \\
    \feh_{\rm surf}
    \end{bmatrix}
    \xrightarrow[]{}
    \begin{bmatrix}
    M_1 \\
    \vdots \\
    M_b
    \end{bmatrix}
\end{equation*}

We choose to apply empirical corrections to $\log R_\star$ and
$\log T_{\rm eff}$ for two reasons. The first is that we expect
both to be strongly affected by magnetic fields, which appear to
``puff up'' stars, making them larger, and contribute to
sunspot activity, making them cooler overall
\citep{berdyugina05,somerspinsonneault15,somers+20}.
Magnetic activity tends to increase at lower masses, leading
to substantial deviations in the observed properties of stars
compared to predictions from models such as {\mist} which does not take these
effects into account. The second reason we choose to apply corrections
in this domain is that detailed modeling of binaries already suggests
that the {\mist} models deviate slightly from the observations
in these two domains \citep{choi+16}.

To keep our empirical corrections as simple as possible,
we only introduce corrections for masses below $M_{\rm init} = 1\,M_\odot$
and ``suppress'' the effects of our derived corrections after
stars evolve off the MS (i.e. after stars have ${\rm EEP} > 454$)
and for sub-solar metallicities (where the fits are relatively unconstrained;
see \S\ref{subsec:calib_benchmark}). We further assume
that our corrections only involve a single parameter, $M_{\rm init}$,
and that they are fully linear. While we experimented with more
complex functional forms, we found that there was not
enough data to warrant using them.
Note that these corrections \textit{do not} deal in any way with
evolved stars, such as the red clump and horizontal giant branch.
While there are known disagreements between the {\mist} models
and observations at those particular stellar evolutionary phases 
\citep{choi+16}, investigating them is beyond the scope of this work.

Altogether, we end up modeling variations in 
$\Delta \log T_{\rm eff} \equiv \log T_{\rm eff}' - \log T_{\rm eff}$ and
$\Delta \log R_\star \equiv \log R_\star' - \log R_\star$ using
\begin{align}
    \Delta \log T_{\rm eff}(\params) &= 
    f_T(M_{\rm init}) \times g({\rm EEP}) \times h(\feh_{\rm init}) \\
    \Delta \log R_\star(\params) &= 
    f_R(M_{\rm init}) \times g({\rm EEP}) \times h(\feh_{\rm init})
\end{align}
We then propagate these modifications to derive relative
corrections $\Delta \log g \equiv \log g' - \log g$ and 
$\Delta \log L_{\rm bol}(T_{\rm eff}') \equiv 
\log L_{\rm bol}'(T_{\rm eff}') - \log L_{\rm bol}(T_{\rm eff}')$
via
\begin{align}
    \Delta \log g &= -2 \times \Delta \log R_{\star} \\
    \Delta \log L_{\rm bol}(T_{\rm eff}') &= 2 \times \Delta \log R_{\star}
\end{align}

Our corrections in $\Delta \log T_{\rm eff}$ 
and $\Delta \log R_\star$ have three components. The first,
$f_T(M_{\rm init})$ and $f_R(M_{\rm init})$, represent the ``baseline'' shifts
in effective temperature and radius, respectively, and are defined to be
piece-wise linear functions of $M_{\rm init}$ such that
\begin{align}
    f_T(M_{\rm init}) &=
    \begin{cases}
    \log\left[1 + c_T \times (M_{\rm init} - 1) \right] 
    & M_{\rm init} < 1\,M_\odot \\
    0 & M_{\rm init} \geq 1\,M_\odot
    \end{cases} \\
    f_R(M_{\rm init}) &=
    \begin{cases}
    \log\left[1 + c_R \times (M_{\rm init} - 1) \right] 
    & M_{\rm init} < 1\,M_\odot \\
    0 & M_{\rm init} \geq 1\,M_\odot
    \end{cases}
\end{align}
where $c_T$ and $c_R$ can be seen as characterizing the (fractional) offset
in $T_{\rm eff}$ and $R_\star$, respectively, as a function of $M_{\rm init}$.
The second, $g({\rm EEP}$, represents a ``suppression'' term that reduces 
the offsets as stars evolve off the MS. We take this to be a modified logistic
function
\begin{equation}
    g({\rm EEP}) = 
    1 - \frac{1}{1 + \exp\left[-({\rm EEP} - 454) / \Delta_{\rm EEP}\right]}
\end{equation}
where $\Delta_{\rm EEP}$ sets the EEP scale over 
which this suppression takes place. This function
strictly decreases from $1 \rightarrow 0$ as a star evolves off the MS.
Finally, $h(\feh_{\rm init})$ represents an additional ``suppression''
term that reduces the offsets at sub-solar metallicities. We take this to be
a simple exponential
\begin{equation}
    h(\feh_{\rm init}) = \exp\left(A_{\rm \feh} \times \feh_{\rm init}\right)
\end{equation}
where $A_{\feh}$ sets the amplitude of the suppression 
($1 / A_{\feh}$ can also be thought of as a scale).
We note that while this function does suppress contributions at sub-solar
metallicities, it actually give a slight \textit{enhancement}
for super-solar metallicites. The tests performed 
in \S\ref{subsec:calib_benchmark} and \S\ref{subsec:calib_field}
did not provide enough evidence to support/refute this effect, and
so we opted to leave it in for simplicity.

While we consider $c_T$ and $c_R$ as free parameters we are
interested in modeling in addition to the cluster parameters
$\params_{\rm cluster}$, we find we do not have enough data
to model $\Delta_{\rm EEP}$ and $A_{\feh}$ reliably. After experimenting
with a variety of values using the data described 
in \S\ref{subsec:calib_benchmark} as well as
observations of Ruprecht 106 taken from \citet{dotter+18},
we ultimately set $\Delta_{\rm EEP}=30$ and $A_{\feh} = 0.5$.
See Table \ref{tab:iso_corr} for a summary of the parameters used to model
empirical corrections used in this work and their final set of values.


A representation of the impact these empirical corrections have on 
our cluster model is shown in the fourth panel of Figure \ref{fig:iso_fit}.
An illustration of the functional forms used for the corrections
themselves is shown in Figure \ref{fig:iso_corr}. We find that the
overall empirical corrections substantially improve behavior down to
$M_{\rm init} \sim 0.5\,M_\odot$, which can be seen more clearly in
Figure \ref{fig:ngc2682} and \S\ref{ap:cluster}.

\subsection{Empirical Photometric Corrections} \label{subsec:cluster_ephot}

Although we find the set of empirical corrections outlined 
in \S\ref{subsec:cluster_eiso} substantially improve the overall
shape and overall offset of the {\mist} isochrones relative to data from
nearby open clusters, they are unable to address overall offsets
in photometry that might come from, e.g., slightly different
photometric calibrations between the synthetic photometry
computed from models versus the real photometry from surveys
(such as in the Vega system) or issues with the {\ctk} stellar 
atmosphere models. Both of these manifest themselves, to first order,
as \textit{photometric offsets} between the model predictions and the
observed data.

We model these offsets explicitly by introducing a set of scale-factors
$\mathbf{s}_{\rm em} = \{ s_{{\rm em}, i} \}_{i=1}^{i=b}$ 
that simply rescale the \textit{data} such that
the new flux density $\hat{F}_{i,j}'$ for a given star $i$ in band $j$ is
\begin{equation}
    \hat{F}_{i,j}' = s_{{\rm em}, j} \times \hat{F}_{i, j}
\end{equation}
We treat $\mathbf{s}_{\rm em}$ as $b$ free parameters that
we are interested in modeling in addition to the cluster parameters
$\params_{\rm cluster}$. We are able to do so thanks to
the large number of available parallax measurements from
\textit{Gaia} DR2 that give independent constraints on the distance,
thereby fixing not just offsets in color but offsets in absolute
magnitude.

Our final cluster model after including these additional empirical corrections 
is shown in the fifth panel of Figure \ref{fig:iso_fit}.

\section{Benchmark Cluster Fits} \label{ap:cluster}

\begin{figure*}
\begin{center}
\includegraphics[width=\textwidth]{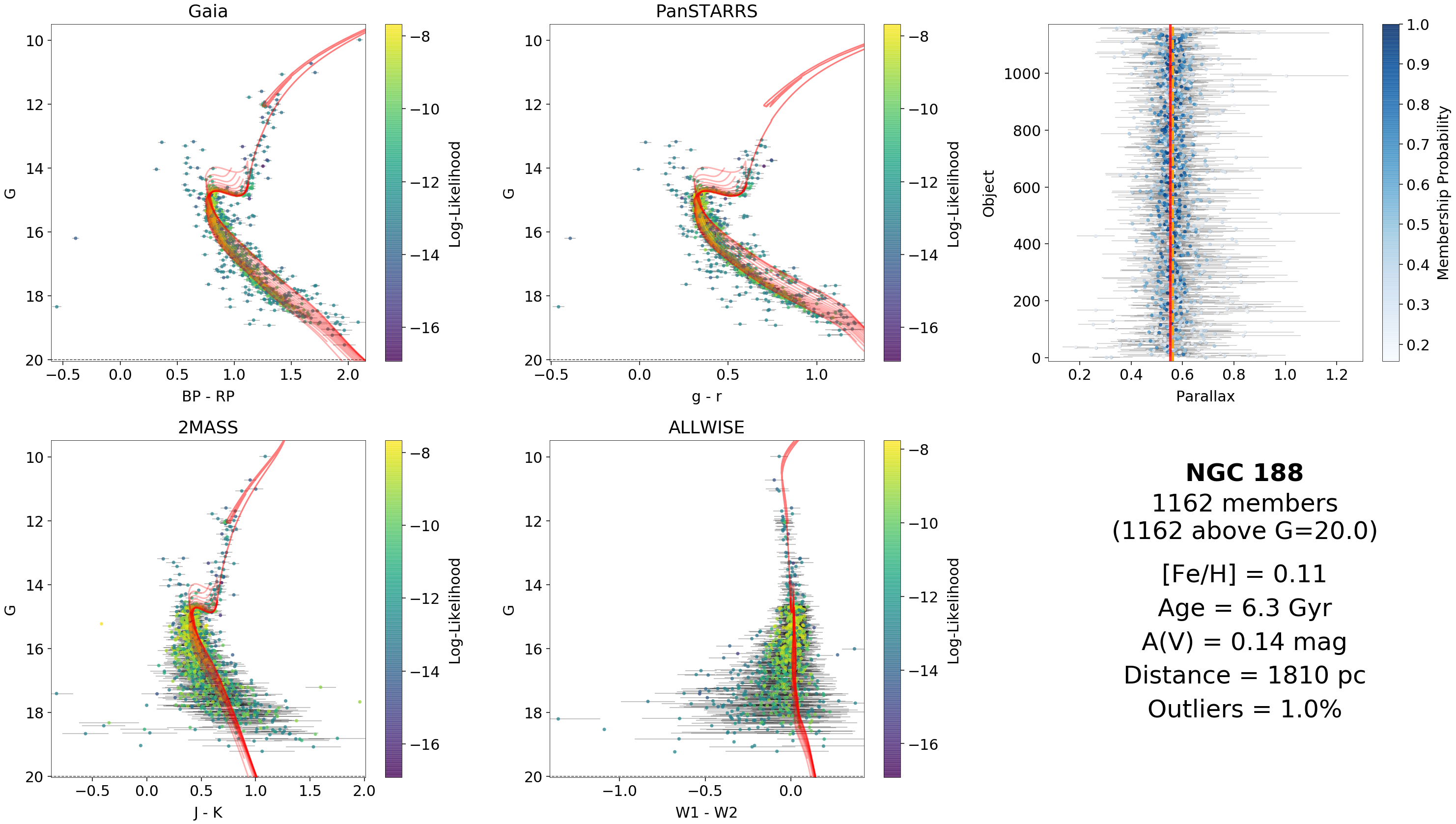}
\end{center}
\caption{As Figure \ref{fig:ngc2682}, but for NGC 188.
}\label{fig:ngc188}
\end{figure*}

\begin{figure*}
\begin{center}
\includegraphics[width=\textwidth]{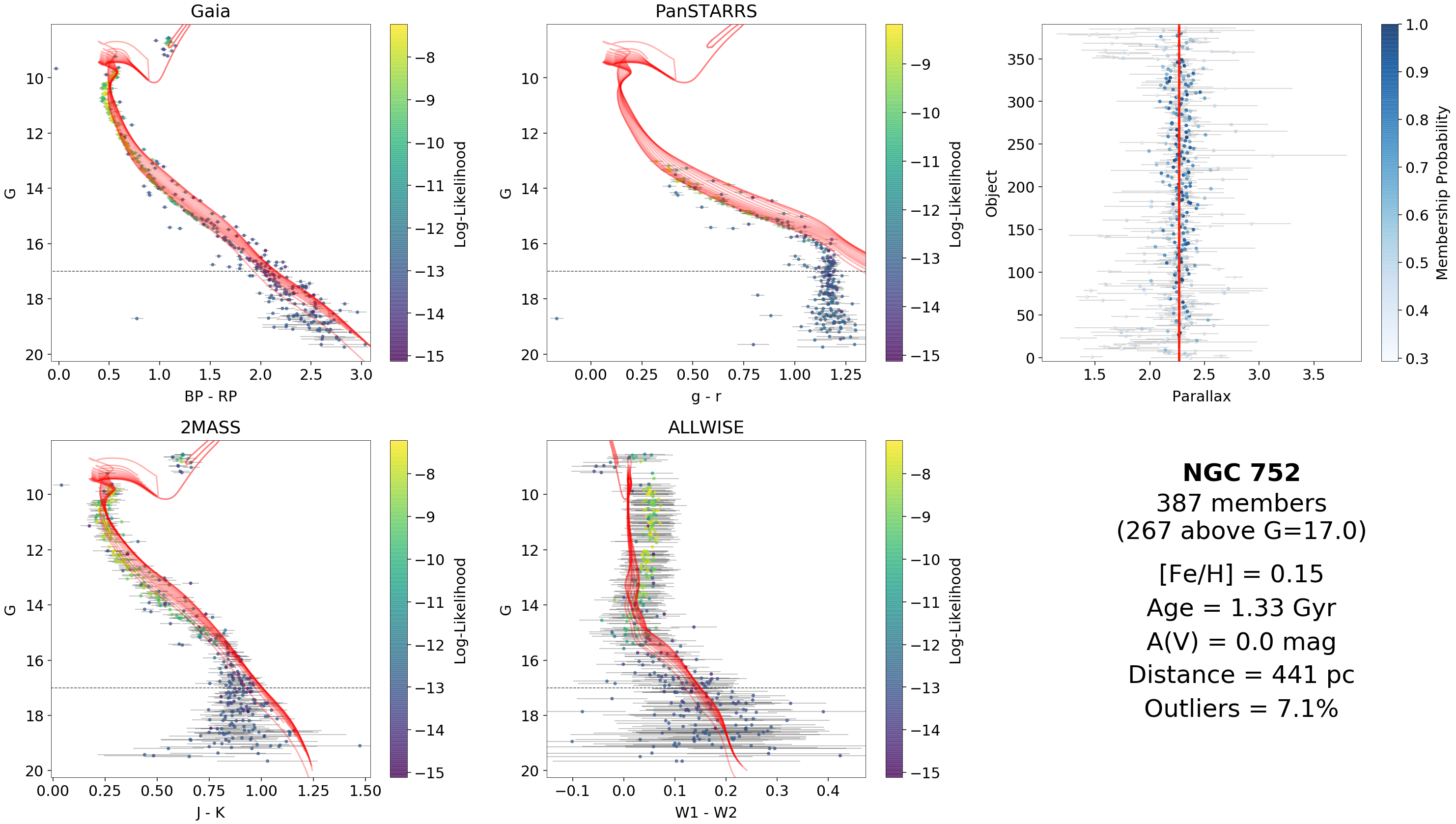}
\end{center}
\caption{As Figure \ref{fig:ngc2682}, but for NGC 752.
}\label{fig:ngc752}
\end{figure*}

\begin{figure*}
\begin{center}
\includegraphics[width=\textwidth]{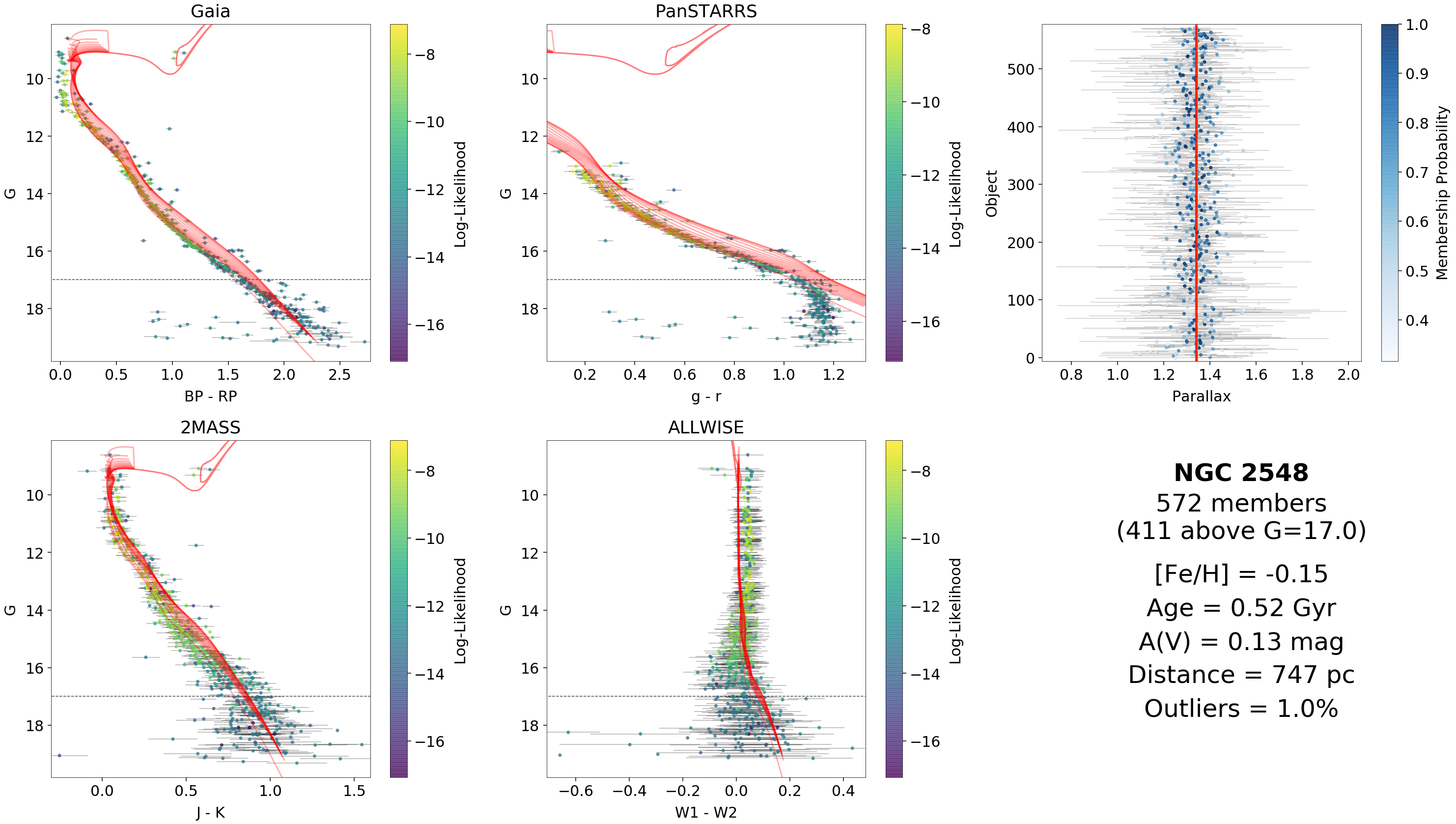}
\end{center}
\caption{As Figure \ref{fig:ngc2682}, but for NGC 2548 (i.e. M48).
}\label{fig:ngc2548}
\end{figure*}

\begin{figure*}
\begin{center}
\includegraphics[width=\textwidth]{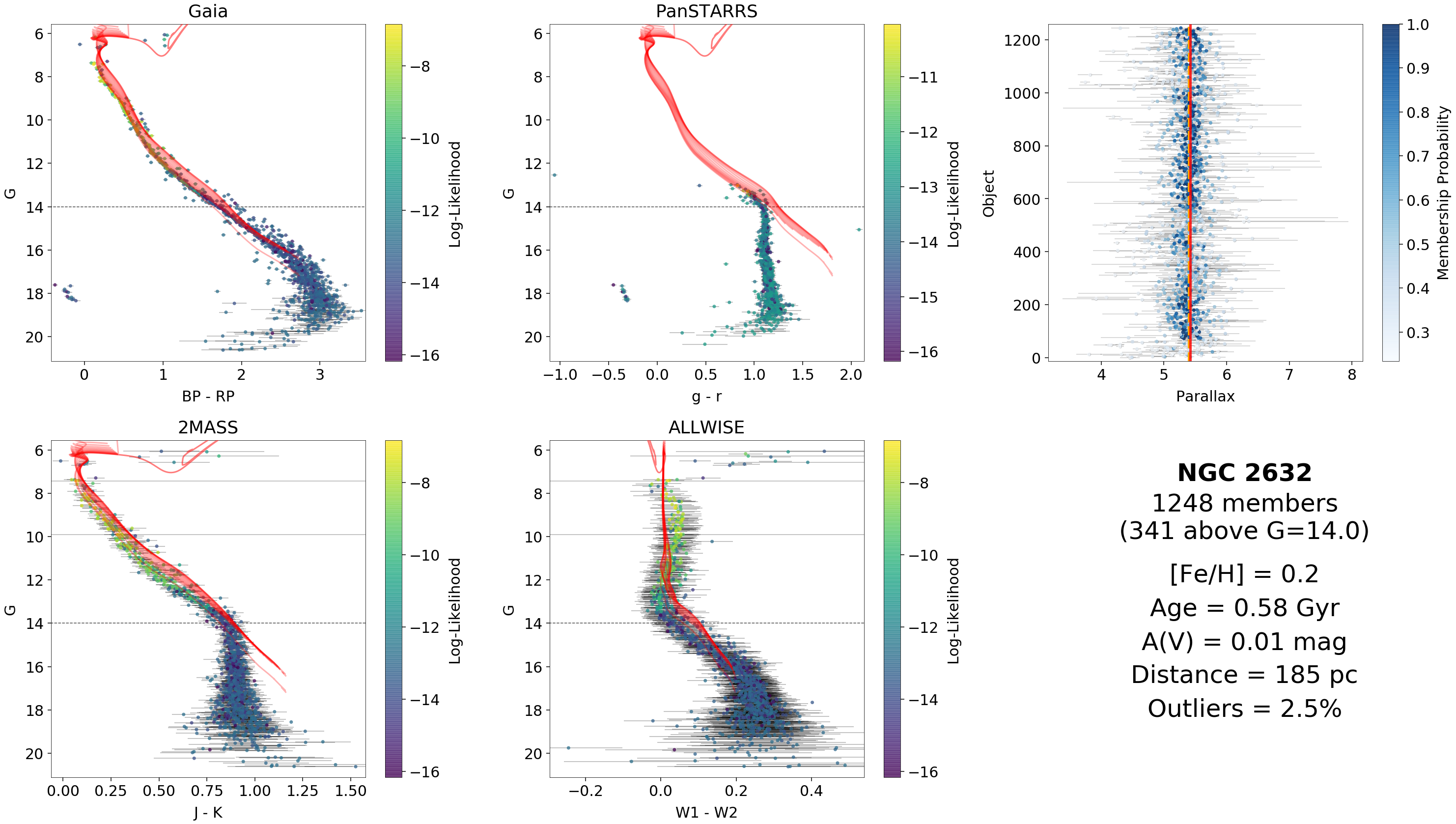}
\end{center}
\caption{As Figure \ref{fig:ngc2682}, but for NGC 2632 (i.e. Praesepe).
}\label{fig:ngc2632}
\end{figure*}

\begin{figure*}
\begin{center}
\includegraphics[width=\textwidth]{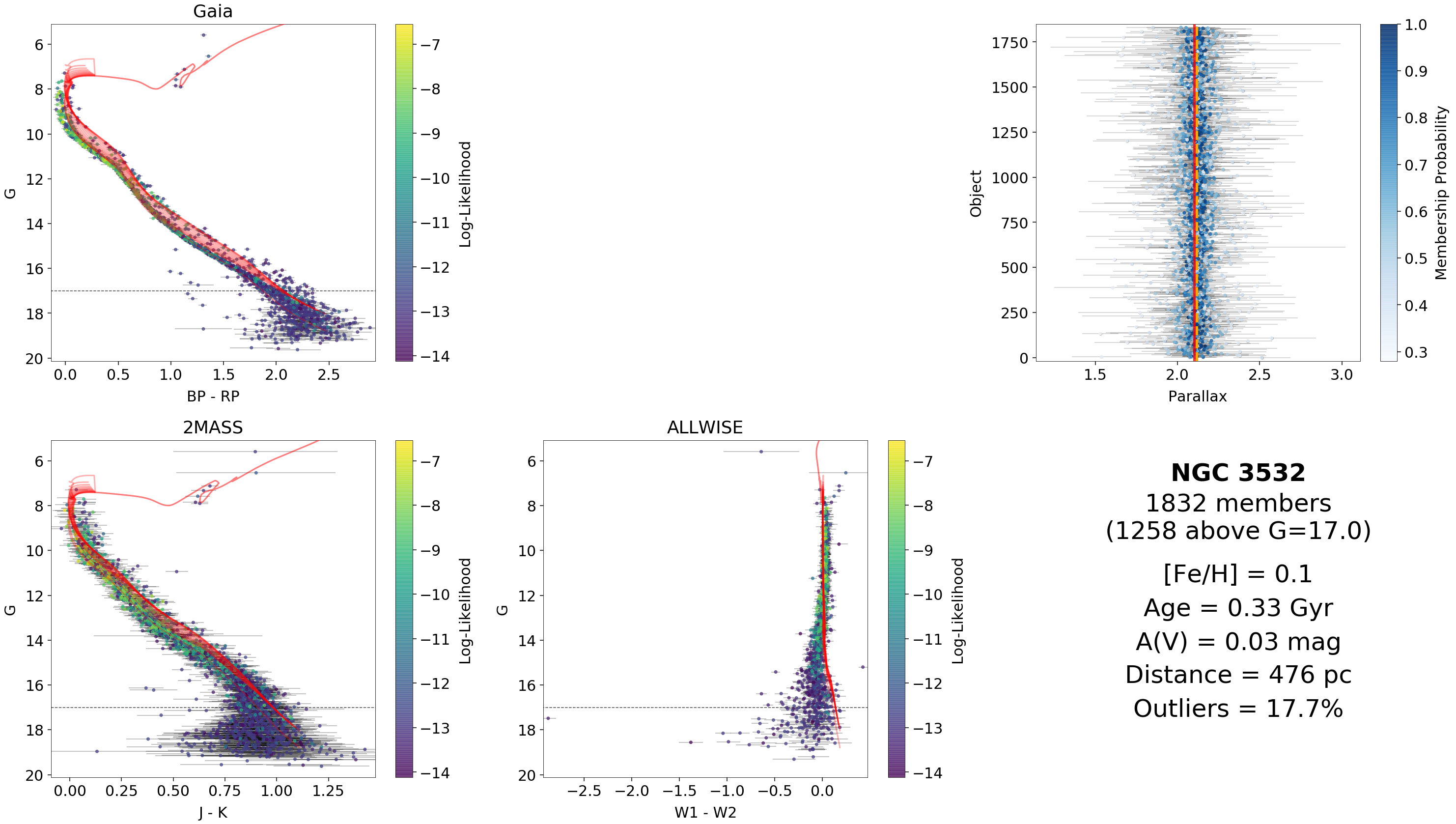}
\end{center}
\caption{As Figure \ref{fig:ngc2682}, but for NGC 3532.
Note that no Pan-STARRS data was available for use in the fit.
}\label{fig:ngc3532}
\end{figure*}

The fits to the remaining five clusters discussed in
\S\ref{subsec:calib_benchmark} are shown in Figures
\ref{fig:ngc188}, \ref{fig:ngc752},
\ref{fig:ngc2548}, \ref{fig:ngc2632},
and \ref{fig:ngc3532}.

\section{Estimating Photometric Offsets from Field Stars} \label{ap:phot_offsets}

We aim to estimate $s_{{\rm em}, j}$ in each band $j$
by computing the ratio of the observed flux
$\hat{F}_{i, j}$ for star $i$ to the 
predicted flux $F_{i, j}(\params, \eparams)$. This will be averaged over
the posterior $P(\params, \eparams|\hat{\flux}_{i, \setminus j}, \hat{\varpi}_i)$,
estimated from the measured flux densities \textit{excluding band $j$}
$\hat{\flux}_{i, \setminus j}$ for each individual source. 
Averaging across all $n$ sources then gives:
\begin{equation}
    s_{{\rm em}, j} \approx 
    \frac{1}{n} \sum_{i=1}^{n} 
    \int \frac{F_{i, j}(\params, \eparams)}{\hat{F}_{i, j}}
    P(\params, \eparams|\hat{\flux}_{i, \setminus j}, \hat{\varpi}_i)
    {\rm d}\params {\rm d}\eparams
\end{equation}

\begin{figure*}
\begin{center}
\includegraphics[width=\textwidth]{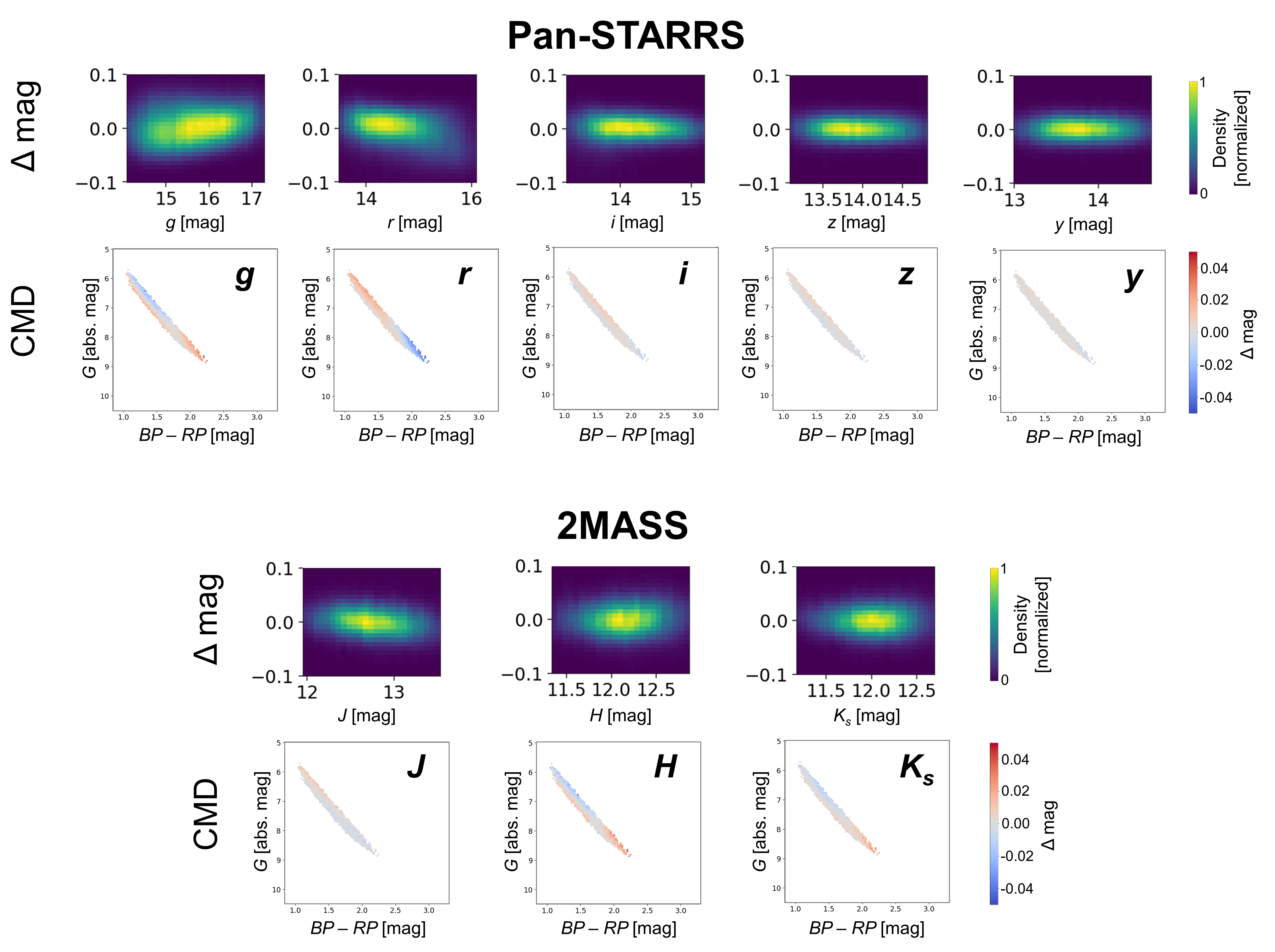}
\end{center}
\caption{Residual magnitude offsets within the {\bayestar} models
in the Pan-STARRS (top) and 2MASS (bottom) data
over the ``benchmark'' field star sample (Figure \ref{fig:gaia_cmd})
after correcting for mean photometric offsets using the procedure
described in \S\ref{subsec:calib_field}. The top row of
each set of figures shows the normalized density of stars
as a function of $\Delta {\rm mag}$ versus observed magnitude 
in each band, while the bottom row shows the
corresponding offset as a function of position on the
\textit{Gaia} $G$ versus $BP-RP$ CMD from Figure \ref{fig:gaia_cmd}.
Since the empirical {\bayestar} models were originally constructed 
using both Pan-STARRS and 2MASS photometry, as expected
the overall offsets are small ($\lesssim 1\%$) and only display 
weak trends as a function of magnitude and position of the CMD.
}\label{fig:offset_bs}
\end{figure*}

Naively, computing this would require generating $b$ sets of
$n$ posteriors (excluding each band in turn). Instead, we opt to
simply reweight the samples from the single set of $n$ posteriors
$P(F_{i, j}|\hat{\flux}_{i}, \params, \eparams)$ computed using \textit{all}
the bands. Assuming that each source $i$ has a set of
$m$ samples $\{ (\params_{i, 1}, \eparams_{i, 1}), 
\dots, (\params_{i, m}, \eparams_{i, m}) \} 
\sim P(\params, \eparams|\hat{\flux}_i, \hat{\varpi}_i)$
from the underlying posterior, our estimate then becomes
\begin{equation}
    s_{{\rm em}, j} \approx 
    \frac{1}{nm} \frac{1}{\hat{F}_{i, j}}
    \sum_{i=1}^{n} \frac{\sum_{k=1}^{m} 
    w_{i, j}(\params_{i, k}, \eparams_{i, k}) 
    \, F_{i, j}(\params_{i, k}, \eparams_{i, k})}
    {\sum_{k=1}^{m} w_{i, j}(\params_{i, k}, \eparams_{i, k})}
\end{equation}
where the importance weights $w_{i, j}(\params_{i, k}, \eparams_{i, k})$ are
defined as
\begin{equation}
    w_{i, j}(\params_{i, k}, \eparams_{i, k}) =
    \frac{P(\params, \eparams|\hat{\flux}_{i, \setminus j}, \hat{\varpi}_i)}
    {P(\params, \eparams|\hat{\flux_{i}}, \hat{\varpi}_i)}
    = \frac{\likelihood_{{\rm phot}, \setminus j}(\params_{i, k}, \eparams_{i, k})}
    {\likelihood_{\rm phot}(\params_{i, k}, \eparams_{i, k})}
\end{equation}
where $\likelihood_{{\rm phot}, \setminus j}(\params, \eparams)$
is the photometric likelihood excluding band $j$. Since computing the
photometric likelihoods with and without band $j$ is trivial,
this procedure offers a much more computationally efficient
way to estimate photometric offsets.

\begin{figure*}
\begin{center}
\includegraphics[width=\textwidth]{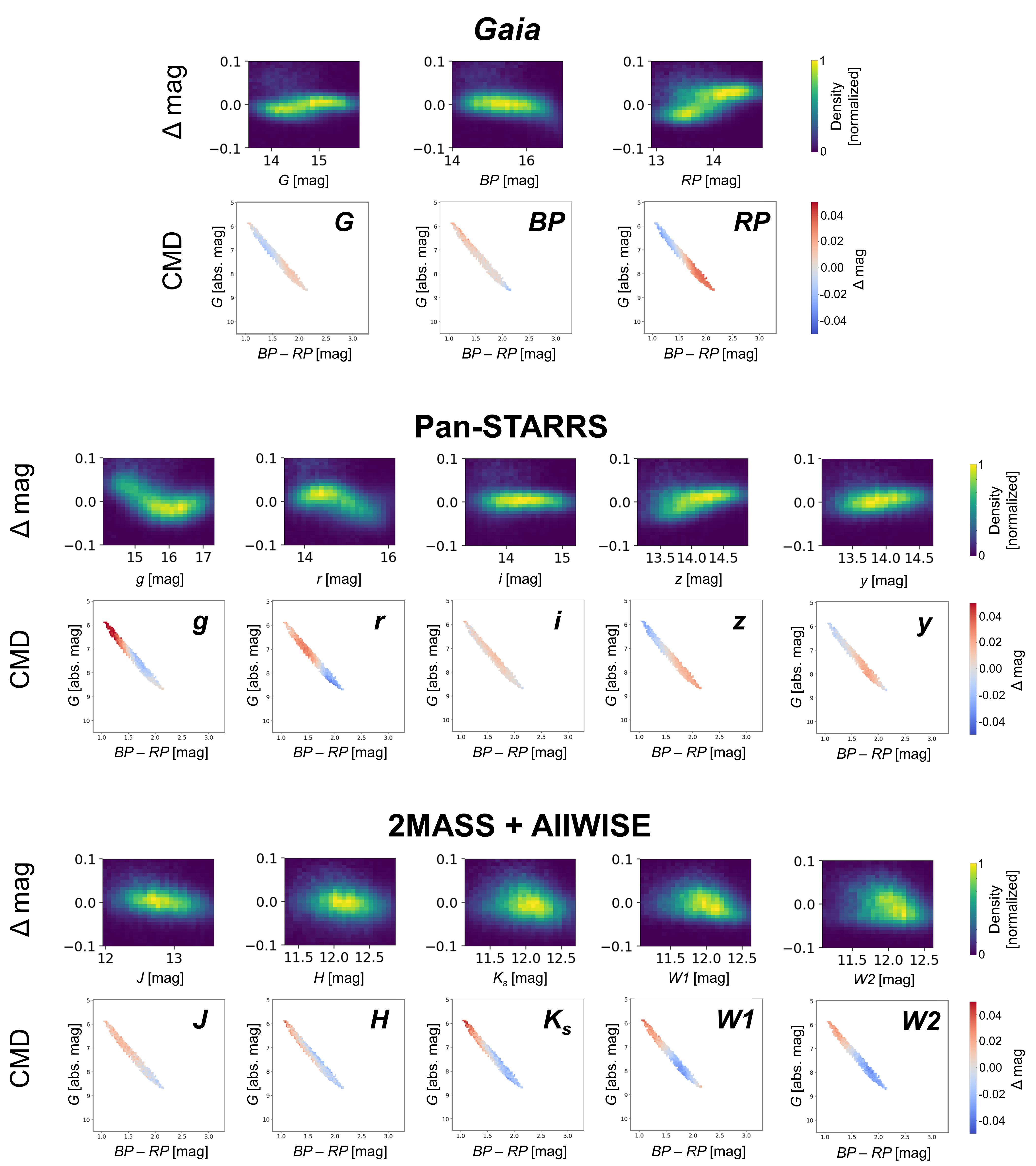}
\end{center}
\caption{As Figure \ref{fig:offset_bs}, but now showing the
residual magnitude offsets within the {\mist} models
in the \textit{Gaia} DR2 (top), Pan-STARRS, and 2MASS+AllWISE (bottom) data.
Compared to the {\bayestar} models (Figure \ref{fig:offset_bs}),
the offsets here are substantially larger ($\sim 3\%$) and display 
strong trends as a function of magnitude and position of the CMD,
especially in the bluest and reddest bands.
}\label{fig:offset_mist}
\end{figure*}

We make three small changes when applying the above strategy in practice:
\begin{enumerate}
    \item Rather than strictly using the weighted mean of the entire
    sample, we compute the median of weighted mean estimates
    from jackknife realizations ($n=1000$)
    in order to be less sensitive to possible outliers.
    \item We apply a Gaussian prior over $\mathbf{s}_{\rm em}$
    with a mean of $1$ and standard deviation of $0.01$ in each band
    to avoid behavior where a combination of (large) offsets can become
    degenerate with adjusting entire intrinsic stellar types.
    We incorporate this into the initial estimate of 
    $\mathbf{s}_{\rm em}$ computed above by using bootstrapping
    to estimate errors in our initial estimate and
    assuming the PDF is Gaussian.
    \item We repeat the process iteratively by applying the
    $\mathbf{s}_{\rm em}$ estimated from the previous stage
    before computing new estimates of the posteriors. We
    consider our results converged when each element of
    $\mathbf{s}_{\rm em}$ changes by less than $\sim 0.01$.
\end{enumerate}

\section{Photometric Offsets Across the Color-Magnitude Diagram}
\label{ap:cmd_resid}

The behavior of our estimated photometric offsets for the {\bayestar} and {\mist}
models as a function of magnitude and position on the \textit{Gaia}
CMD is shown in Figures \ref{fig:offset_bs} and \ref{fig:offset_mist},
respectively. As expected, the {\bayestar} models 
do not exhibit large variations
since they were calibrated on the same datasets (Pan-STARRS
and 2MASS) used in this work. By contrast, we see that the {\mist}
models still exhibit large variation ($\sim 0.05\,{\rm mag}$)
across both magnitude and position on the CMD, especially on the blue
side. This is in line with expectations from \citet{choi+16}.
A combination of this observed behavior, expected systematics from
each survey, and the disagreement between estimates of $\mathbf{s}_{\rm em}$
derived in \S\ref{subsec:calib_benchmark} and \S\ref{subsec:calib_field}
is then used to set the combined systematic errors in each band listed
in Table \ref{tab:phot_off}.

\section{A/B Tests} \label{ap:test_cases}

\begin{figure*}
\begin{center}
\includegraphics[width=0.8\textwidth]{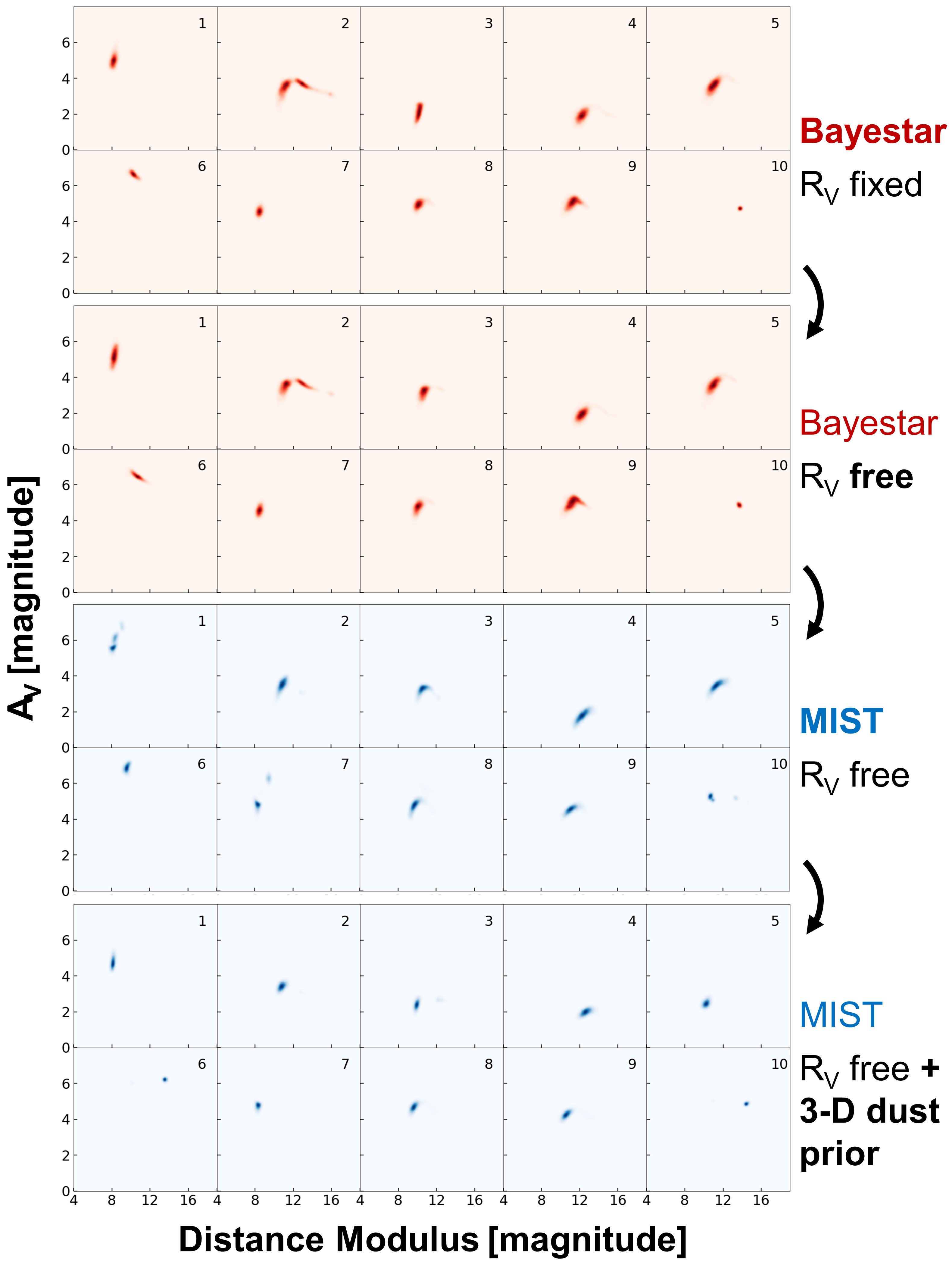}
\end{center}
\caption{Probability density functions (shaded regions)
in extinction $A_V$ and distance modulus $\mu$ for ten representative
stars along sightlines to the Orion star-forming region 
$(\ell, b) = (204.7^\circ, -19.2^\circ)$ under various
modeling assumptions. The top panels show the results for the
{\bayestar} stellar models (red) assuming $R_V$ is fixed (far top)
or allowed to vary (middle top). The bottom panels show the
results for the {\mist} models before (middle bottom) and after
(far bottom) applying the prior from the {\bayestarmap} 3-D dust map.
While some stars do not show much change, a number of sources
show marked changes in inferred distance and/or reddening, 
illustrating the impact the underlying priors
and stellar models when inferring \textit{extrinsic} parameters $\eparams$
for each source (in addition to the \textit{intrinsic} parameters $\params$).
}\label{fig:case_all}
\end{figure*}

We investigate the impact various assumptions can
have on recovery of extrinsic parameters $\eparams$, in particular
distance $d$ and extinction $A_V$. To do this, we decide to model
a subsample of stars in sightlines towards the Orion star forming
complex taken from \citet{zuckerspeagle+19}. For each star,
we model the 8-band SED (Pan-STARRS and 2MASS) along with the measured
parallax from \textit{Gaia} DR2 under four different assumptions:
\begin{itemize}
    \item \textit{Case 1}: We emulate the setup used in
    \citet{green+19} when deriving the 3-D {\bayestarmap} dust map,
    where the underlying stellar models are the {\bayestar} models 
    the $R_V$ is fixed to a value of $3.3$, and there is no 3-D dust
    prior applied.
    \item \textit{Case 2}: As above, but now allowing $R_V$
    to vary subject to the prior described in \S\ref{subsec:prior_rv}.
    This emulates the setup from \citet{zuckerspeagle+19} and
    \citet{zucker+20}.
    \item \textit{Case 3}: As above, but now using the {\mist} models
    instead of the {\bayestar} models. This approximates the setup in
    \citet{cargile+20}, where the 3-D dust extinction prior is
    extremely weak.
    \item \textit{Case 4}: As above, but now with the full 3-D
    {\bayestarmap} dust map prior applied. This is the default setup
    in {\brutus}.
\end{itemize}

The results of each of these tests for a representative subsample
of stars that were well-fit in all four cases 
is shown in Figure \ref{fig:case_all}.
While we find some stars do not exhibit many changes in their
inferred distance modulus $\mu$ or extinction $A_V$,
we do find large variations for many sources. Some of these
can be attributed to $R_V$ variation, such as with source 3,
which shows a substantial change in $A_V$ linked to shifts in
$R_V$ and changes in the inferred intrinsic parameters $\params$
as a result. We also see substantial changes for a number of sources
when moving from the {\bayestar} to the {\mist} models, many of
which are either caused by changes in stellar type or by additional
solutions now made available due to the increased diversity of
allowed stellar models (see Figure \ref{fig:mist_vs_bayestar}).
Finally, we see that the 3-D dust extinction prior
serves as a strong constraint in many cases, with the inferred
$\mu$ and $A_V$ constraints both shifting around and tightening
considerably. These results, along with the degeneracies 
discussed in \S\ref{subsec:test_mock}, highlight the importance
of obtaining strong constraints on the dust extinction towards
various sources in order to be confident that the 
inferred $\params$ for various stars are both well-constrained
and accurate.

We also perform an additional test with an alternate codebase designed to more
closely emulate the setup (modeling, priors, etc.) used in \citet{schlafly+14} 
and \citet{green+19}.
The differences between the two approaches were minimal and could
primarily be traced back to differences in the Galactic prior.

\end{document}